\documentstyle[12pt,epsf]{article}

\renewcommand{\thefootnote}{\fnsymbol{footnote}}

\newcommand{\lnmum}{L_{\mu m}}

\begin{document}

\title{\vskip-3cm{\baselineskip14pt
\centerline{\normalsize\hfill MPI/PhT/97--029}
\centerline{\normalsize\hfill TTP97--18\footnote{The 
  complete postscript file of this
  preprint, including figures, is available via anonymous ftp at
  www-ttp.physik.uni-karlsruhe.de (129.13.102.139) as /ttp97-18/ttp97-18.ps 
  or via www at http://www-ttp.physik.uni-karlsruhe.de/cgi-bin/preprints.}}
\centerline{\normalsize\hfill hep-ph/9705254}
\centerline{\normalsize\hfill May 1997}
}
\vskip1.5cm
Heavy Quark Current Correlators to ${\cal O}(\alpha_s^2)$\thanks{
  Supported by BMBF under Contract 057KA92P, DFG under Contract Ku 502/8-1 
  and INTAS under Contract INTAS-93-744-ext.}
}
\author{
 K.G.~Chetyrkin$^{a,b}$,
 J.H.~K\"uhn$^{c}$ 
 and
 M.~Steinhauser$^{a}$
}
\date{}
\maketitle

\begin{center}
$^a${\it Max-Planck-Institut f\"ur Physik,
    Werner-Heisenberg-Institut,\\ D-80805 Munich, Germany\\ }
  \vspace{3mm}
$^b${\it Institute for Nuclear Research, Russian Academy of Sciences,\\
   Moscow 117312, Russia\\}
  \vspace{3mm}
$^c${\it Institut f\"ur Theoretische Teilchenphysik,
    Universit\"at Karlsruhe,\\ D-76128 Karlsruhe, Germany.\\ }
\end{center}

\begin{abstract}
\noindent
In this paper the three-loop polarization functions $\Pi(q^2)$
are calculated for the cases of an external vector, axial-vector, scalar or
pseudo-scalar current. Results are presented for the
imaginary part which directly leads to the cross section
$\sigma(e^+e^-\to Z\to\mbox{hadrons})$ and to the Higgs
decay rates, respectively.

\end{abstract}

\thispagestyle{empty}
\newpage
\setcounter{page}{1}


\renewcommand{\thefootnote}{\arabic{footnote}}
\setcounter{footnote}{0}

\section{Introduction and notation}

A variety of important observables are essentially given by the
correlators of currents with different tensorial structure. The cross
section for $e^+e^-$ annihilation into hadrons is governed by the vector
current correlator, the $Z$ decay rate by a combination of vector and
axial correlators \cite{CKKRep}.
Higgs decays can be expressed by
the correlators of the corresponding scalar or pseudo-scalar current
densities. Theoretical predictions for all these quantities in one-
and two-loop approximation, corresponding to the Born level and the
order $\alpha_s$ corrections, are available since long. However,
in view of the present or the foreseeable experimental precision
improved calculations of these quantities are required, at least up to
order $\alpha_s^2$, if possible even up to order $\alpha_s^3$. For
massless quarks the NNLO result is available both for the vector
\cite{GorKatLar91SurSam91} and the scalar correlators
\cite{Che96}. Quark mass
effects are often incorporated by inclusion of the lowest one or
two terms in an expansion in $m^2/s$. This approach is well justified
for many applications \cite{CKKRep}. Even for relatively  low energy
values, down to about three to four times the quark mass this approach
leads to an adequate result if sufficiently many terms are included in
the expansion \cite{CheHarKueSte97,HarSte97}. 
Nevertheless it is desirable to calculate
the correlator for arbitrary values  of $m^2/s$ without 
directly invoking the
high momentum expansion. Recently this was achieved for both real and
imaginary parts of the three-loop polarization function 
\cite{CheKueSte96} with a heavy quark coupled to an external current.
In a first step the method was applied to the case of the vector current 
correlator.
In this paper also the axial-vector, scalar and pseudo-scalar cases
are considered. 
This completes the relevant ${\cal O}(\alpha_s^2)$ corrections
to polarization functions for neutral gauge bosons 
induced by heavy quarks, more specifically to their ``non-singlet'' parts. 
Contributions from the double-triangle diagram, giving rise to 
``singlet'' contributions, are not considered in this work and will
be treated elsewhere.

It is useful to define dimensionless variables:
\begin{eqnarray}
z\,\,=\,\,\frac{q^2}{4m^2},
&&
x\,\,=\,\,\frac{2m}{\sqrt{s}},
\end{eqnarray}
where $q$ is the external momentum of the polarization function
and $s$ is the center of mass energy in the 
process $e^+e^-\to\mbox{hadrons}$.
Then the velocity, $v$, of one of the produced quarks reads
\begin{eqnarray}
v&=&\sqrt{1-x^2}.
\end{eqnarray}
Every time the generic index $\delta$ appears without further explanation
it is understood that
$\delta$ represents one of the letters $a,v,s$ or $p$.

The polarization functions for the four cases of interest are defined by
\begin{eqnarray}
\left(-q^2g_{\mu\nu}+q_\mu q_\nu\right)\,\Pi^\delta(q^2)
+q_\mu q_\nu\,\Pi^\delta_L(q^2)
&\!\!=\!\!&
i\int dx\,e^{iqx}\langle 0|Tj^\delta_\mu(x) j^\delta_\nu(0)|0 \rangle
\,\,\,\mbox{for}\,\,\, \delta=v,a\,,
\label{eqpivadef}
\\
q^2\,\Pi^\delta(q^2)
&\!\!=\!\!&
i\int dx\,e^{iqx}
\langle 0|Tj^\delta(x)j^\delta(0)|0 \rangle
\,\,\,\mbox{for}\,\,\, \delta=s,p,
\label{eqpispdef}
\end{eqnarray}
with the currents
\begin{eqnarray}
j_\mu^v = \bar{\psi}\gamma_\mu \psi,\qquad
j_\mu^a = \bar{\psi}\gamma_\mu\gamma_5 \psi,\qquad
j^s = \bar{\psi}\psi,\qquad
j^p = i \bar{\psi}\gamma_5 \psi.
\end{eqnarray}
In Eqs.~(\ref{eqpivadef}) and (\ref{eqpispdef}) two powers
of $q$ are factored out in order to end up with dimensionless
quantities $\Pi^\delta(q^2)$. 
As we are only interested in the imaginary part the overall 
renormalization can be performed in such a way that this is 
possible.
Furthermore we ensure that $\Pi^\delta(0)=0$.
The transformation from this scheme to other
(overall) renormalization conditions is discussed in
App.~\ref{appmsbar}.
Concerning the renormalization it should be mentioned that 
for the scalar and pseudo-scalar current the combinations
$mj^s$ and $mj^p$, where $m$ is the pole mass,
have to be considered in order to arrive at finite results.
Note that $\Pi^v_L=0$ and $\Pi_L^a$ is trivially obtained from 
$\Pi^p$ through the axial Ward identity
$(q^2)^2\Pi_L^a(q^2)=4m^2q^2\,(\Pi^p(q^2)
-q^2(\partial\Pi^p(q^2)/\partial q^2)|_{q^2=0})$.

The physical observable $R(s)$ is related to $\Pi(q^2)$ by
\begin{eqnarray}
R^\delta (s)&=&12\pi\,\mbox{Im}\,\Pi^\delta(q^2=s+i\epsilon)
\qquad\qquad \mbox{for  } \delta=v,a\,,
\label{eqrtopiva}
\\
R^\delta (s)&=&8\pi\,\,\,\,\mbox{Im}\,\Pi^\delta(q^2=s+i\epsilon)
\qquad\qquad \mbox{for  } \delta=s,p\,.
\label{eqrtopisp}
\end{eqnarray}
It is convenient to define 
\begin{eqnarray}
\Pi^\delta(q^2) &=& \Pi^{(0),\delta}(q^2) 
         + \frac{\alpha_s(\mu^2)}{\pi} C_F \Pi^{(1),\delta}(q^2)
         + \left(\frac{\alpha_s(\mu^2)}{\pi}\right)^2\Pi^{(2),\delta}(q^2)
         + \ldots\,\,,
\\
\Pi^{(2),\delta} &=&
                C_F^2       \Pi_A^{(2),\delta}
              + C_A C_F     \Pi_{\it NA}^{(2),\delta}
              + C_F T   n_l \Pi_l^{(2),\delta}
              + C_F T       \Pi_F^{(2),\delta}
              + C_F T       \Pi_S^{(2),\delta},
\label{eqpi2}
\end{eqnarray}
and similarly for $R^\delta(s)$.
The abelian contribution $\Pi_A^{(2),\delta}$ is already present in 
(quenched) QED and $\Pi_{NA}^{(2),\delta}$ originates from the 
non-abelian structure
specific for QCD. The polarization functions containing a second
massless or massive quark loop are denoted 
by $\Pi_l^{(2),\delta}$ and $\Pi_F^{(2),\delta}$, respectively.
$\Pi_S^{(2),\delta}$ represents the double-triangle contribution.
Our procedure will be applied to the first three terms in 
Eq.~(\ref{eqpi2}), the last two terms will be studied elsewhere.

The paper is organized as follows: In the next section
the expressions for the polarization functions in the different
kinematical regions are provided. In Section \ref{secapprox} the 
approximation method is described and 
the results are given in Section \ref{secresults}. 
The conclusions are finally presented in Section~\ref{seccon}.

\section{Discussion of the kinematical regions}
\label{seckinreg}

This section provides a discussion of three kinematical regions
where analytical results are available and contains the 
input data required for the approximation method.

\vspace{1em}
\noindent
{\bf High energy region}
\vspace{1em}

The input from the kinematical region where $-q^2\gg m^2$ puts
stringent constraints on the form of the polarization function and
plays an important r\^ole for our procedure.
In the limit of large external momentum the
polarization function can be cast into the following form:
\begin{eqnarray}
\Pi^\delta(q^2) &=& \frac{3}{16\pi^2}\sum_{n\ge0} D^\delta_n \frac{1}{z^n}.
\end{eqnarray}
The coefficients $D_n^\delta$ contain $\ln(-z)$-terms up to third order.
For the subsequental discussion only the terms with $n=0$ and $1$
are needed. These first two terms can be calculated by simply
Taylor expanding the
diagrams in the mass
\footnote{Recently terms up to order $n=4$ for the scalar and
  pseudo-scalar correlator \cite{HarSte97} and up to $n=6$ for the 
  vector correlator \cite{CheHarKueSte97}
  have been calculated. This requires also the 
  knowledge of massive tadpole integrals.}. 
This leads to massless 
three-loop integrals for which the technique is known since long
\cite{CheTka81}.
The following decomposition of the coefficients $D^\delta_n$
is adopted:
\begin{eqnarray}
D^\delta_n &=& D^{(0),\delta}_n 
         + \frac{\alpha_s(\mu^2)}{\pi} C_F D^{(1),\delta}_n 
         + \left(\frac{\alpha_s(\mu^2)}{\pi}\right)^2 D^{(2),\delta}_n
         + \ldots\,\,,
\nonumber\\
D^{(2),\delta}_n &=& 
C_F^2 D^{(2),\delta}_{A,n}
+C_AC_F D^{(2),\delta}_{NA,n}
+C_FTn_l D^{(2),\delta}_{l,n}
+C_FT D^{(2),\delta}_{F,n},
\nonumber\\
D^{(j),\delta}_{x,n} &=& \sum_{k=0}^3\, d_{x,n,k}^{(j),\delta}
\left(\ln\frac{-q^2}{m^2}\right)^k
\qquad\qquad
j\in\{0,1,2\},\,
x\in\{A,NA,l,F\},
\end{eqnarray}
where $d_{x,n,k}^{(j),\delta}$ are numerical constants.
For $D^{(0),\delta}_{n}$ and $D^{(1),\delta}_{n}$ the sum runs only
up to $k=1$ and $k=2$, respectively.
The results for the four cases are available in the literature
\cite{GorKatLar91SurSam91,axpssc,Che96}.
\begin{table}[t]
{\footnotesize
\renewcommand{\arraystretch}{1.3}
\begin{center}
\begin{tabular}{|l||r|r|r|r||r|r|r|r|}
\hline
&\multicolumn{4}{c||}{$n=0$}
&\multicolumn{4}{c|}{$n=1$}
\\
\hline
$k$  &0&1&2&3&0&1&2&3\\ 
\hline
\hline
$D_n^{(0),v}$
&  2.2222& -1.3333&  0.0000&  0.0000&  2.0000&  0.0000&  0.0000&  0.0000\\
$D_n^{(1),v}$
& -3.9749& -1.0000&  0.0000&  0.0000&  0.0000& -3.0000&  0.0000&  0.0000\\
$D_{A,n}^{(2),v}$
&  1.3075&  0.1250&  0.0000&  0.0000&  2.7727& -0.3750&  2.2500&  0.0000\\
$D_{NA,n}^{(2),v}$
& -9.5651& -0.7175&  0.4583&  0.0000& -3.9903& -7.7083&  1.3750&  0.0000\\
$D_{l,n}^{(2),v}$
&  2.9723&  0.2306& -0.1667&  0.0000&  2.2899&  2.1667& -0.5000&  0.0000\\
$D_{F,n}^{(2),v}$
& -1.2583&  0.2306& -0.1667&  0.0000& -5.1048&  2.1667& -0.5000&  0.0000\\
\hline
\hline
$D_n^{(0),a}$
&  3.5556& -1.3333&  0.0000&  0.0000& -2.0000&  2.0000&  0.0000&  0.0000\\
$D_n^{(1),a}$
& -2.0860& -1.0000&  0.0000&  0.0000&  1.9623&  1.5000& -1.5000&  0.0000\\
$D_{A,n}^{(2),a}$
& -1.0387&  0.1250&  0.0000&  0.0000& -8.7928&  0.2672& -1.3125&  0.7500\\
$D_{NA,n}^{(2),a}$
& -9.8007& -0.7175&  0.4583&  0.0000& 14.8142& -1.5224& -4.5417&  0.4583\\
$D_{l,n}^{(2),a}$
&  3.6266&  0.2306& -0.1667&  0.0000& -8.5657&  2.3606&  1.3333& -0.1667\\
$D_{F,n}^{(2),a}$
& -0.2639&  0.2306& -0.1667&  0.0000&  6.5500& -4.5090&  1.3333& -0.1667\\
\hline
\hline
$D_n^{(0),s}$
&  5.3333& -2.0000&  0.0000&  0.0000& -3.0000&  3.0000&  0.0000&  0.0000\\
$D_n^{(1),s}$
& -3.9623& -4.5000&  1.5000&  0.0000&  7.8185&  3.0000& -4.5000&  0.0000\\
$D_{A,n}^{(2),s}$
&  2.0416&  0.8578&  3.5625& -0.7500& -6.0543&-10.0450& -5.0625&  4.5000\\
$D_{NA,n}^{(2),s}$
&-28.2865& -5.2693&  5.9167& -0.4583& 34.4008& -7.3623&-12.9375&  1.3750\\
$D_{l,n}^{(2),s}$
& 15.5016& -0.5273& -1.8333&  0.1667&-21.3855&  9.3091&  3.7500& -0.5000\\
$D_{F,n}^{(2),s}$
&-10.1876&  6.3423& -1.8333&  0.1667& 18.5432&-14.2997&  3.7500& -0.5000\\
\hline
\hline
$D_n^{(0),p}$
&  4.0000& -2.0000&  0.0000&  0.0000&  1.0000&  1.0000&  0.0000&  0.0000\\
$D_n^{(1),p}$
& -2.9623& -4.5000&  1.5000&  0.0000&  2.6062& -3.0000& -1.5000&  0.0000\\
$D_{A,n}^{(2),p}$
&  1.0691&  0.8578&  3.5625& -0.7500& -5.6088& -6.8483&  4.3125&  1.5000\\
$D_{NA,n}^{(2),p}$
&-17.6658& -5.2693&  5.9167& -0.4583&  6.0218&-13.6485& -2.4792&  0.4583\\
$D_{l,n}^{(2),p}$
& 10.9195& -0.5273& -1.8333&  0.1667& -0.9696&  6.3253&  0.5833& -0.1667\\
$D_{F,n}^{(2),p}$
&-11.5909&  6.3423& -1.8333&  0.1667&  0.8079& -3.5443&  0.5833& -0.1667\\
\hline
\end{tabular}
\end{center}
}
\caption{\label{tabdn}
Numerical values for the coefficients $D_n^\delta$ 
at one-, two- and three-loop level for $\mu^2=m^2$.
}
\end{table}
We have independently repeated this calculation. 
The results in the on-shell scheme are listed in Tab.~\ref{tabdn}.
Due to our renormalization condition ($\Pi^\delta(0)=0$)
for a comparison with \cite{GorKatLar91SurSam91,axpssc,Che96}
the terms presented in App.~\ref{appmsbar} have to be taken into
account.
The analytic expressions for $D_n^v$ can be found in
\cite{CheHarKueSte97} 
and for $D_n^s$ and $D_n^p$ in
\cite{HarSte97}. $D_0^a$ and $D_1^a$ are listed in 
App.~\ref{appdn}.

This information will serve as input for the procedure described in
Section \ref{secapprox}. It should be noted that in contrast to the
vector case treated in 
\cite{CheKueSte96}
the axial-vector, scalar and pseudo-scalar correlator 
also develop cubic logarithms $\ln^3(-z)$. If we had adopted
the $\overline{\mbox{MS}}$ definition for the mass and $\mu^2=q^2$,
these cubic logarithms would vanish.

\vspace{1em}
\noindent
{\bf Behaviour at $q^2=0$}
\vspace{1em}

An important input to the behaviour of the polarization function
originates from the Taylor expansion around $q^2=0$. In this case the
three-loop diagrams have to be expanded in the external momentum leading to 
massive tadpole integrals. The calculation is performed with the help of the
algebraic program MATAD written in FORM
\cite{VerFORM}.
It automatically expands in $q$ up to the desired order, performs the 
traces and applies recurrence relations \cite{Bro92}
to reduce the many different diagrams to a small set of master integrals.
The structure of $\Pi(q^2)$ is as follows:
\begin{eqnarray}
\Pi^\delta(q^2) &=& \frac{3}{16\pi^2}\sum_{n>0} C^\delta_n z^n, 
\nonumber\\
C^\delta_n &=& C^{(0),\delta}_n 
         + \frac{\alpha_s(\mu^2)}{\pi} C_F C^{(1),\delta}_n 
         + \left(\frac{\alpha_s(\mu^2)}{\pi}\right)^2 C^{(2),\delta}_n
         + \ldots\,\,,
\nonumber\\
C^{(2),\delta}_n &=& 
C_F^2 C^{(2),\delta}_{A,n}
+C_AC_F C^{(2),\delta}_{NA,n}
+C_FTn_l C^{(2),\delta}_{l,n}
+C_FT C^{(2),\delta}_{F,n}.
\end{eqnarray}
Although the calculation is performed analytically the results 
are listed in numerical form in Tab.~\ref{tabcn}
with the choice $\mu^2=m^2$.
The analytic expressions are given in App.~\ref{appcn}.

\begin{table}[ht]
{\footnotesize
\renewcommand{\arraystretch}{1.15}
\begin{center}
\begin{tabular}{|l|l|r|r|r|r|r|r|}
\hline
&&&&&&&\\[-4mm]
  $\delta$  &  $n$  &  $C_n^{(0),\delta}$  &  $C_n^{(1),\delta}$  
            &  $C_{A,n}^{(2),\delta}$ 
            &  $C_{{\it NA},n}^{(2),\delta}$  &  $C_{l,n}^{(2),\delta}$  
            &  $C_{F,n}^{(2),\delta}$ \\
&&&&&&&\\[-4mm]
\hline
\hline
$v$
&1&  1.06667&  4.04938&  5.07543&  7.09759& -2.33896&  0.72704\\
&2&  0.45714&  2.66074&  6.39333&  6.31108& -2.17395&  0.26711\\
&3&  0.27090&  2.01494&  6.68902&  5.39768& -1.89566&  0.14989\\
&4&  0.18470&  1.62997&  6.68456&  4.69907& -1.67089&  0.09947\\
&5&  0.13640&  1.37194&  6.57434&  4.16490& -1.49436&  0.07230\\
&6&  0.10609&  1.18616&  6.42606&  3.74591& -1.35348&  0.05566\\
&7&  0.08558&  1.04568&  6.26672&  3.40886& -1.23871&  0.04459\\
&8&  0.07094&  0.93558&  6.10789&  3.13175& -1.14341&  0.03677\\
\hline
\hline
$a$
&1&  0.53333&  1.70123&  2.31402&  2.71368& -1.04643&  0.34325\\
&2&  0.15238&  0.71577&  1.64824&  1.69284& -0.66383&  0.06509\\
&3&  0.06772&  0.39678&  1.18988&  1.09011& -0.43131&  0.02238\\
&4&  0.03694&  0.25243&  0.89978&  0.75793& -0.30145&  0.01013\\
&5&  0.02273&  0.17477&  0.70769&  0.55831& -0.22280&  0.00540\\
&6&  0.01516&  0.12819&  0.57409&  0.42925& -0.17167&  0.00320\\
&7&  0.01070&  0.09805&  0.47720&  0.34095& -0.13656&  0.00205\\
&8&  0.00788&  0.07743&  0.40449&  0.27780& -0.11138&  0.00139\\
\hline
\hline
$s$
&1&  0.80000&  0.45185&  0.03484& -2.51105&  0.88148&  0.71856\\
&2&  0.22857&  0.77651&  1.42576&  0.89546& -0.35912&  0.19112\\
&3&  0.10159&  0.52152&  1.52607&  0.99889& -0.40023&  0.07144\\
&4&  0.05541&  0.35707&  1.32709&  0.82750& -0.33359&  0.03362\\
&5&  0.03410&  0.25650&  1.11491&  0.66459& -0.26913&  0.01832\\
&6&  0.02273&  0.19215&  0.93714&  0.53778& -0.21845&  0.01103\\
&7&  0.01605&  0.14892&  0.79516&  0.44174& -0.17982&  0.00715\\
&8&  0.01182&  0.11862&  0.68238&  0.36853& -0.15024&  0.00489\\
\hline
\hline
$p$
&1&  1.33333&  2.33333&  2.71218& -1.85805&  0.92593&  1.31106\\
&2&  0.53333&  2.61481&  7.03952&  3.57843& -1.23162&  0.49637\\
&3&  0.30476&  2.12783&  8.27333&  4.10409& -1.50248&  0.25889\\
&4&  0.20317&  1.75361&  8.46632&  3.93931& -1.47479&  0.16180\\
&5&  0.14776&  1.48309&  8.33148&  3.65904& -1.38655&  0.11237\\
&6&  0.11366&  1.28241&  8.09038&  3.38071& -1.29093&  0.08353\\
&7&  0.09093&  1.12869&  7.82133&  3.13004& -1.20154&  0.06509\\
&8&  0.07488&  1.00753&  7.55403&  2.91007& -1.12143&  0.05249\\
\hline
\end{tabular}
\end{center}
}
\caption{\label{tabcn} 
Numerical values for the coefficients $C_n^\delta$ at one-, two-
and three-loop level for $\mu^2=m^2$.
}
\end{table}

For the vector correlator the first seven coefficients 
are already listed in
\cite{CheKueSte96}, whereas all other results are new.

\vspace{1em}
\noindent
{\bf Threshold behaviour}
\vspace{1em}

At threshold it is most convenient to consider first the
information about $R^\delta(v)$ and transform this subsequently
into the
corresponding expression for $\Pi^\delta(q^2)$ via 
dispersion relations.
Whereas the treatment
of the four cases 
in the high energy and small $q^2$ region is quite similar 
there is a big difference at threshold
between the vector and pseudo-scalar correlators
on one hand and the axial-vector and scalar correlators
on the other hand.
The latter are suppressed by a factor $v^2$ w.r.t. the former.
This can already be seen by considering the Born results:
\begin{eqnarray}
R^{(0),v}\,\,=\,\,3\frac{v\left(3-v^2\right)}{2},
\quad
R^{(0),a}\,\,=\,\,3v^3,
\quad
R^{(0),s}\,\,=\,\,3v^3,
\quad
R^{(0),p}\,\,=\,\,3v.
\end{eqnarray}
At ${\cal O}(\alpha_s)$ an expansion for $v\to 0$ is helpful
for the considerations below. It reads:
\begin{eqnarray}
R^{(1),v} &=& R^{(0),v}\left(\frac{\pi^2\left(1+v^2\right)}{2v}-4\right)
             +{\cal O}(v^3),
\label{eqrv1l}
\\
R^{(1),a} &=& R^{(0),a}\left(\frac{\pi^2\left(1+v^2\right)}{2v}-2\right)
             +{\cal O}(v^5),
\\
R^{(1),s} &=& R^{(0),s}\left(\frac{\pi^2\left(1+v^2\right)}{2v}-1\right)
             +{\cal O}(v^5),
\\
R^{(1),p} &=& R^{(0),p}\left(\frac{\pi^2\left(1+v^2\right)}{2v}-3\right)
             +{\cal O}(v^3).
\label{eqrp1l}
\end{eqnarray}
The exact results can be found in
\cite{JerLaeZer82,DreHik90}.
The analytical evaluation of the double-bubble diagrams with massless
fermion loop insertions indicates
\cite{HoaKueTeu95}
that the characteristic scale of the first two terms 
proportional to $\pi^2$
is given by
the relative momentum, the last term (which is due to hard
transversal gluon exchange) by the mass of the heavy fermion. This 
motivates the decomposition adopted in Eqs.~(\ref{eqrv1l}-\ref{eqrp1l}).

Let us at ${\cal O}(\alpha_s^2)$ first discuss the abelian
contribution proportional to $C_F^2$. The ratio between
$R^{(2),\delta}_A$ and $R^{(0),\delta}$ is proportional to the Sommerfeld
factor $y/(1-e^{-y})$ with $y=C_F\pi\alpha_s/v$, which resums
contributions of the form $(\alpha_s/v)^n$. Axial-vector
and scalar 
correlators follow the P-wave scattering solution for the Coulomb
potential. Hence an additional
factor $(1+y^2/(4\pi^2))$ has to be
taken into account
in order to obtain the correct
leading term of ${\cal O}(\alpha_s^2)$
\cite{FadKho91}.
For the vector and pseudo-scalar contribution also the next-to-leading
term in $v$ can be determined by taking into account the correction
factor arising from the exchange of transversal gluons which reads 
$(1-C_F 4\alpha_s/\pi)$ for the vector and
$(1-C_F 3\alpha_s/\pi)$ for the pseudo-scalar case.
As $R_A^{(2),a}$ and $R_A^{(2),s}$ 
already start at ${\cal O}(v)$ the corresponding factors are not 
considered.
Finally we arrive at
\begin{eqnarray}
R_A^{(2),v}\,\,=\,\,3\left(\frac{\pi^4}{8v}-3\pi^2\right) +{\cal O}(v),
&&
R_A^{(2),a}\,\,=\,\, 3\left(\frac{\pi^2(3+\pi^2)}{12}v\right) +{\cal O}(v^2),
\label{eqAthr}\\
R_A^{(2),s} \,\,=\,\, 3\left(\frac{\pi^2(3+\pi^2)}{12}v\right) +{\cal O}(v^2),
&&
R_A^{(2),p} \,\,=\,\, 
      3\left(\frac{\pi^4}{12v}-\frac{3}{2}\pi^2\right) +{\cal O}(v).
\end{eqnarray}
It is, of course, necessary to incorporate
the strong $1/v$ singularity into the approximation method
for the vector and pseudo-scalar case.
In contrast the axial-vector and scalar current correlator are 
very smooth at threshold
so that these terms are not used for our procedure. The comparison
of these Pad\'e results with the exact terms at threshold will be 
performed in Section \ref{secresults} and demonstrates that the threshold
behaviour is well reproduced --- an independent test of our method. 

The analytical results for the three-loop diagrams where a massless 
quark loop is inserted into the
gluon propagator plus the corresponding real corrections
are available for the vector current correlator
\cite{HoaKueTeu95}
and the remaining correlators as well\footnote{We would like to thank
the authors of \cite{HoaTeu96} for providing their results prior to 
publication.}
\cite{HoaTeu96}
in analytic form. 
The scalar case can also be found in
\cite{Mel96}.
Expanding the results near threshold leads to
\begin{eqnarray}
R^{(2),v}_l &=& R^{(0),v}\,\,\frac{\pi^2\left(1+v^2\right)}{v}
                \left(\frac{1}{6}\ln\frac{v^2s}{\mu^2}-\frac{5}{18}\right)
           +{\cal O}(v),
\\
R^{(2),a}_l &=& R^{(0),a}\,\,\frac{\pi^2\left(1+v^2\right)}{v}
                \left(\frac{1}{6}\ln\frac{v^2s}{\mu^2}-\frac{11}{18}\right)
           +{\cal O}(v^3),
\\
R^{(2),s}_l &=& R^{(0),a}\,\,\frac{\pi^2\left(1+v^2\right)}{v}
                \left(\frac{1}{6}\ln\frac{v^2s}{\mu^2}-\frac{11}{18}\right)
           +{\cal O}(v^3),
\\
R^{(2),p}_l &=& R^{(0),p}\,\,\frac{\pi^2\left(1+v^2\right)}{v}
                \left(\frac{1}{6}\ln\frac{v^2s}{\mu^2}-\frac{5}{18}\right)
           +{\cal O}(v).
\end{eqnarray}
We include subleading terms proportional to $\ln v$ in this expansion,
since the agreement of our approximation improves visibly in those cases 
where the analytical result is known.

In order to get the threshold behaviour for the 
non-abelian part it is either
possible to use the QCD potential and the perturbative relation
between $\alpha_V({\vec{q}}\,^2)$ and $\alpha_s(\mu^2)$ or to 
proceed as demonstrated in
\cite{CheHoaKueSteTeu96}
and deduce the gluonic double-bubble diagram, $R_g^{(2),\delta}$,
from the corresponding fermionic contribution and evaluate it for 
a special choice of the gauge parameter $\xi$.
This is based on the observation that 
the terms proportional to $C_A$ in the relation between
$\alpha_V({\vec{q}}\,^2)$ and $\alpha_s(\mu^2)$ are covered by
the (one-loop) gluon propagator choosing $\xi=4$.
We will choose the second method since
this trick is used also for the actual calculation.
Following 
\cite{CheKueSte96} 
the expansion of the ``double-bubble'' result for $\xi=4$ is taken
to represent the expansion of the full non-abelian part. For the
four correlators it is given by:
\begin{eqnarray}
R^{(2),v}_{NA} &=& R^{(0),v}\,\,\frac{\pi^2\left(1+v^2\right)}{v}
                \left(-\frac{11}{24}\ln\frac{v^2s}{\mu^2}+\frac{31}{72}\right)
           +{\cal O}(v),
\label{eqthrvna}
\\
R^{(2),a}_{NA} &=& R^{(0),a}\,\,\frac{\pi^2\left(1+v^2\right)}{v}
                \left(-\frac{11}{24}\ln\frac{v^2s}{\mu^2}+\frac{97}{72}\right)
           +{\cal O}(v^3),
\\
R^{(2),s}_{NA} &=& R^{(0),s}\,\,\frac{\pi^2\left(1+v^2\right)}{v}
                \left(-\frac{11}{24}\ln\frac{v^2s}{\mu^2}+\frac{97}{72}\right)
           +{\cal O}(v^3),
\\
R^{(2),p}_{NA} &=&R^{(0),p}\,\,\frac{\pi^2\left(1+v^2\right)}{v}
                \left(-\frac{11}{24}\ln\frac{v^2s}{\mu^2}+\frac{31}{72}\right)
           +{\cal O}(v).
\label{eqthrpna}
\end{eqnarray}

To combine the results from different 
kinematical regions the above expressions for the imaginary part have to be
transformed into analytical functions for $\Pi^\delta(q^2)$ which respect
Eqs.~(\ref{eqrtopiva}) and (\ref{eqrtopisp}). 
This can be done in close analogy to 
\cite{CheKueSte96}.


\section{The approximation procedure}
\label{secapprox}

This section is devoted to the description of the approximation
method. In order to save space we will not present explicit
formulae. They look very similar to the ones for the vector case
discussed in
\cite{CheKueSte96}.
The treatment of the abelian part of the pseudo-scalar correlator is in
close analogy to \cite{BaiBro95}.

In a first step a function $\tilde{\Pi}^\delta(q^2)$ is constructed which 
contains no high energy singularities and no logarithmic
terms at threshold. This is achieved with the help of the function
\begin{eqnarray}
G(z)=\frac{2u\ln u}{u^2-1},\,\,\,\,  
&&
u=\frac{\sqrt{1-1/z}-1}{\sqrt{1-1/z}+1}.
\end{eqnarray}
The combination $(1-z)G(z)$ has a polynomial behaviour for $z\to0$
and vanishes at threshold ($z\to1$). For the case $z\to -\infty$ the
expansion of $G(z)$ develops logarithms starting with
$\ln(-1/(4z))/(2z)$. This property is exploited and a function of the
form
\begin{eqnarray}
\sum_{n,m,l}\,c_{nml}\,z^n(1-z)^m \left(G(z)\right)^l,
\end{eqnarray}
where $n$ is an integer and $m,l\ge0$ is constructed in order to remove
the $\ln(-z)$ terms of $\Pi^\delta(q^2)$. Whereas for the vector case 
described in
\cite{CheKueSte96}
no cubic logarithms appear (see Tab.~\ref{tabdn}) and therefore
quadratic combinations in $G(z)$ are sufficient, for the 
other three cases this is not true: The axial-vector correlator
develops $\ln^3(-z)/z$ terms and combinations like 
$(1-z)^2(G(z))^3$ are required. For the scalar and pseudo-scalar case
also cubic logarithms appear which are not suppressed by powers of $z$,
whence terms like $z(1-z)^2(G(z))^3$ must appear in 
the expression subtracted from $\Pi^\delta(q^2)$.

If logarithmic terms are present near threshold
($\Pi_{NA}^{(2),\delta}$ and $\Pi_{l}^{(2),\delta}$)
they are first subtracted and then the high energy singularities
are removed. For the abelian polarization function where either
$1/v$ singularities 
($\Pi_{A}^{(2),v}$ and $\Pi_{A}^{(2),p}$)
or just constant terms
($\Pi_{A}^{(2),a}$ and $\Pi_{A}^{(2),s}$)
are present for $z\to1$
only the high energy logarithms are removed.

In a second step we perform a variable change. Via the 
conformal mapping 
\begin{eqnarray}
\omega = \frac{1-\sqrt{1-q^2/4m^2}}{1+\sqrt{1-q^2/4m^2}},\,\,\,\,
&&
z = \frac{q^2}{4m^2} = \frac{4\omega}{(1+\omega)^2}.
\label{omega}
\end{eqnarray}
the complex $q^2$-plane is mapped into the interior of the unit circle
and the upper (lower) part of the cut starting at $z=1$ is mapped
onto the upper (lower) perimeter of the circle.
The special points
$q^2=0,4m^2,-\infty$ correspond to $\omega=0,1,-1$, respectively.
In this new variable we construct a function $P(\omega)$
for which the Pad\'e approximation is performed. According to
the different behaviour of $\tilde{\Pi}^\delta(q^2)$ near threshold 
actually two different functions have to be defined:
\begin{eqnarray}
P^{I}(\omega)&=&\frac{1-\omega}{(1+\omega)^2}\left(
               \tilde{\Pi}(q^2) - \tilde{\Pi}(-\infty)
                                             \right),
\\
P^{II}(\omega)&=&\frac{1}{(1+\omega)^2}\left(
               \tilde{\Pi}(q^2) - \tilde{\Pi}(-\infty)
                                             \right).
\end{eqnarray}
Thereby $P^{I}(\omega)$ takes care of the cases where a
$1/v$ singularity is present
($\Pi_{A}^{(2),v}$ and $\Pi_{A}^{(2),p}$).
The factor $(1-\omega)$ corresponds effectively to a multiplication with $v$.
At this point we should mention that in order to incorporate also the 
constant term at threshold 
which has its origin in the correction factor introduced 
before Eq.~(\ref{eqAthr}) 
the combinations
$\Pi_{A}^{(2),v}+4\Pi^{(1),v}$
and
$\Pi_{A}^{(2),p}+3\Pi^{(1),p}$
are considered.
The two-loop results 
$\Pi^{(1),v}$ and $\Pi^{(1),p}$ may be found in
\cite{pi1v}
and
\cite{pi1p},
respectively.

$P^{II}(\omega)$ treats all other cases where
$\tilde{\Pi}^\delta(q^2)$ is just a constant for $z=1$. Note,
that this constant is unknown and consequently $P^{II}(1)$
may not be used for the construction of the Pad\'e approximation.
$P^{I}(1)$ is directly connected with the $1/v$ singularity
and, of course, known.
The high energy terms are treated in the same way for 
$P^{I}(\omega)$ and
$P^{II}(\omega)$: Due to the subtraction of $\tilde{\Pi}^\delta(-\infty)$
the constant terms transform to $P(0)$ and the difference together
with the prefactor $1/(1+\omega)^2$ projects out the $1/z$ suppressed
terms in the limit $\omega\to -1$.
Finally the moments from $z\to0$ transform into derivatives of 
$P(\omega)$ at $\omega=0$.
In total the following information is 
available for $P^{I}(\omega)$:
$\{P^{I}(-1),P^{I}(0),P^{I,(1)}(0),\ldots,P^{I,(8)}(0),P^{I}(1)\}$.
These eleven data points allow the construction of 
Pad\'e approximations like $[5/5]$, $[6/4]$ or $[4/6]$.
For $P^{II}(\omega)$ the threshold information $P^{II}(1)$ is not
available which means that at most Pad\'e approximations like
$[5/4]$ or $[4/5]$ may be constructed.

For the non-abelian contributions proportional to $C_AC_F$ there is
an alternative approach.
Following the method outlined in
\cite{CheHoaKueSteTeu96}
the imaginary part of the gluonic double-bubble contributions,
$R_g^{(2),\delta}(s)$,
can be computed analytically from the knowledge of the 
fermionic contribution 
$R_{l}^{(2),\delta}(s)$.
$R_g^{(2),\delta}(s)$ 
is, of course, gauge dependent. However, for 
the special choice $\xi=4$, where $\xi$ is defined via the 
gluon propagator 
$(-\,g^{\mu\nu} + \xi\,q^\mu\,q^\nu/q^2)/(q^2+i\epsilon)$
the threshold behaviour of the non-abelian contribution 
(see Eqs.~(\ref{eqthrvna}-\ref{eqthrpna}))
and the leading high energy logarithms are covered by
$R_g^{(2),\delta}(s)$.
Therefore it is promising to apply the procedure described above
to the difference
$\Pi_{NA}^{(2),\delta}(q^2) - \Pi_g^{(2),\delta}(q^2)|_{\xi=4}$ 
which has a less singular behaviour than 
$\Pi_{NA}^{(2),\delta}(q^2)$.
The results for the non-abelian contribution presented in the next 
section are based on this method.


\section{Results}
\label{secresults}

\begin{figure}[ht]
 \begin{center}
 \begin{tabular}{cc}
   \leavevmode
   \epsfxsize=7.0cm
   \epsffile[110 290 460 540]{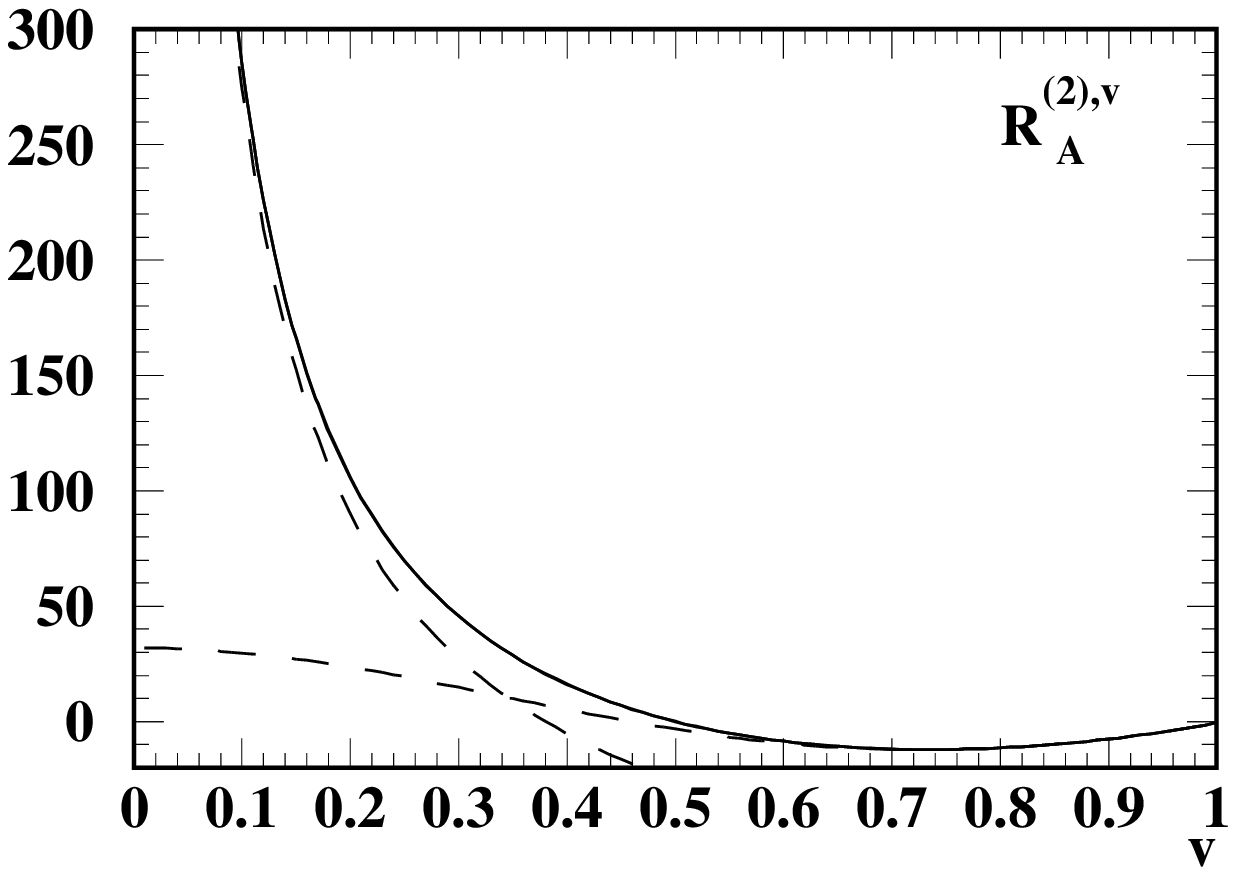}
   &
   \epsfxsize=7.0cm
   \epsffile[110 290 460 540]{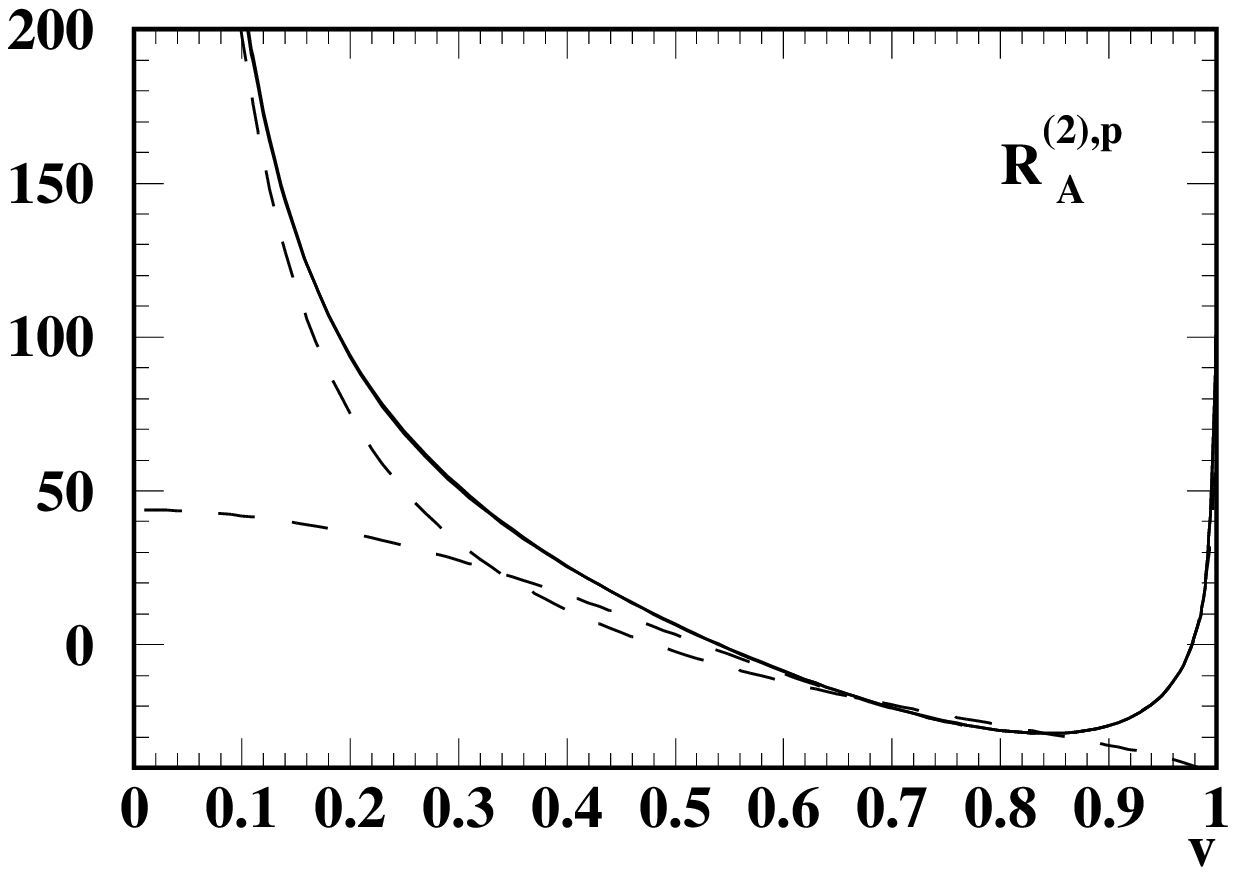}
   \\
   \epsfxsize=7.0cm
   \epsffile[110 290 460 540]{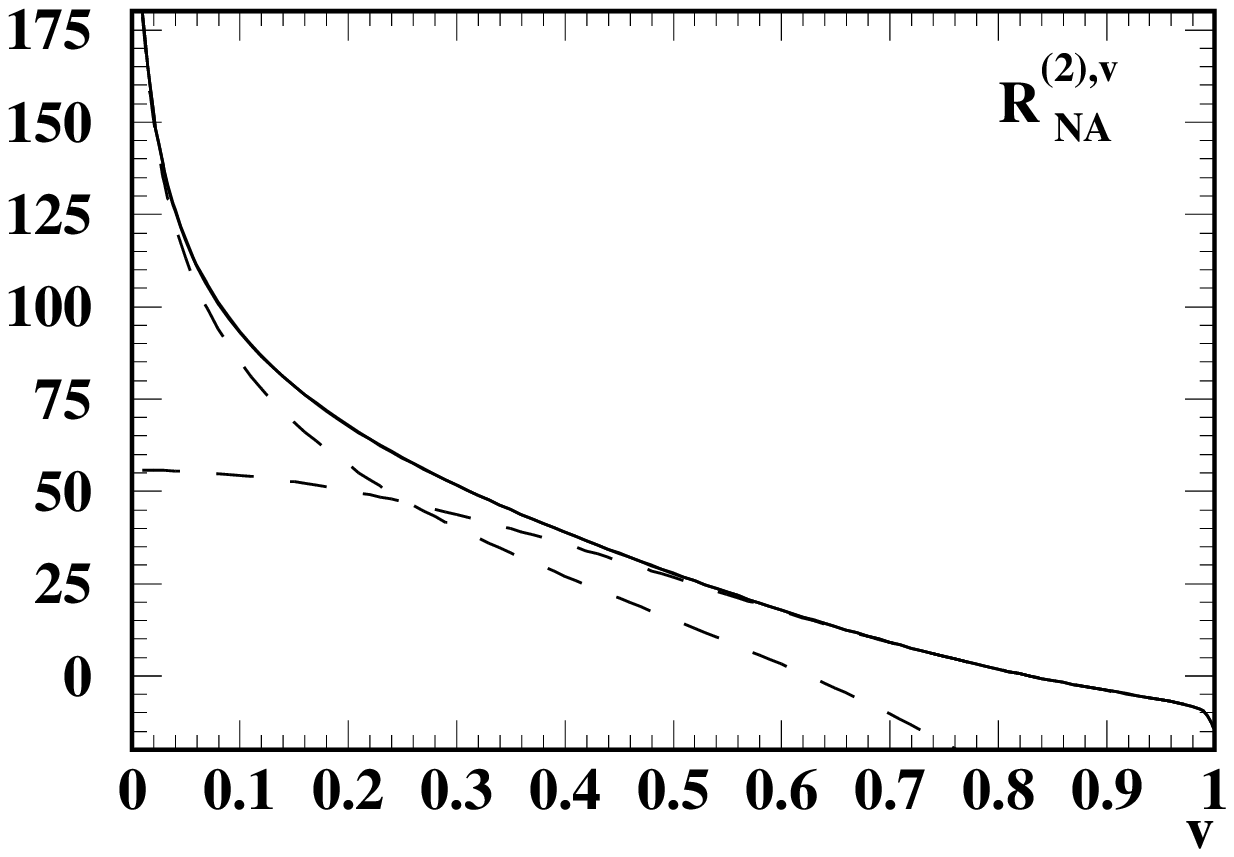}
   &
   \epsfxsize=7.0cm
   \epsffile[110 290 460 540]{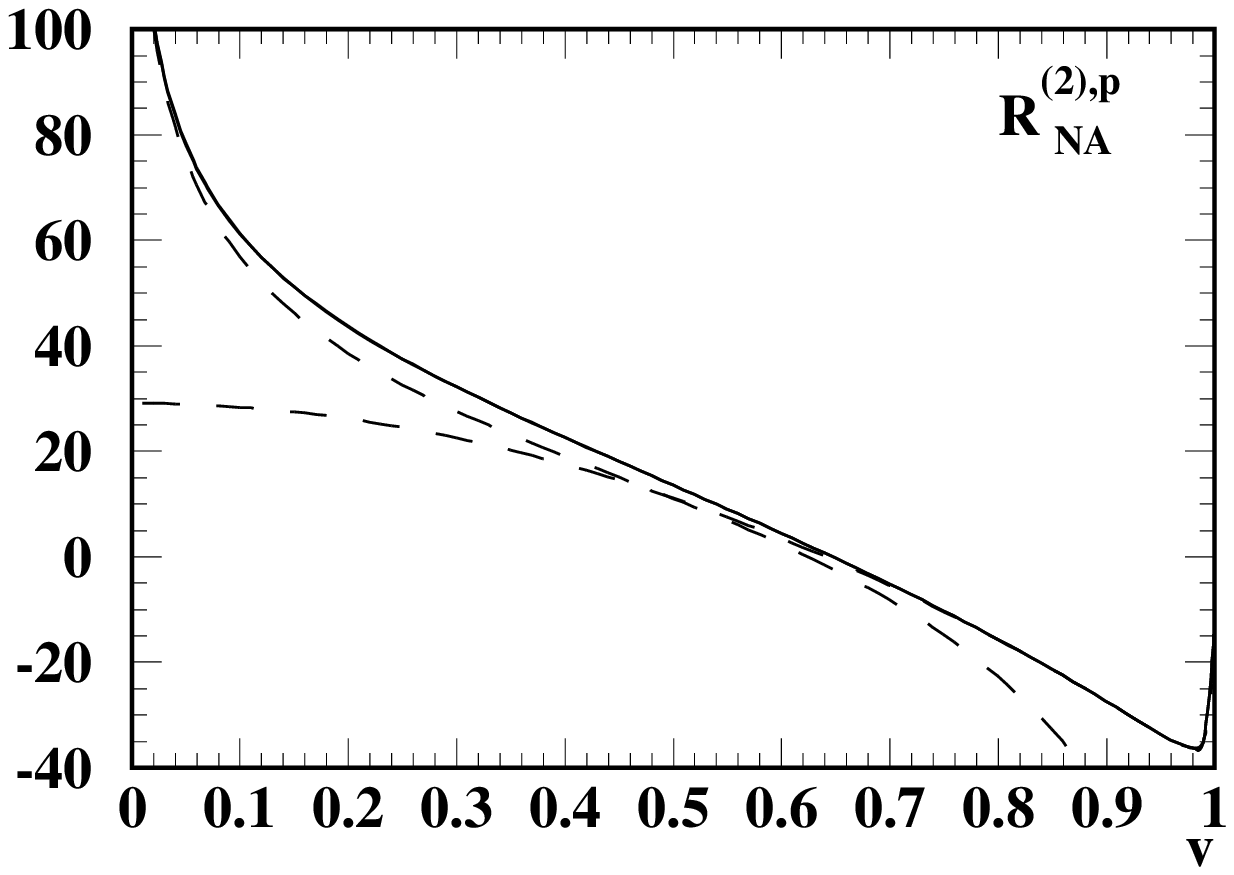}
   \\
   \epsfxsize=7.0cm
   \epsffile[110 290 460 540]{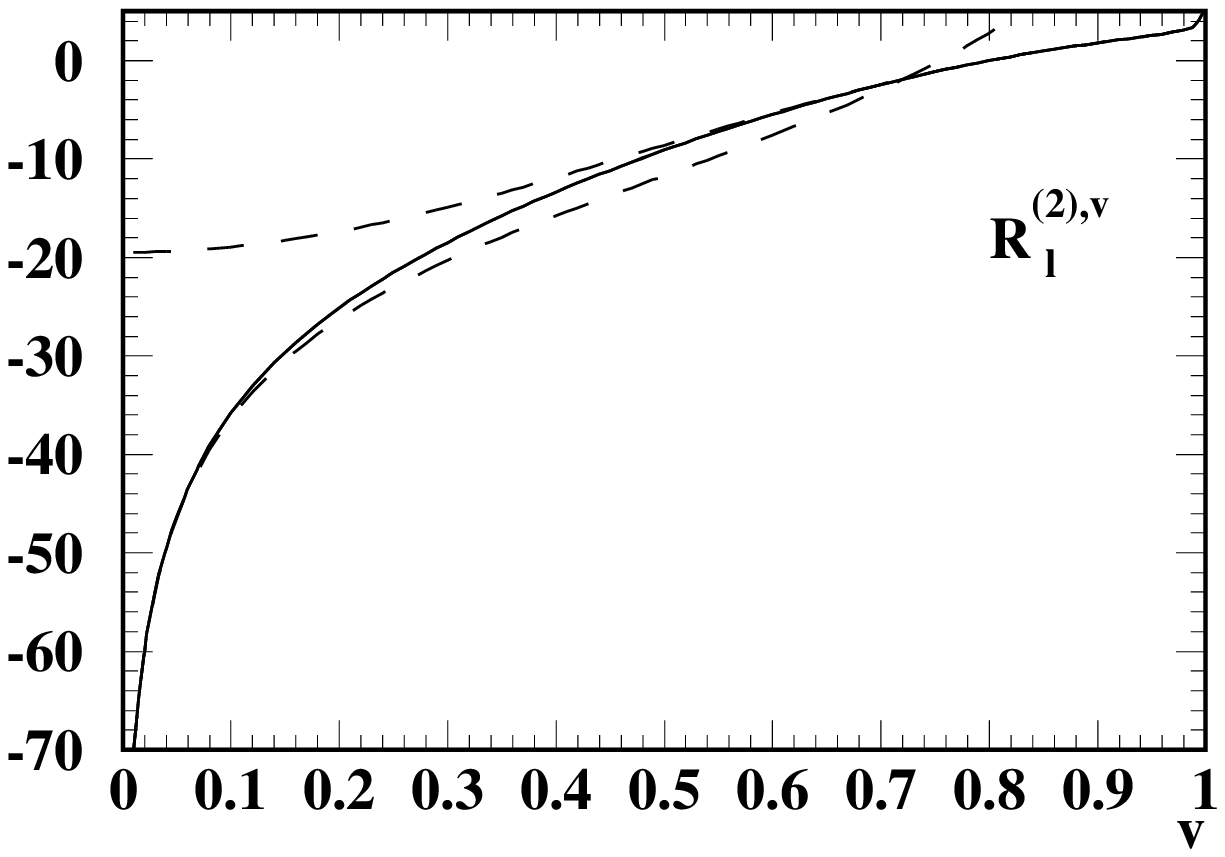}
   &
   \epsfxsize=7.0cm
   \epsffile[110 290 460 540]{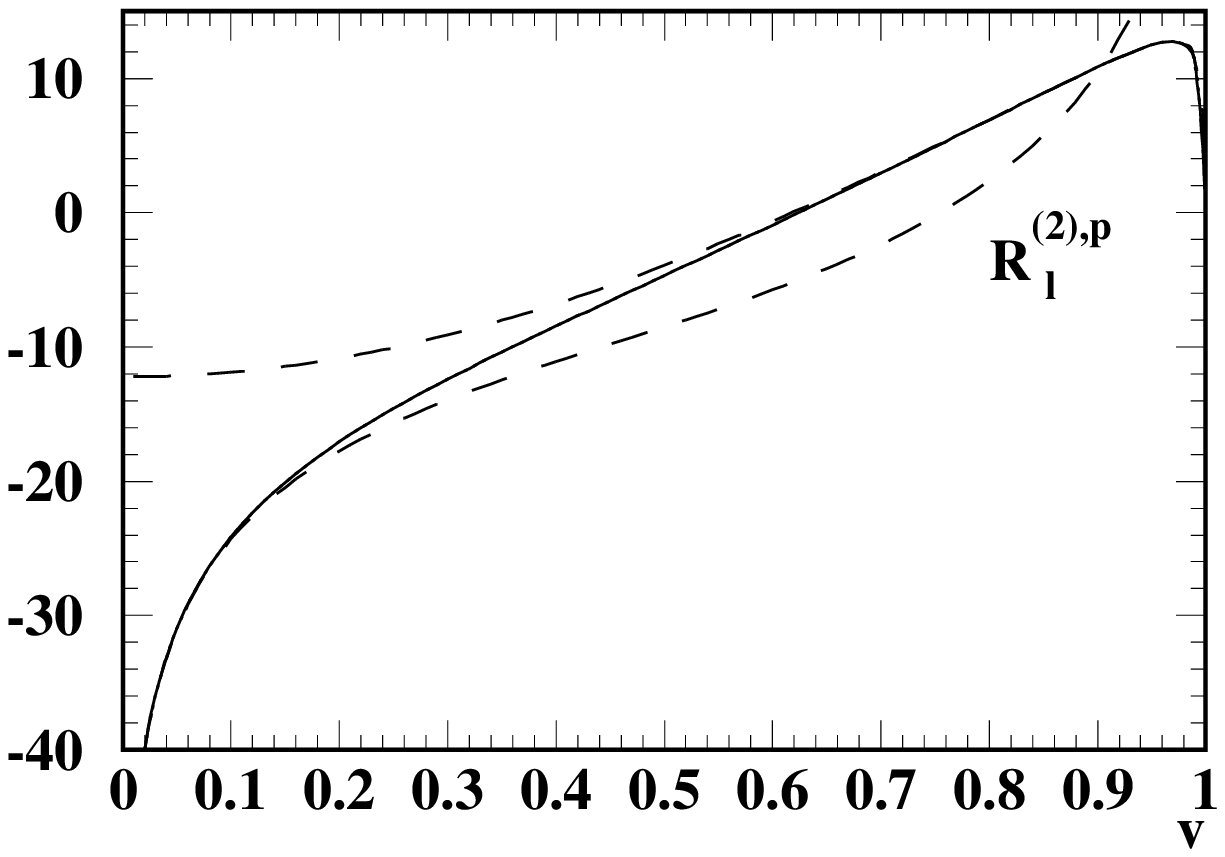}
 \end{tabular}
 \caption{\label{figvpv} $R^{(2),v}$ and $R^{(2),p}$ plotted against $v$. 
          The dashed curves represent the threshold and the 
          high energy approximations, respectively. Whereas for the 
          vector case also the terms of order $(m^2/s)^6$ are available 
          \protect\cite{CheHarKueSte97} for the 
          pseudo-scalar correlator terms of order
          $(m^2/s)^4$ \protect\cite{HarSte97} are plotted.}
 \end{center}
\end{figure}

\begin{figure}[ht]
 \begin{center}
 \begin{tabular}{cc}
   \leavevmode
   \epsfxsize=7.0cm
   \epsffile[110 290 460 540]{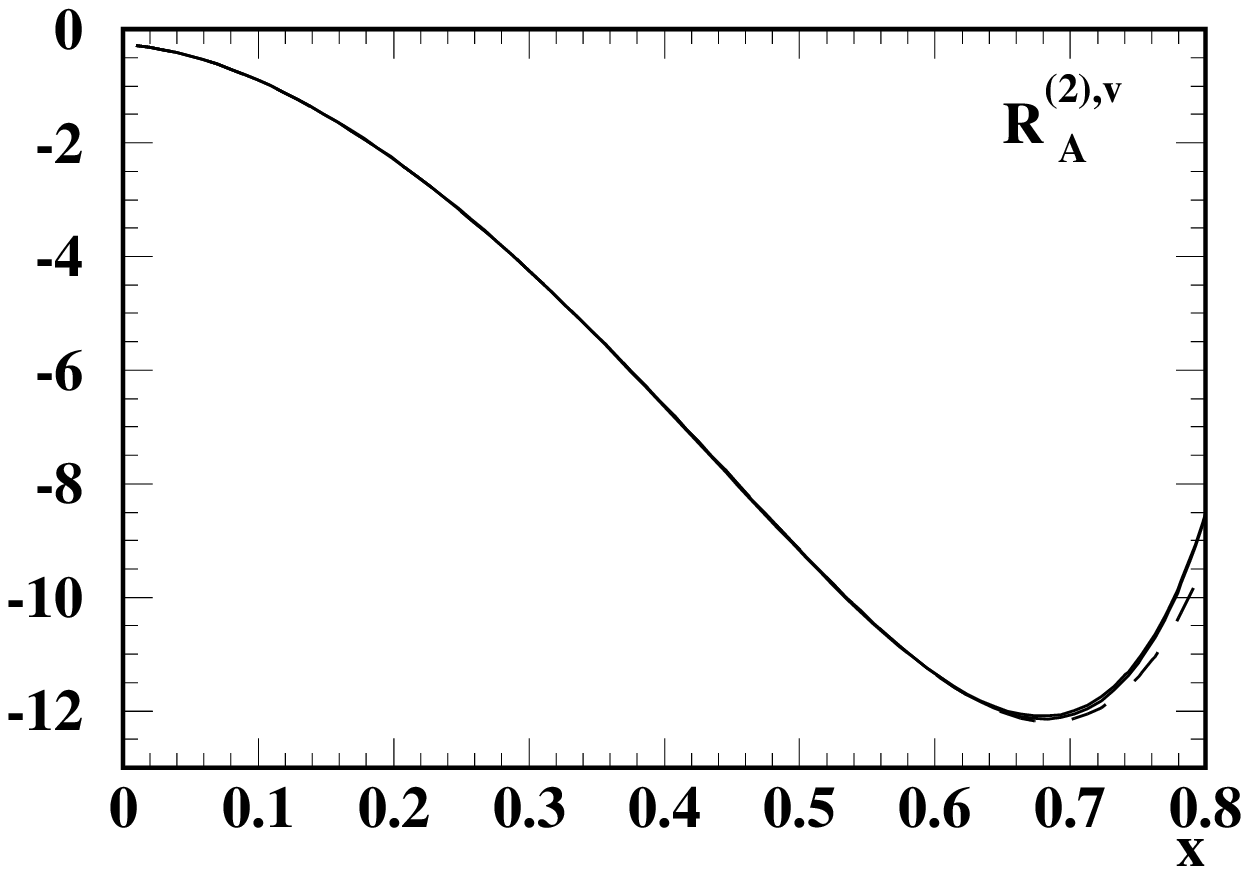}
   &
   \epsfxsize=7.0cm
   \epsffile[110 290 460 540]{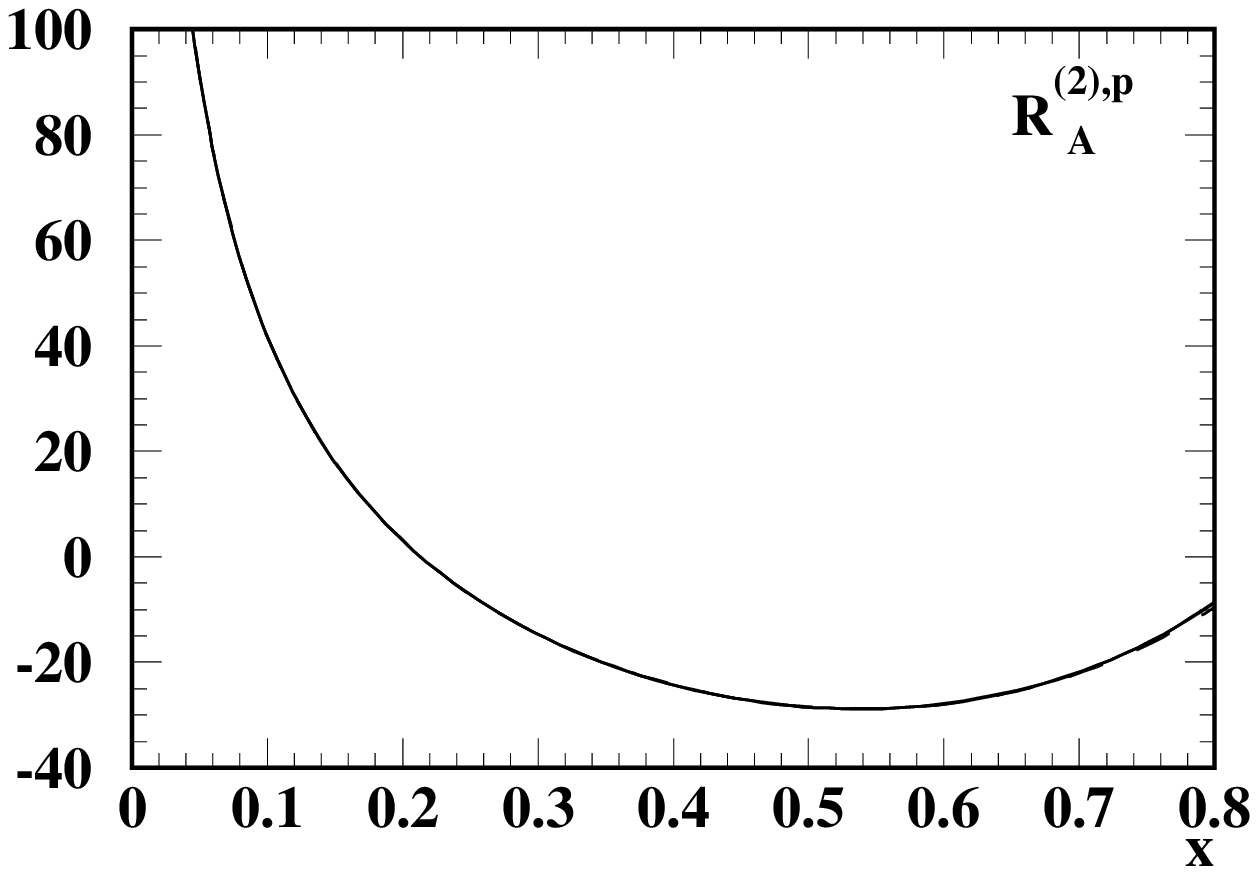}
   \\
   \epsfxsize=7.0cm
   \epsffile[110 290 460 540]{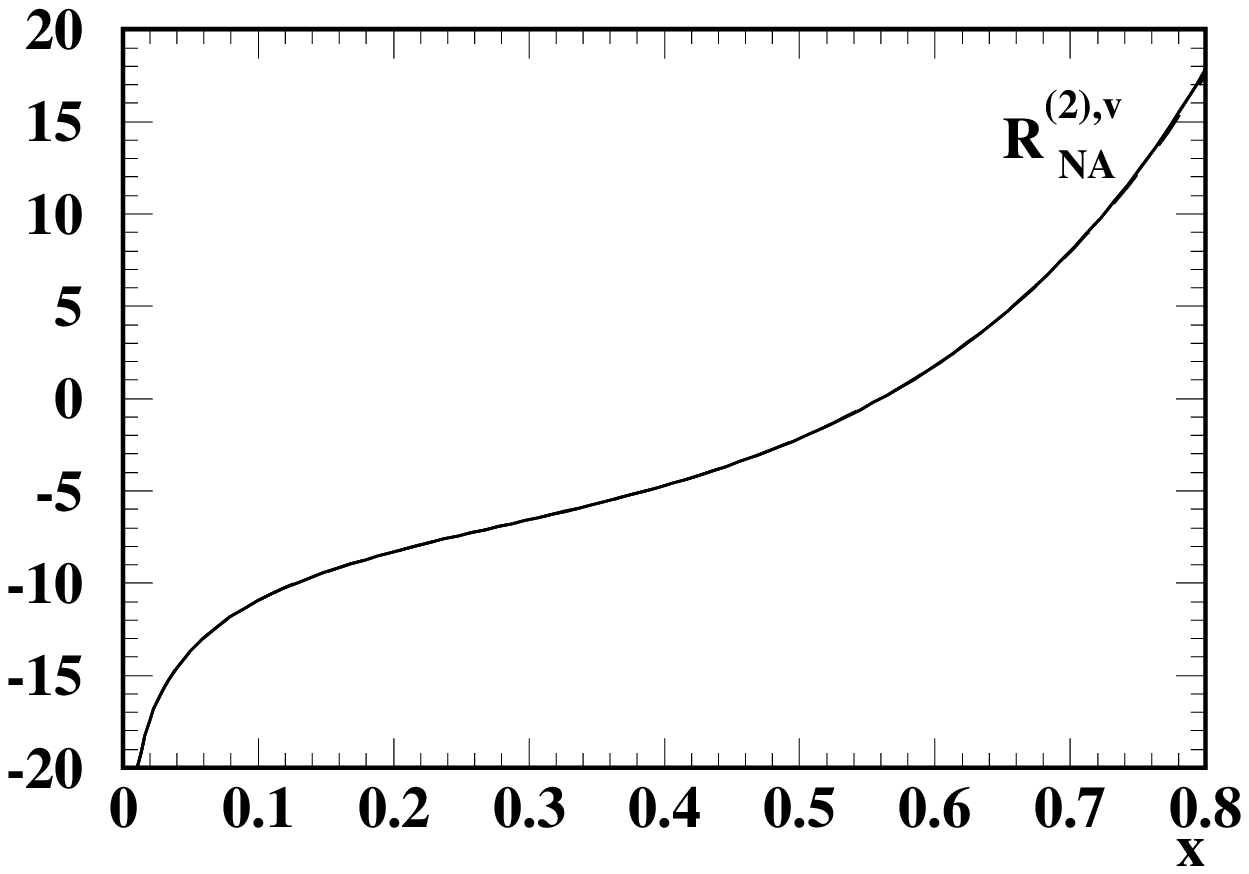}
   &
   \epsfxsize=7.0cm
   \epsffile[110 290 460 540]{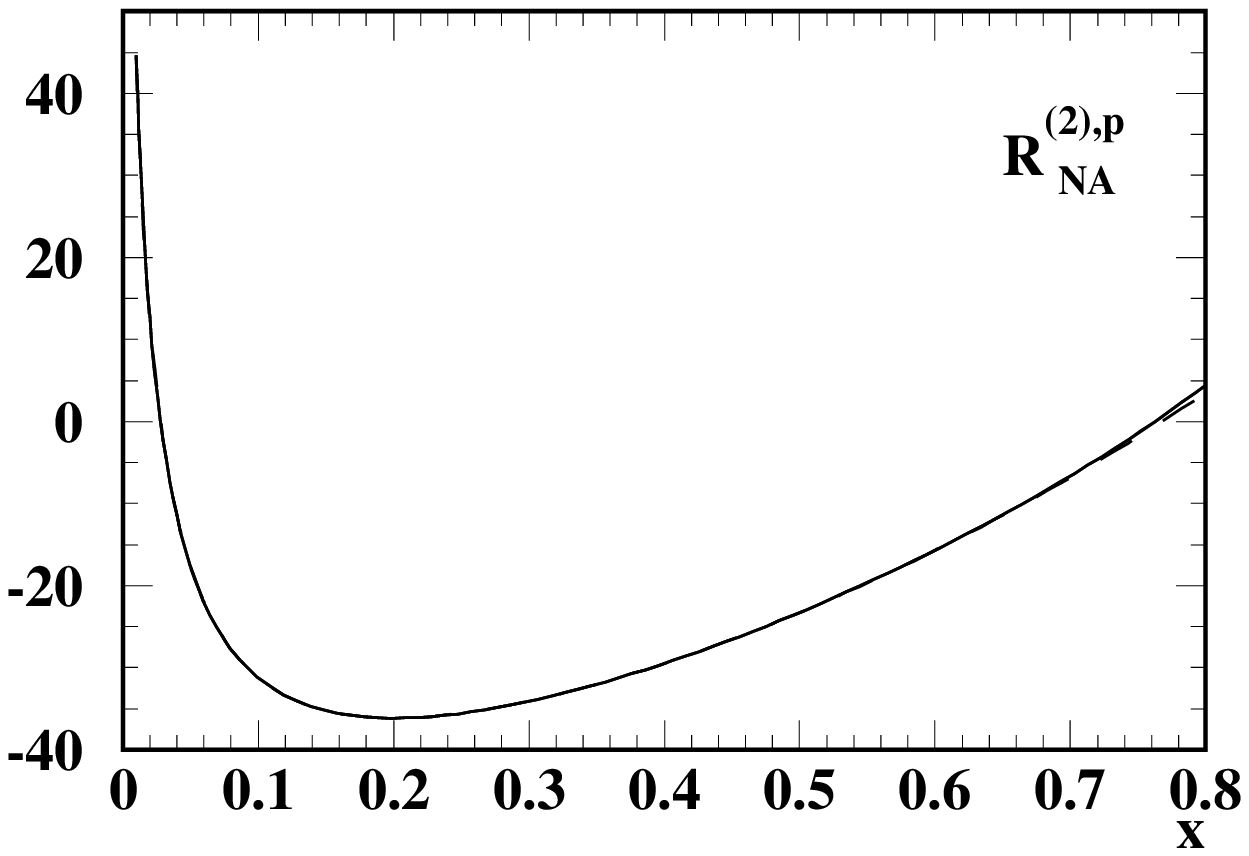}
   \\
   \epsfxsize=7.0cm
   \epsffile[110 290 460 540]{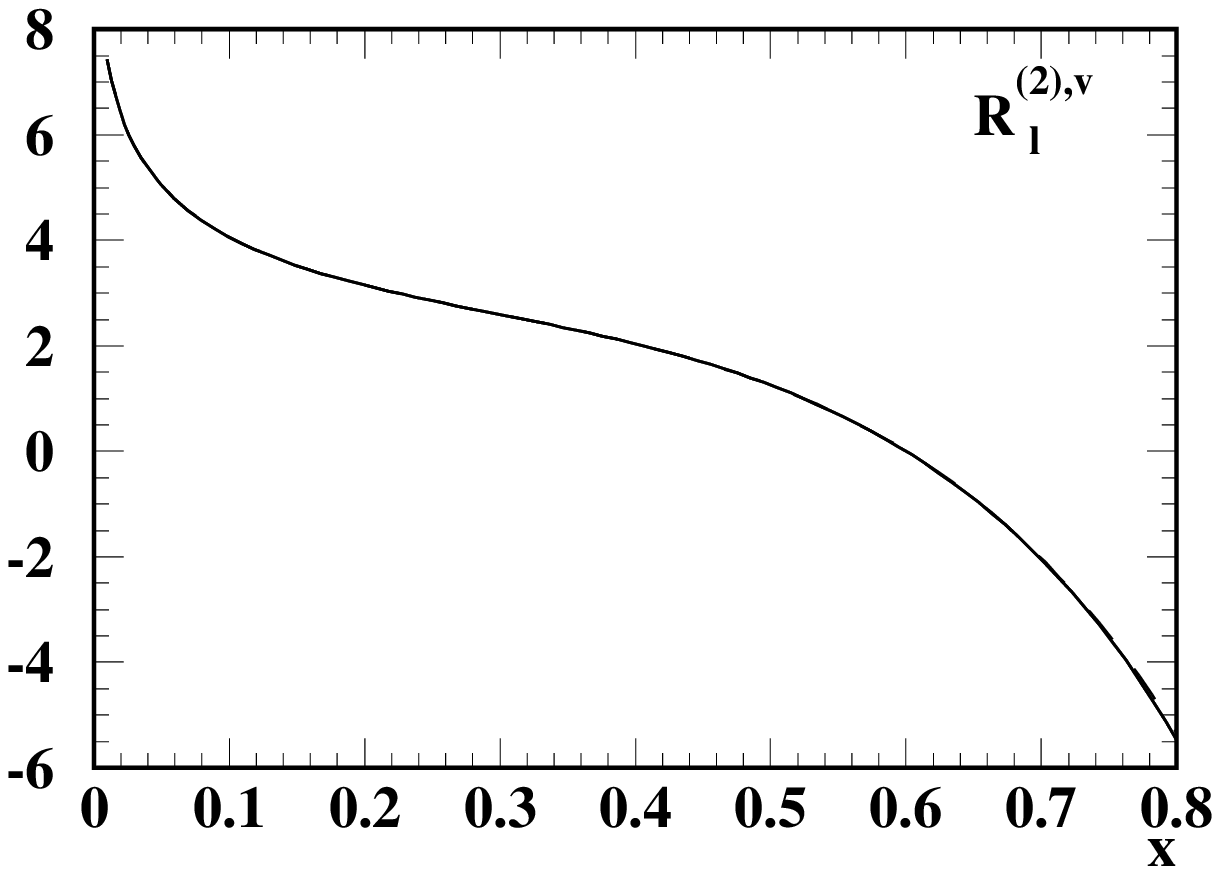}
   &
   \epsfxsize=7.0cm
   \epsffile[110 290 460 540]{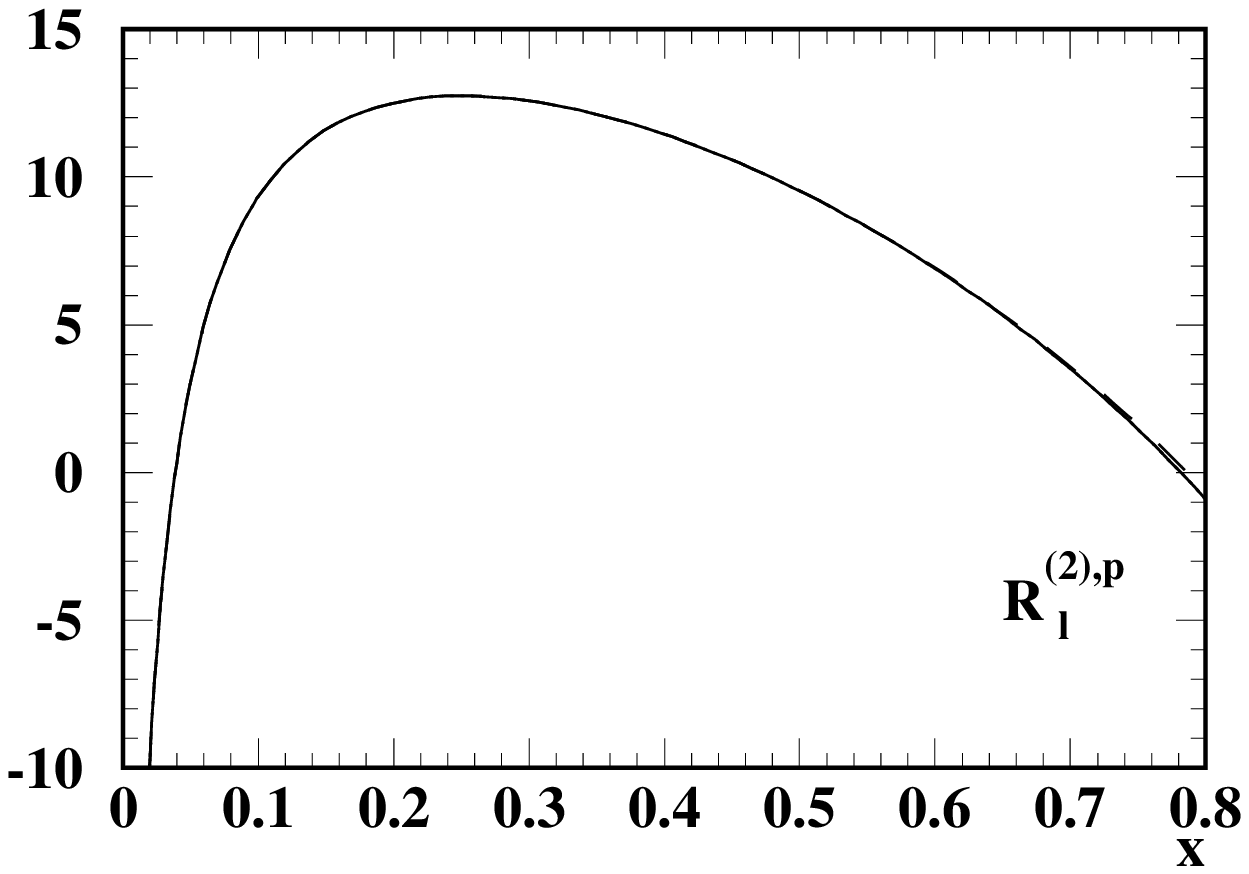}
 \end{tabular}
 \caption{\label{figvpx} $R^{(2),v}$ and $R^{(2),p}$ plotted against $x$.
          The dashed curves represent the high energy approximations
          including terms up to ${\cal O}(x^{12})$
          for the vector correlator and ${\cal O}(x^8)$ for the
          pseudo-scalar case.}
 \end{center}
\end{figure}

\begin{figure}[ht]
 \begin{center}
 \begin{tabular}{cc}
   \leavevmode
   \epsfxsize=7.0cm
   \epsffile[110 290 460 540]{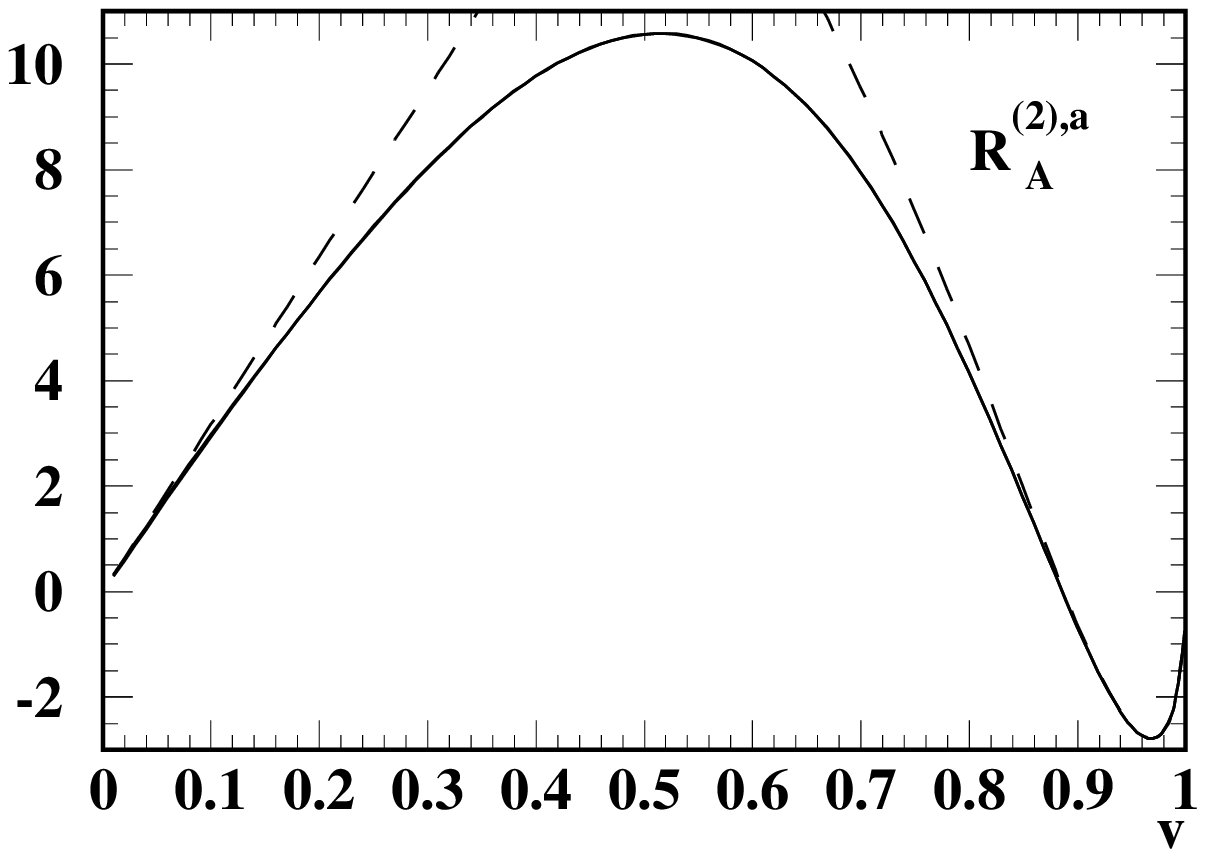}
   &
   \epsfxsize=7.0cm
   \epsffile[110 290 460 540]{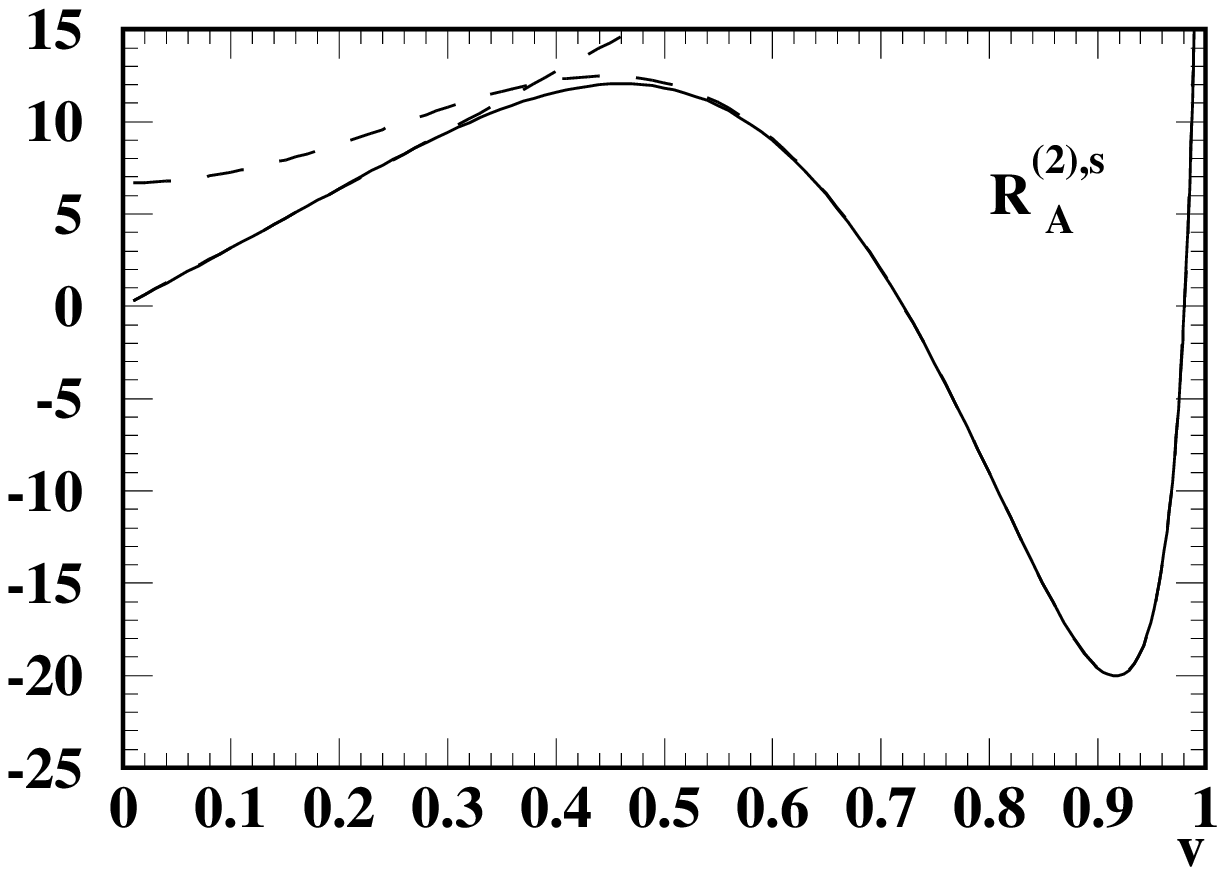}
   \\
   \epsfxsize=7.0cm
   \epsffile[110 290 460 540]{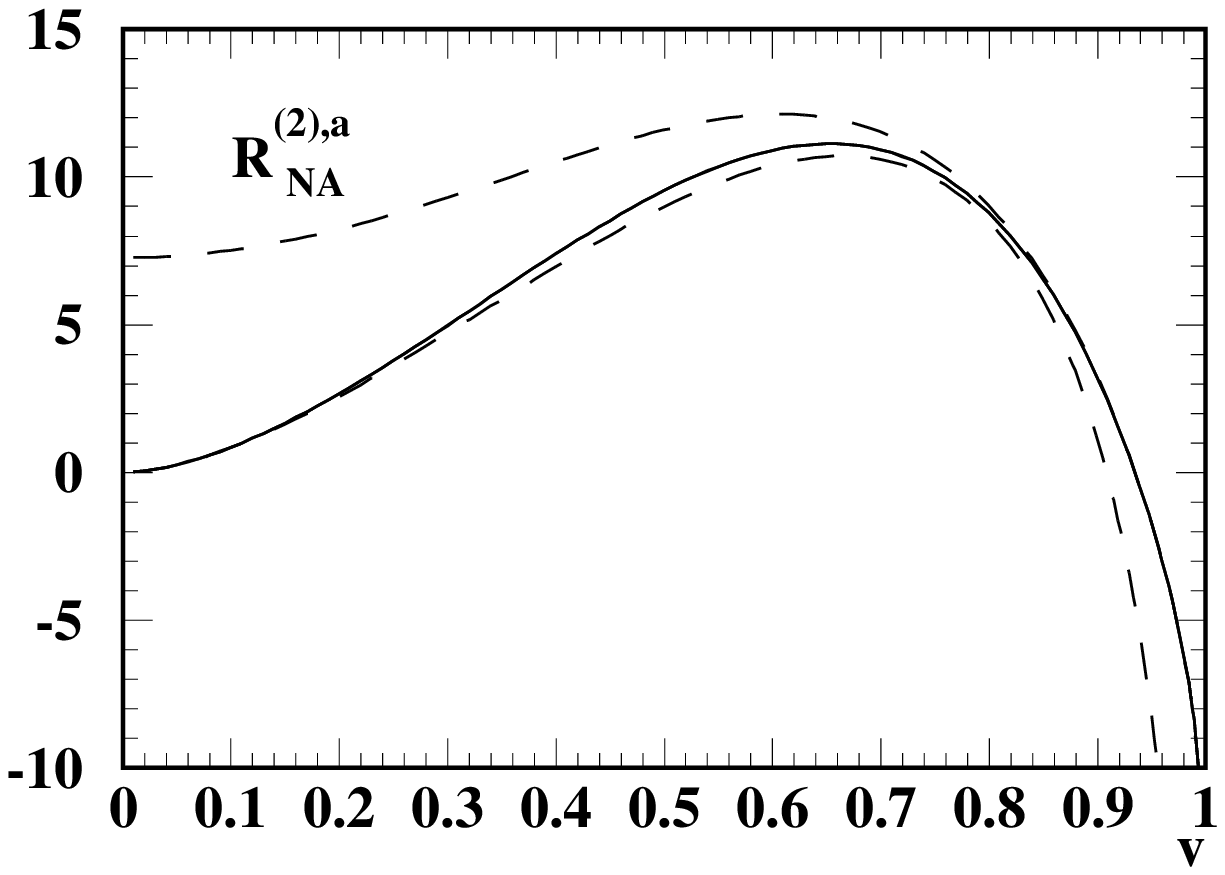}
   &
   \epsfxsize=7.0cm
   \epsffile[110 290 460 540]{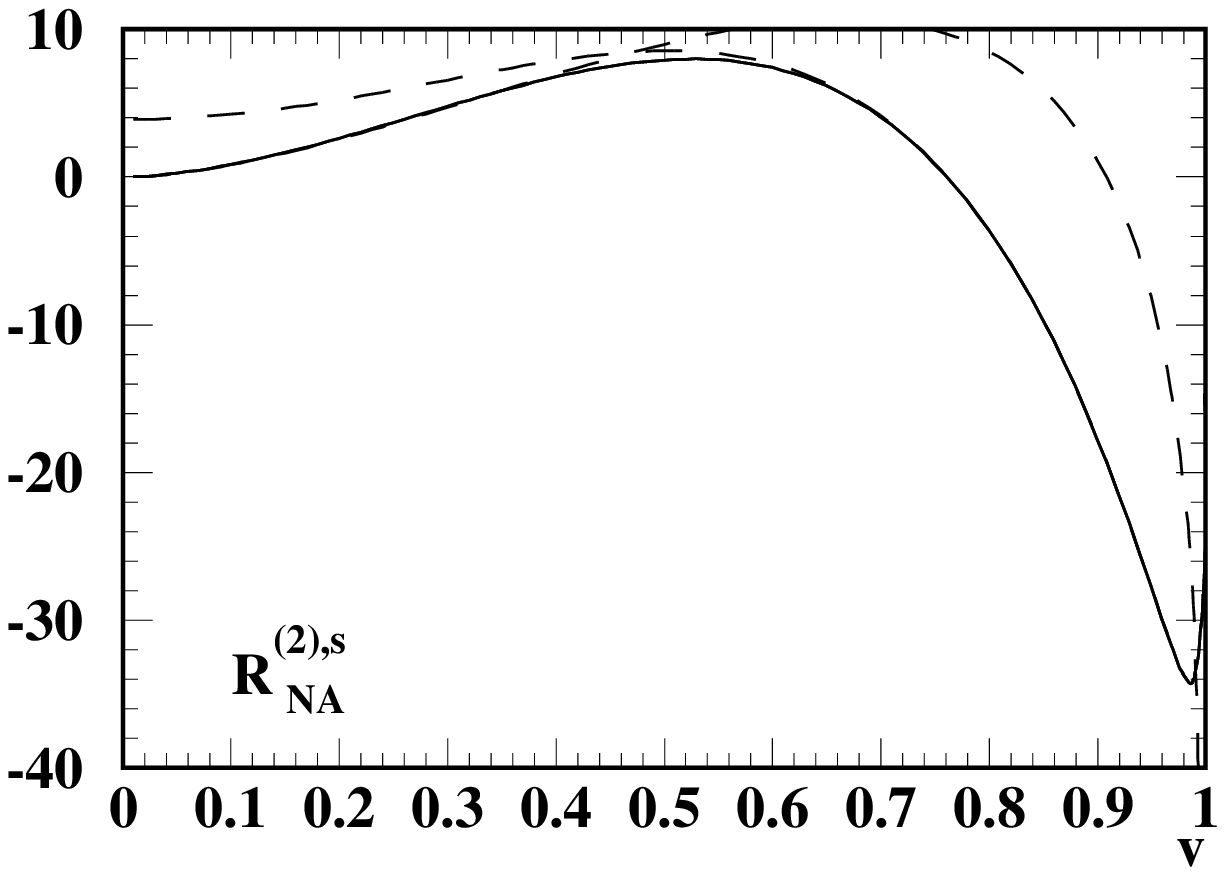}
   \\
   \epsfxsize=7.0cm
   \epsffile[110 290 460 540]{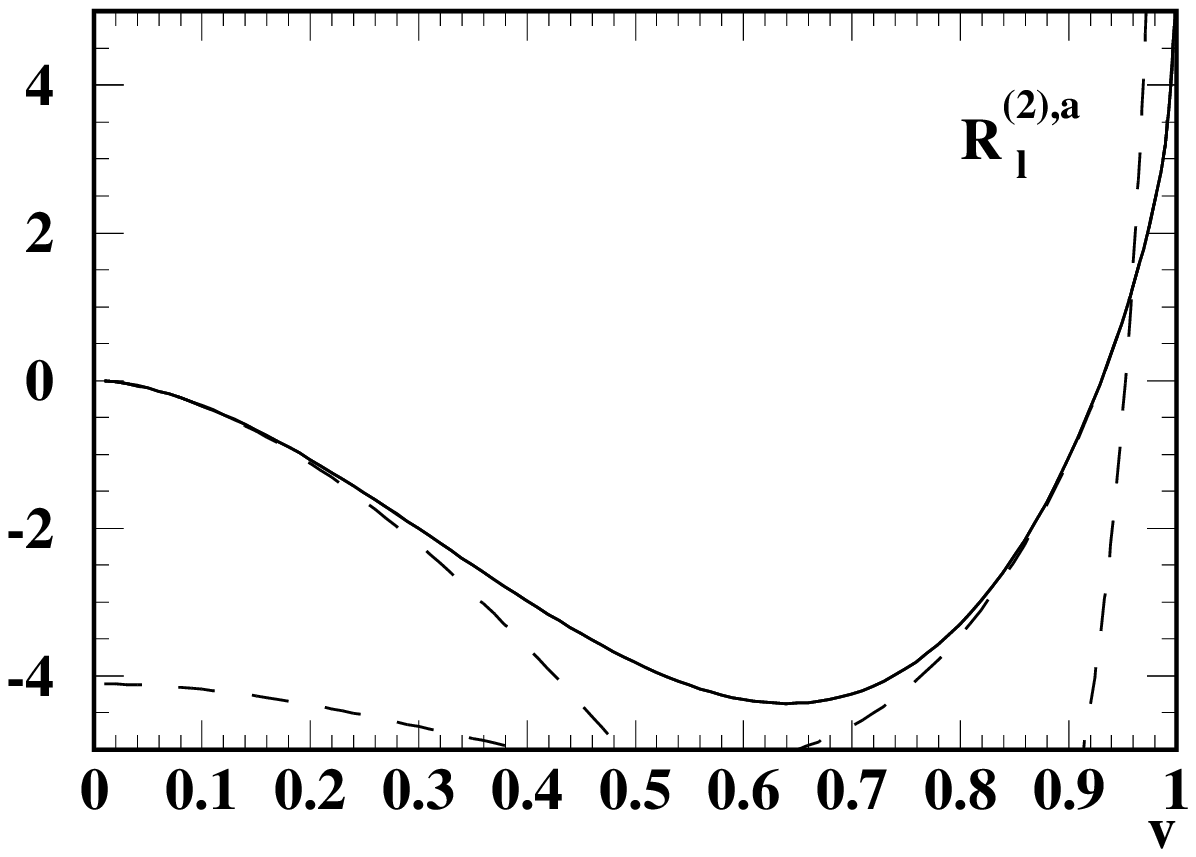}
   &
   \epsfxsize=7.0cm
   \epsffile[110 290 460 540]{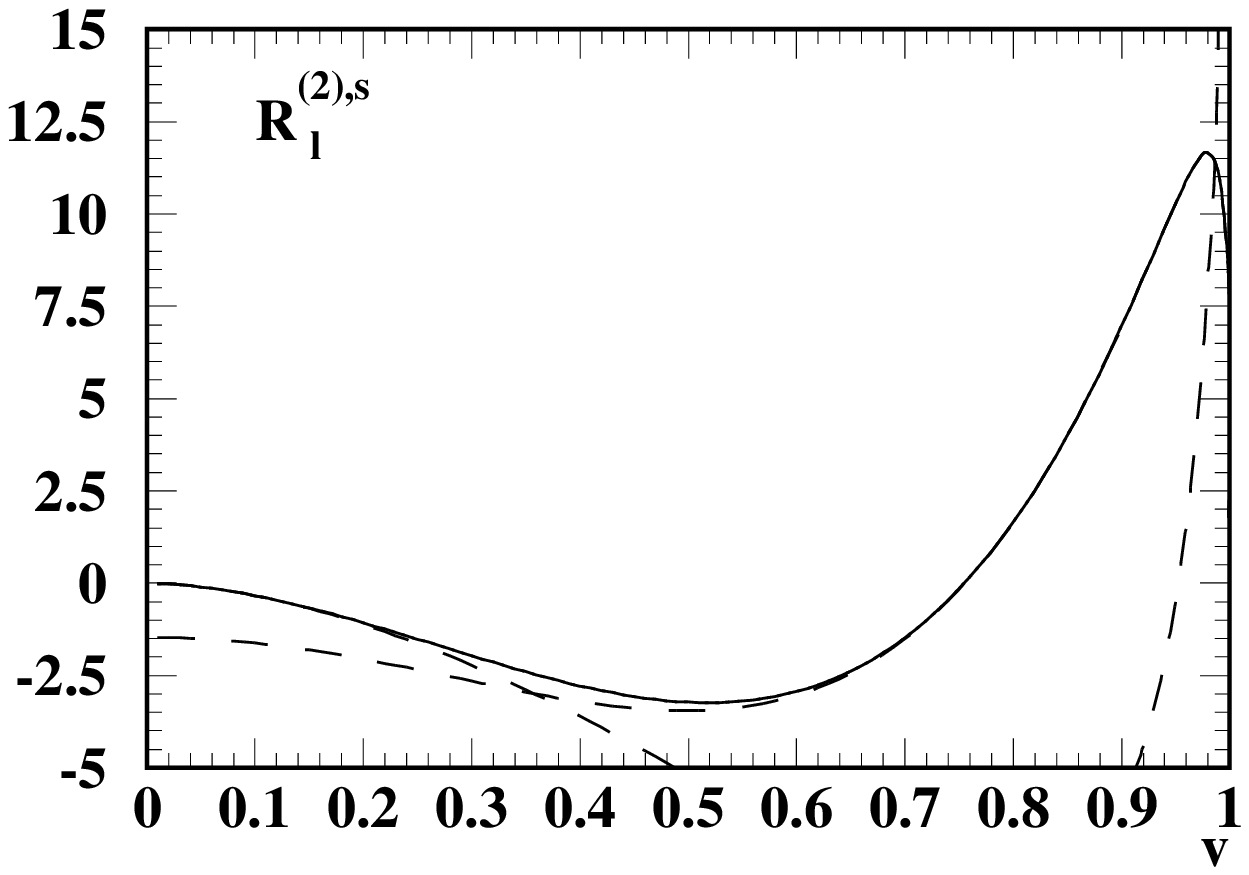}
 \end{tabular}
 \caption{\label{figasv} $R^{(2),a}$ and $R^{(2),s}$ plotted against $v$. 
          The dashed curves represent the threshold and the 
          high energy approximations, respectively. For the 
          axial-vector case terms of order $(m^2/s)^4$ are available 
          \protect\cite{CheKue94} and for the scalar
          correlator terms of order
          $(m^2/s)^4$ \protect\cite{HarSte97} are plotted.}
 \end{center}
\end{figure}

\begin{figure}[ht]
 \begin{center}
 \begin{tabular}{cc}
   \leavevmode
   \epsfxsize=7.0cm
   \epsffile[110 290 460 540]{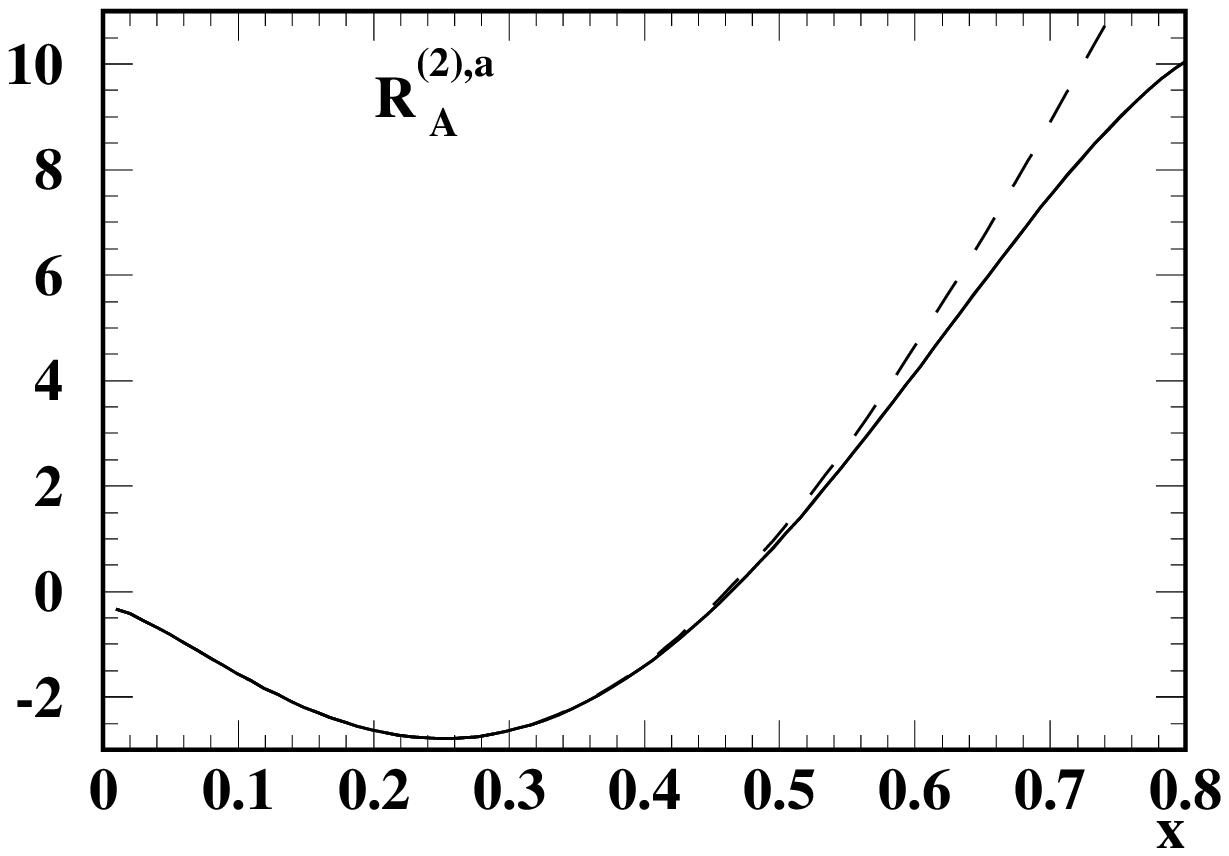}
   &
   \epsfxsize=7.0cm
   \epsffile[110 290 460 540]{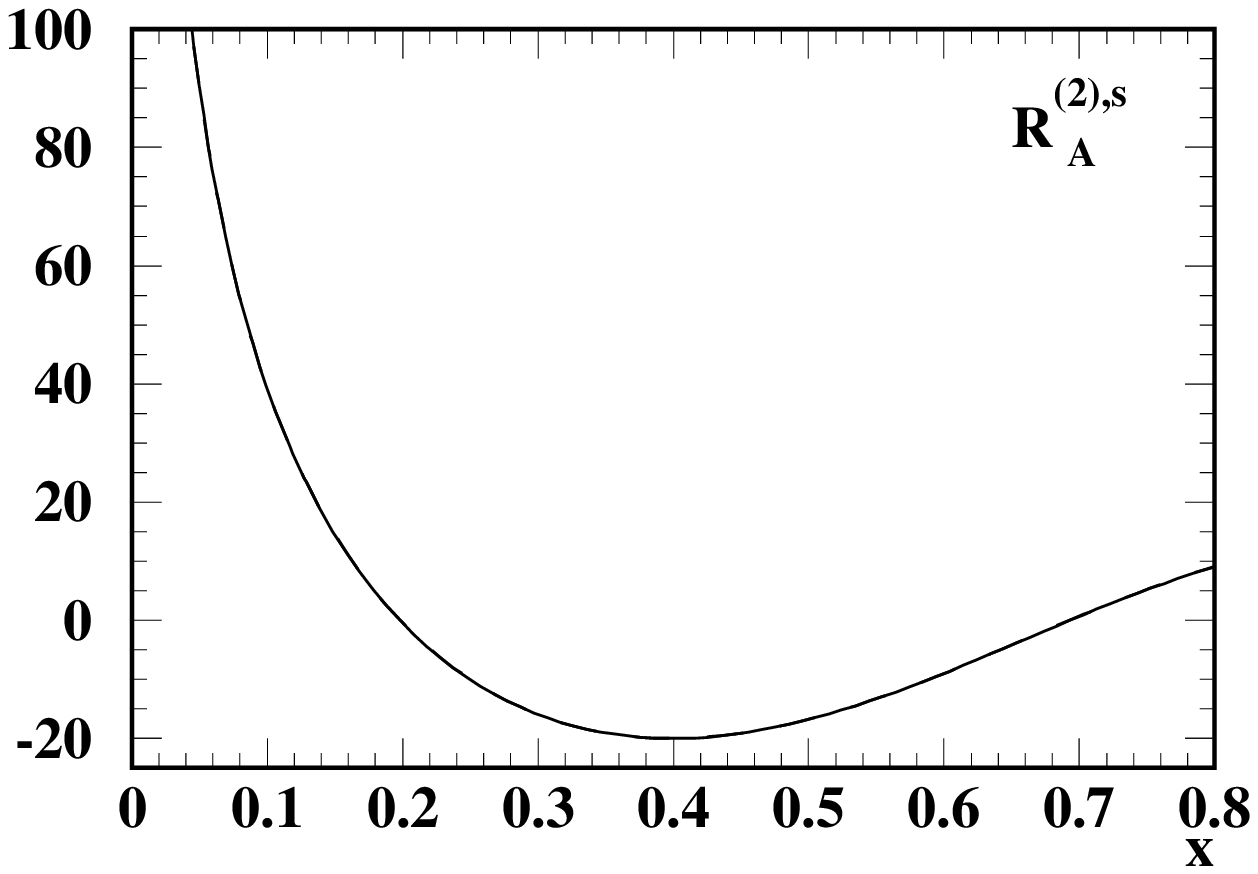}
   \\
   \epsfxsize=7.0cm
   \epsffile[110 290 460 540]{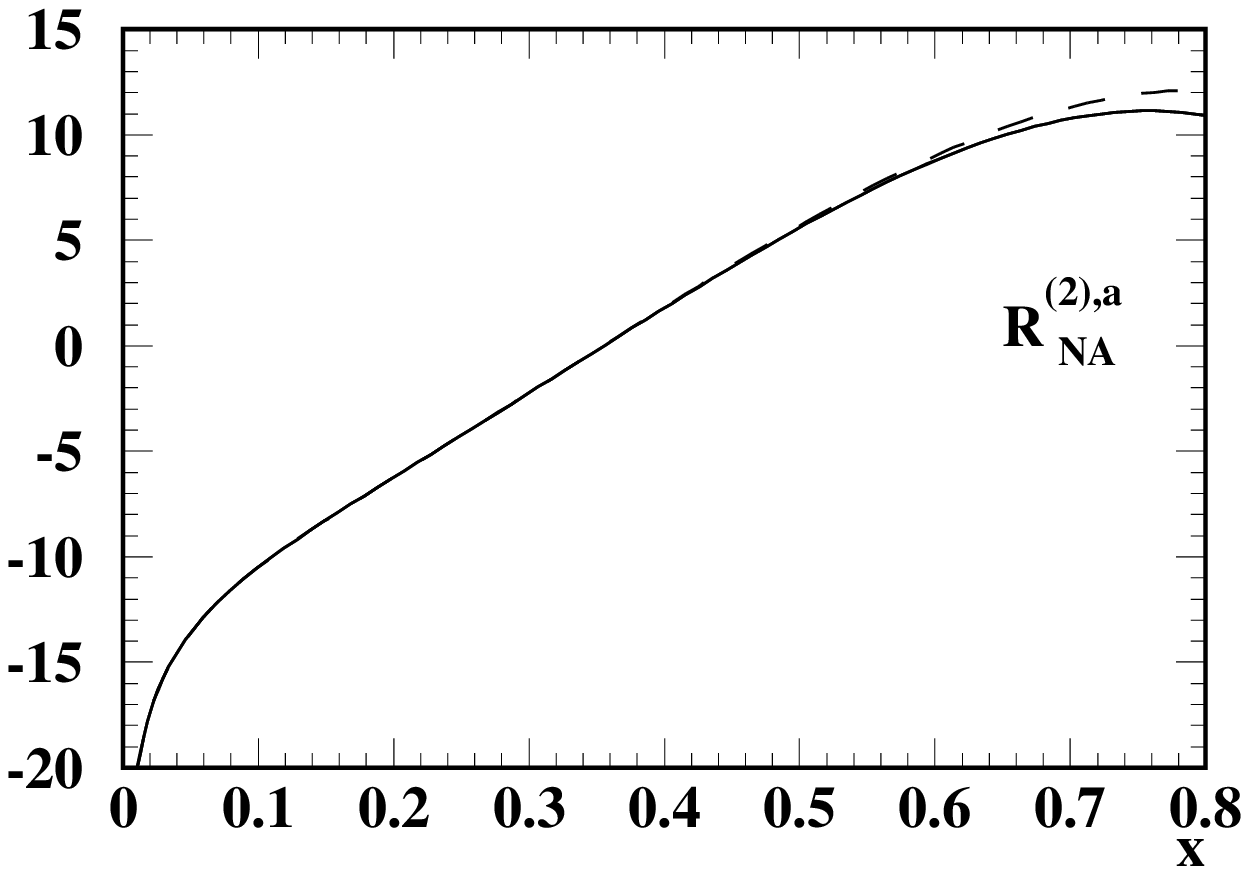}
   &
   \epsfxsize=7.0cm
   \epsffile[110 290 460 540]{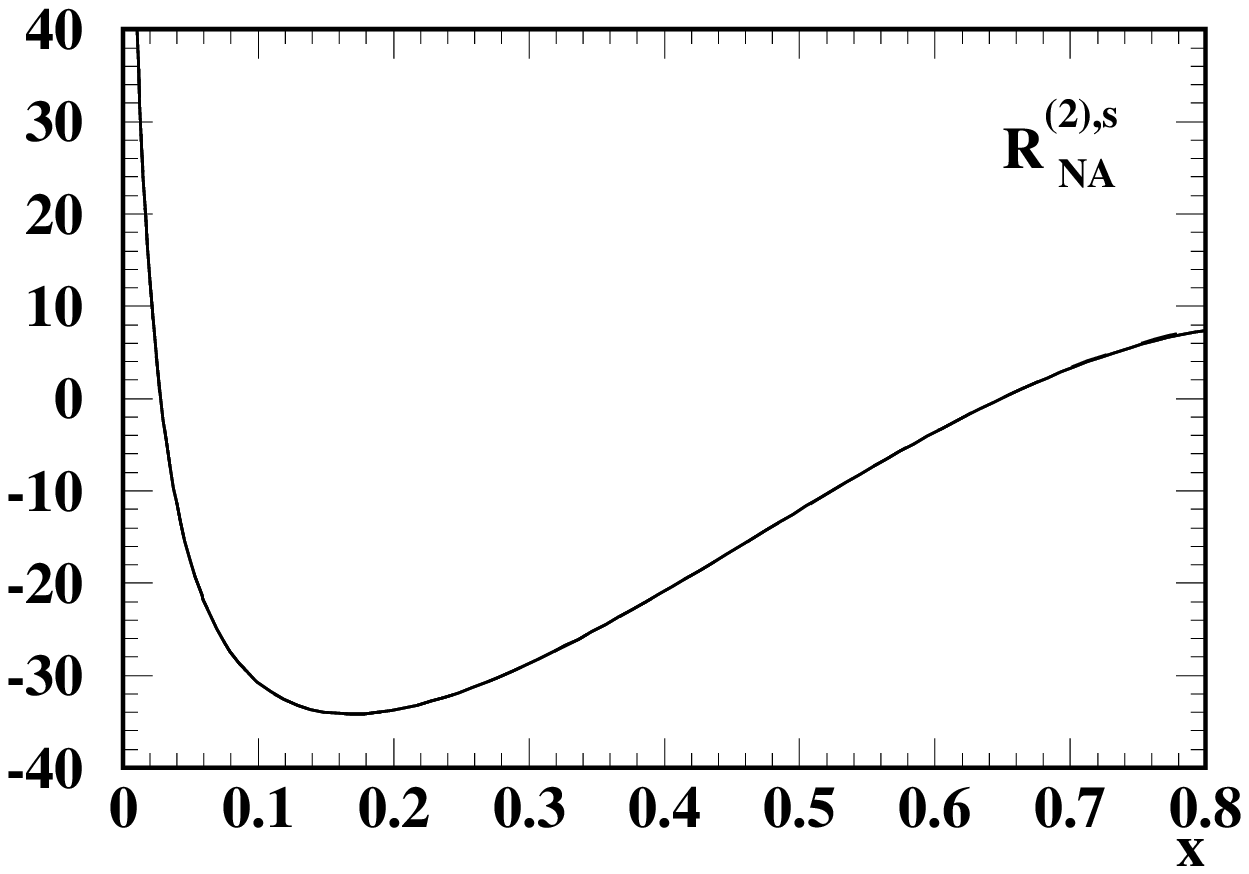}
   \\
   \epsfxsize=7.0cm
   \epsffile[110 290 460 540]{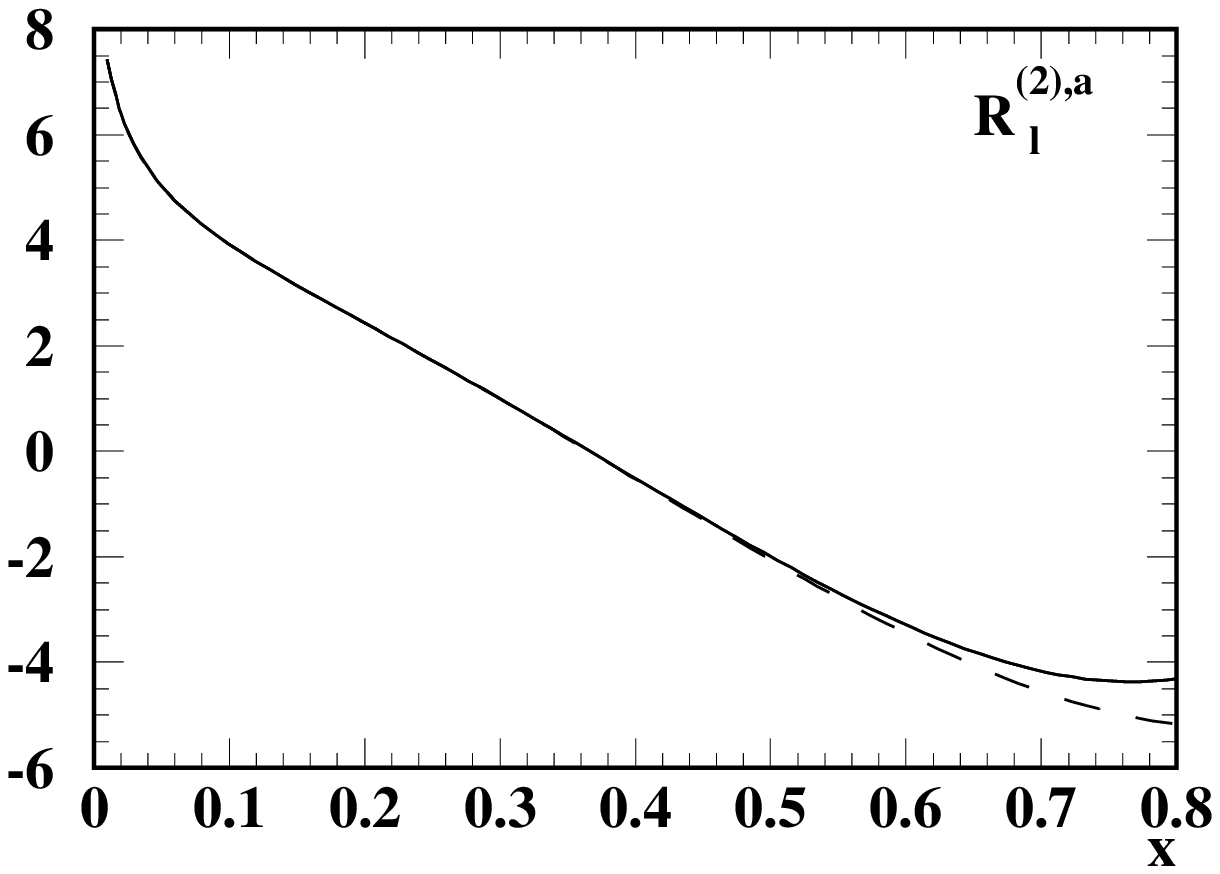}
   &
   \epsfxsize=7.0cm
   \epsffile[110 290 460 540]{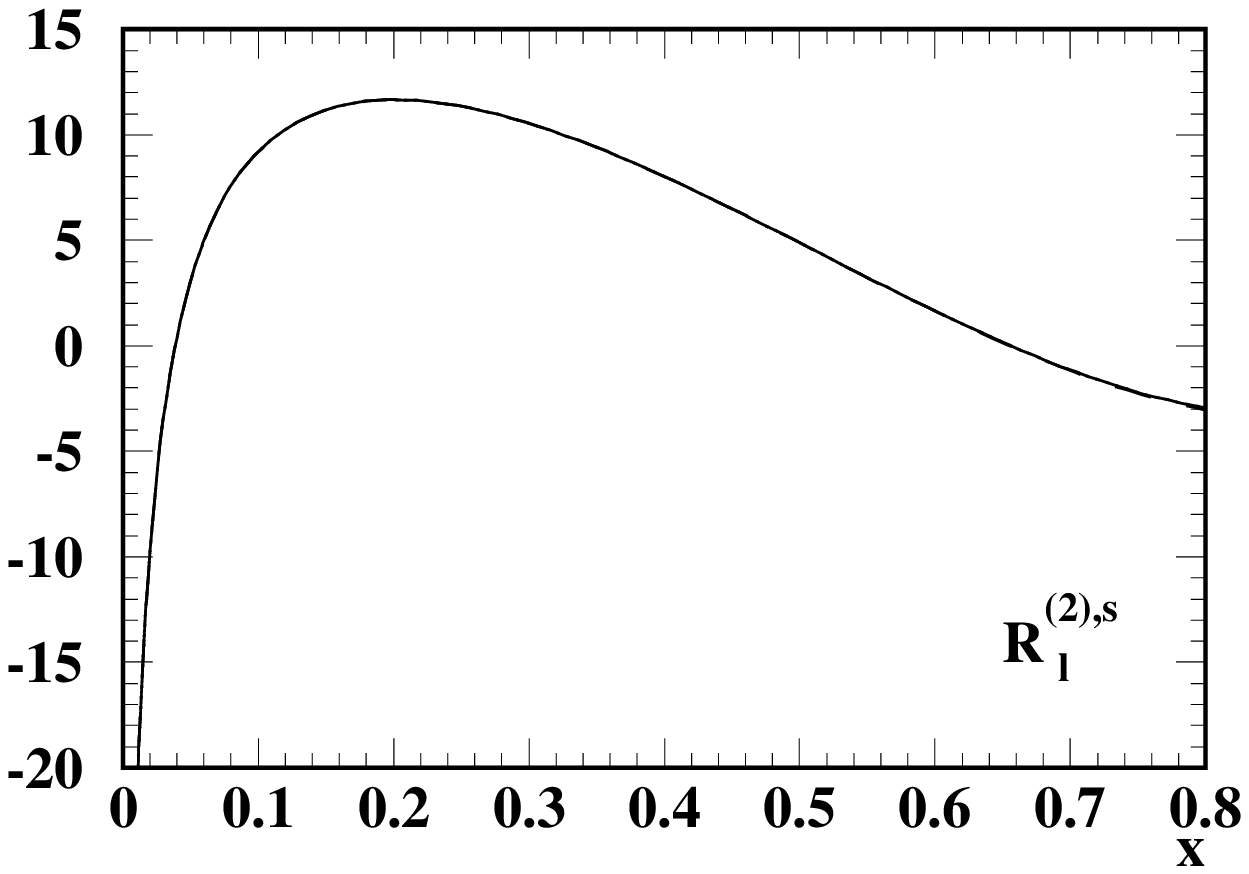}
 \end{tabular}
 \caption{\label{figasx} $R^{(2),a}$ and $R^{(2),s}$ plotted against $x$.
          The dashed curves represent the high energy approximations
          including terms up to ${\cal O}(x^{4})$
          for the axial-vector correlator and ${\cal O}(x^8)$ for the
          scalar case.}
 \end{center}
\end{figure}

After the Pad\'e approximation is performed for the function
$P(\omega)$ the corresponding equations are inverted in order to
get $\Pi^\delta(q^2)$. 
In Figs.~\ref{figvpv}--\ref{figasx} the results are presented
grouped according to the threshold behaviour.
In Fig.~\ref{figvpv} and \ref{figasv} $R(s)$ is plotted against
the velocity for the vector and pseudo-scalar and
the axial-vector and scalar case, respectively.
The Figs.~\ref{figvpx} and \ref{figasx} contain the corrections
plotted versus $x$.
Also the threshold and high energy approximations are shown
(dashed lines).
For the vector correlator terms up to ${\cal O}(x^{12})$ are 
available
\cite{CheHarKueSte97}.
Although in our procedure only terms up to 
${\cal O}(x^2)$ are incorporated the higher order terms
are very well reproduced. For the scalar and pseudo-scalar
polarization function terms up to ${\cal O}(x^8)$
are available
\cite{HarSte97}.
Again only the quadratic terms are build into the approximation method.
However, the numerical coincidence with the high energy approximations
is very good. Actually it is hardly possible to detect a difference 
between the high energy terms and the Pad\'e results when $x$ is used as
abscissa.
A similar behaviour is observed for the axial-vector case where only quartic 
terms are available
\cite{CheKue94}.

\begin{figure}[ht]
 \begin{center}
 \begin{tabular}{cc}
   \leavevmode
   \epsfxsize=7.0cm
   \epsffile[110 290 460 540]{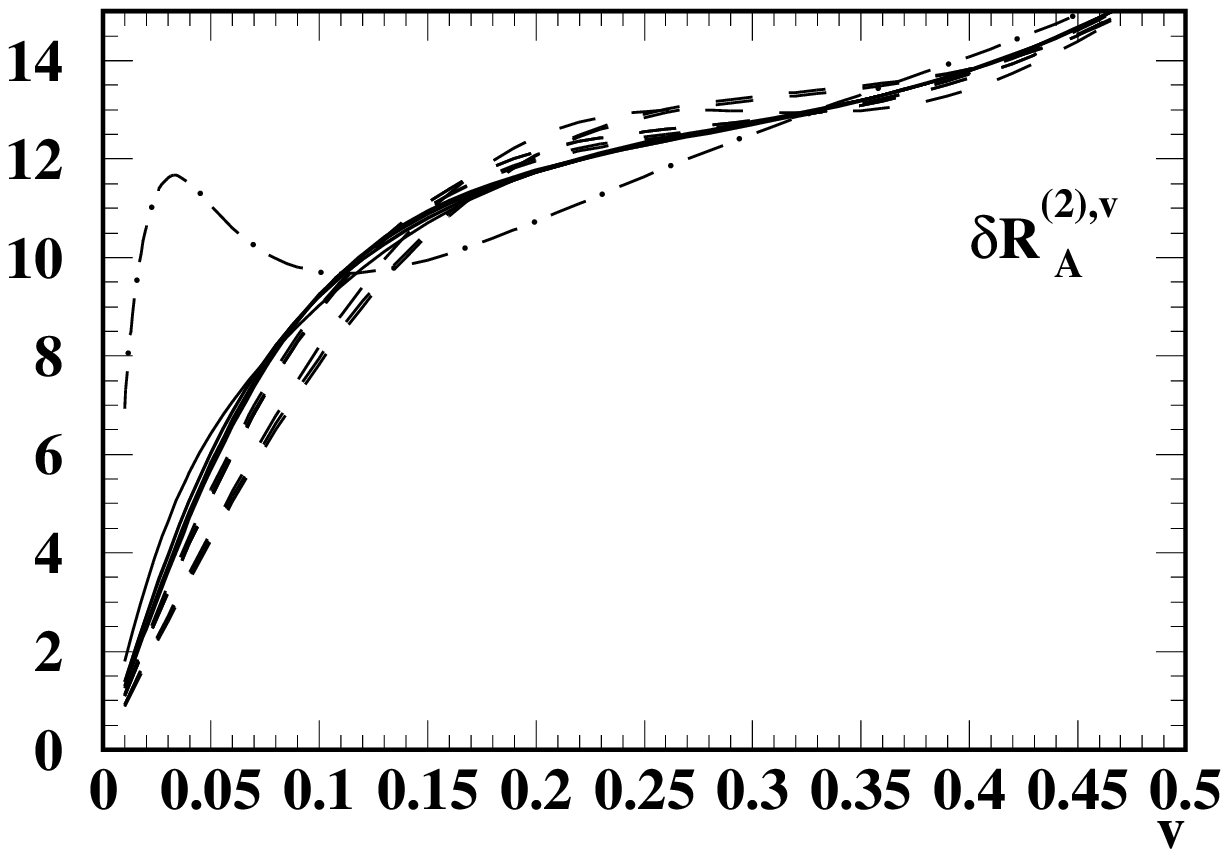}
   &
   \epsfxsize=7.0cm
   \epsffile[110 290 460 540]{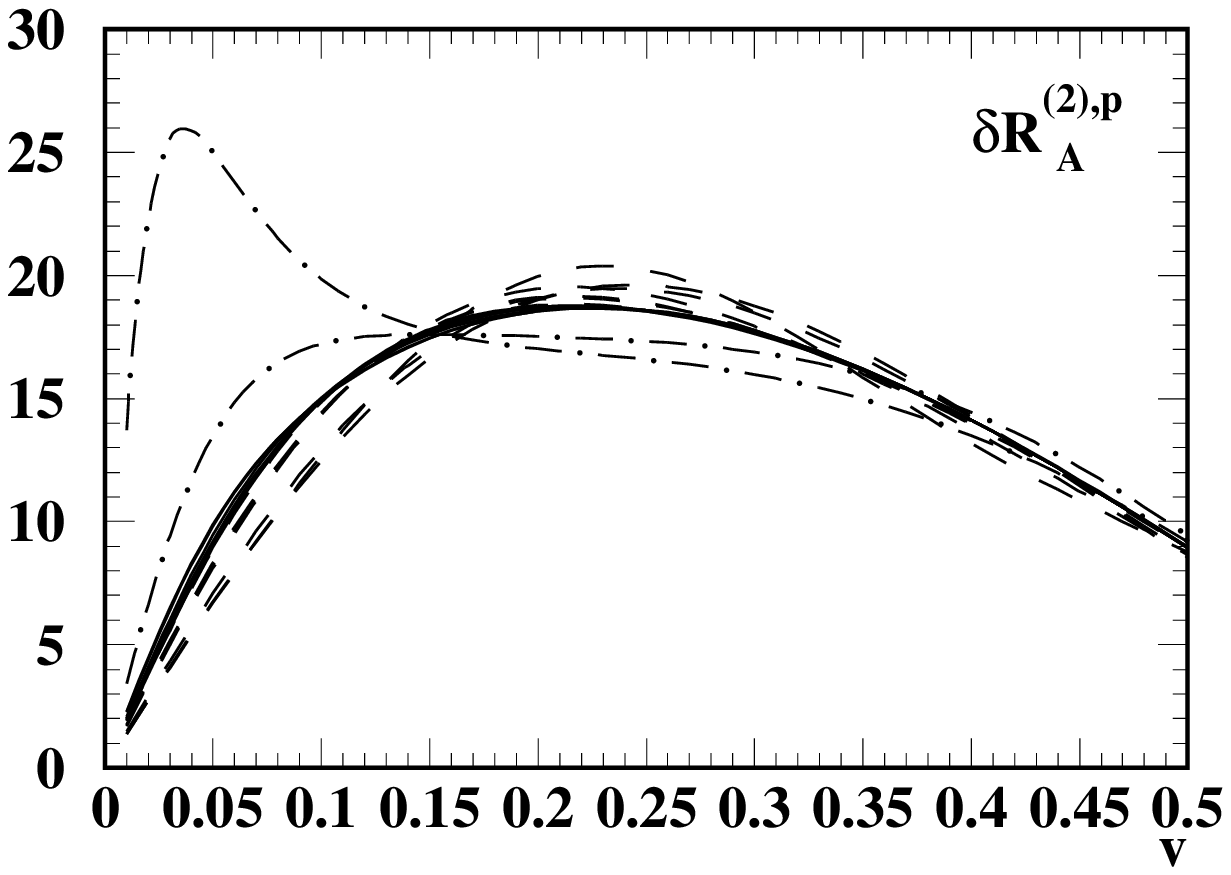}
   \\
   \epsfxsize=7.0cm
   \epsffile[110 290 460 540]{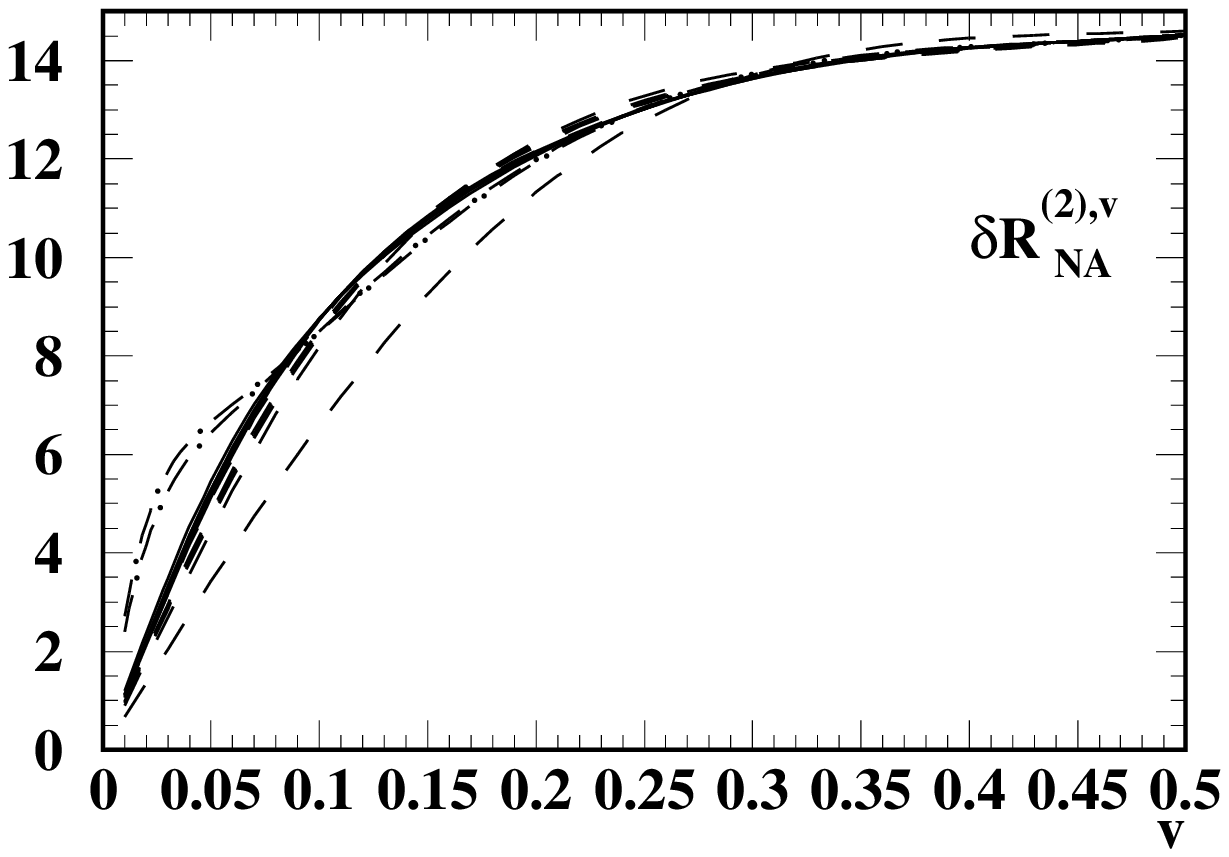}
   &
   \epsfxsize=7.0cm
   \epsffile[110 290 460 540]{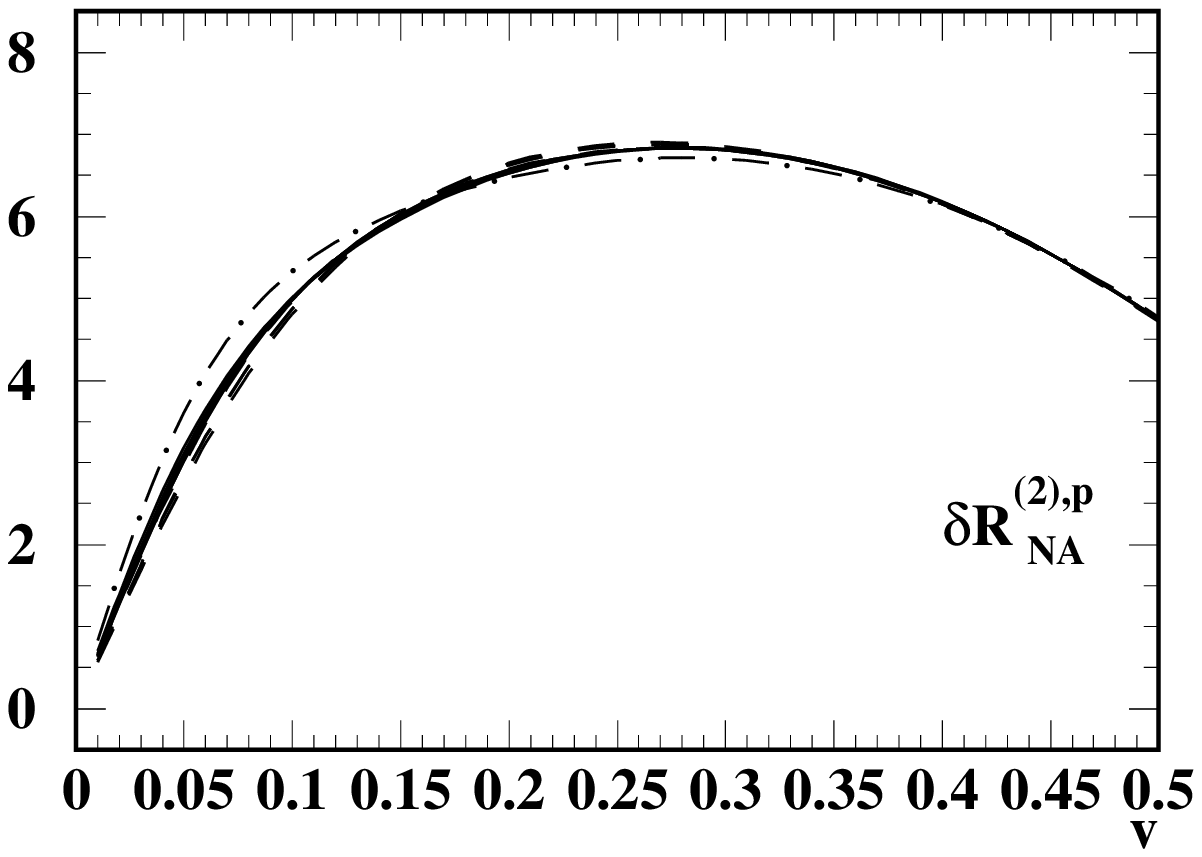}
   \\
   \epsfxsize=7.0cm
   \epsffile[110 290 460 540]{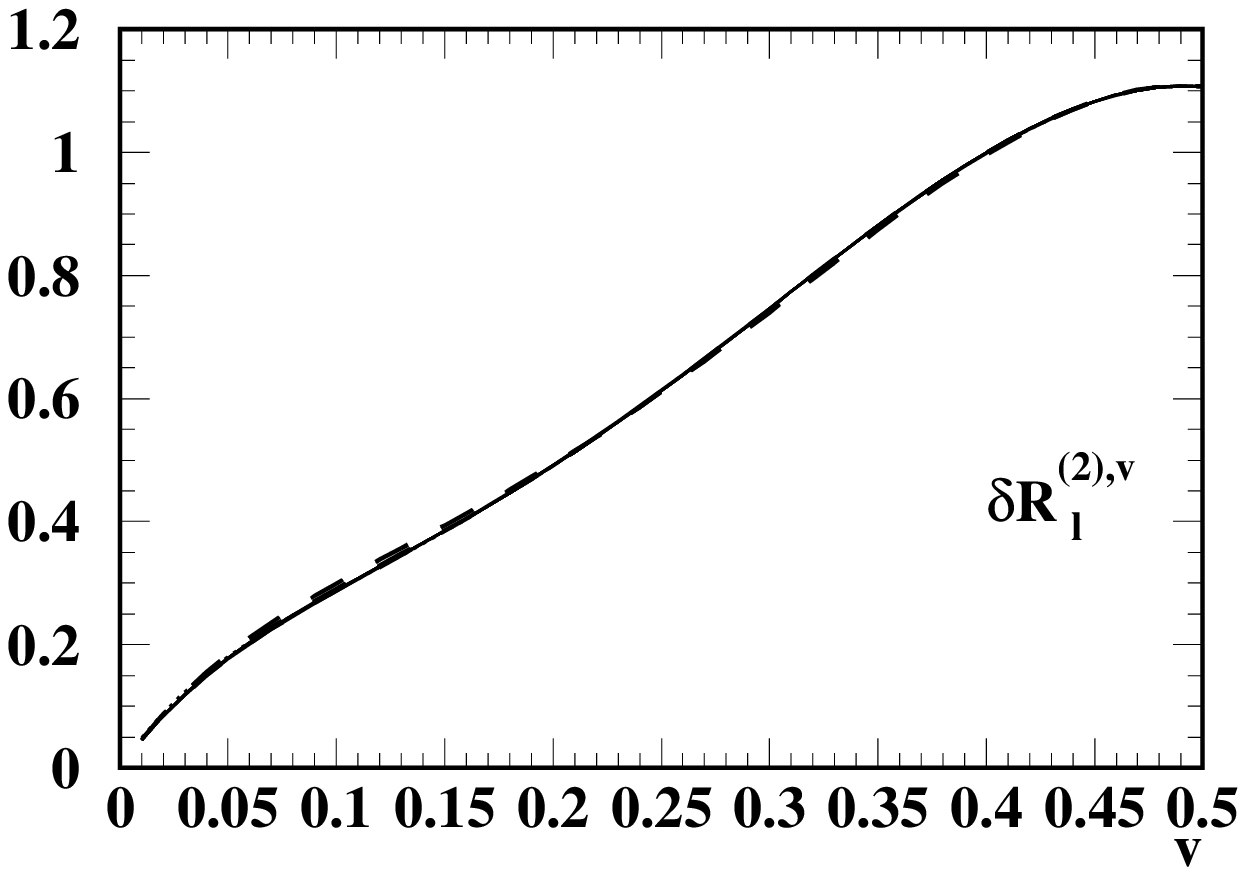}
   &
   \epsfxsize=7.0cm
   \epsffile[110 290 460 540]{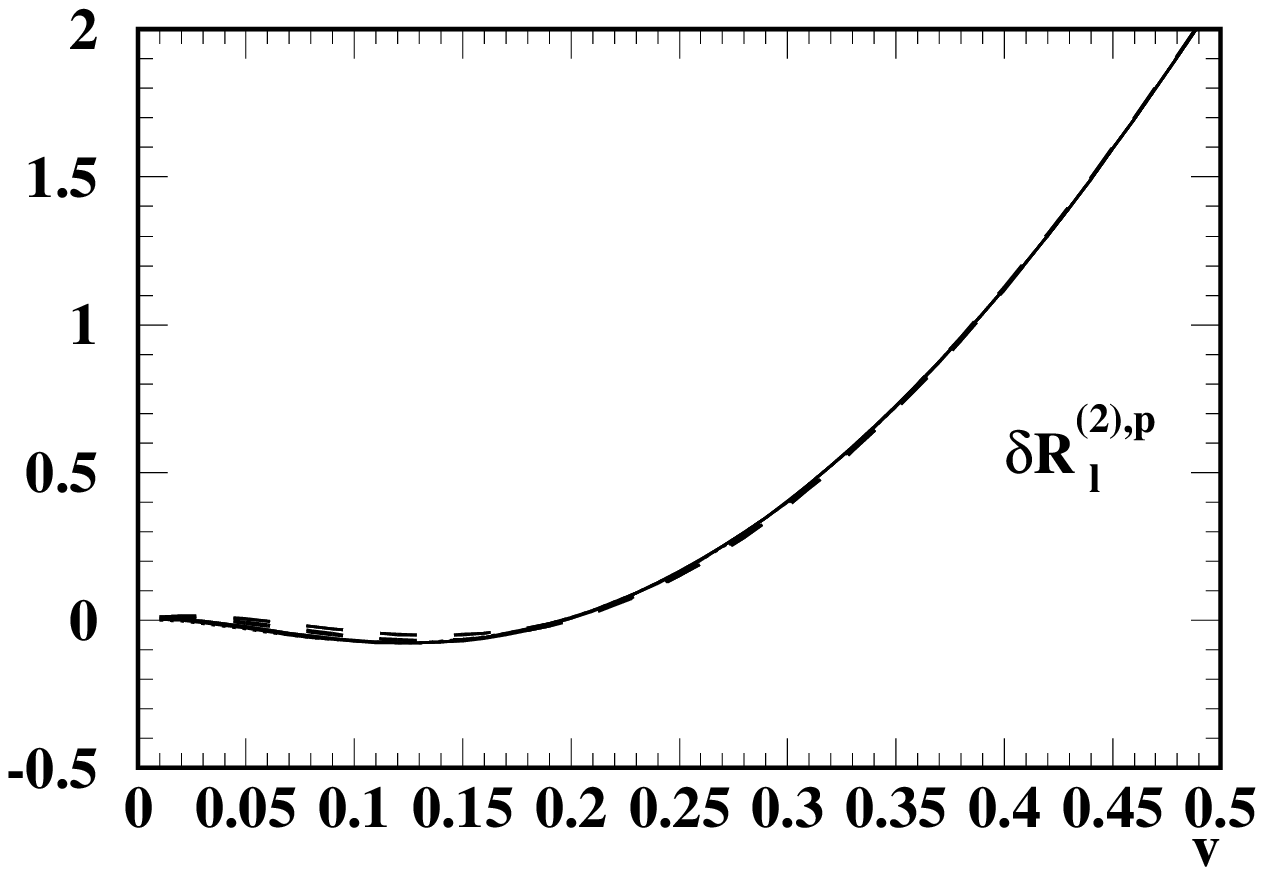}
 \end{tabular}
 \caption{\label{figvpvsub} $R^{(2),v}$ and $R^{(2),p}$ plotted against $v$.
          The leading threshold terms are subtracted. The dashed lines 
          contain only information up to $C_6$ whereas for the
          full curves also $C_7$ and $C_8$ is used. 
          The obvious exceptions are represented by the dash-dotted curves.
          They are described in the text.
          }
 \end{center}
\end{figure}

\begin{figure}[ht]
 \begin{center}
 \begin{tabular}{cc}
   \leavevmode
   \epsfxsize=7.0cm
   \epsffile[110 290 460 540]{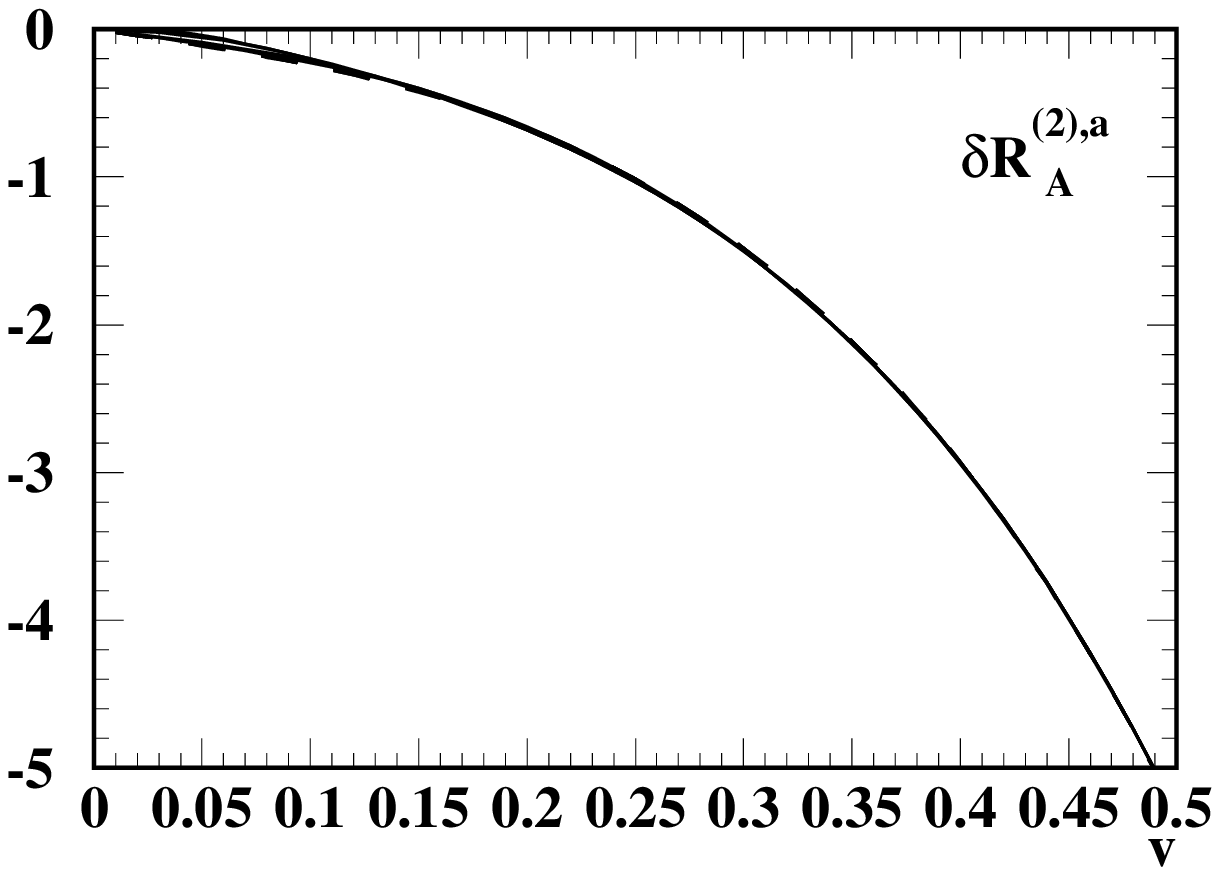}
   &
   \epsfxsize=7.0cm
   \epsffile[110 290 460 540]{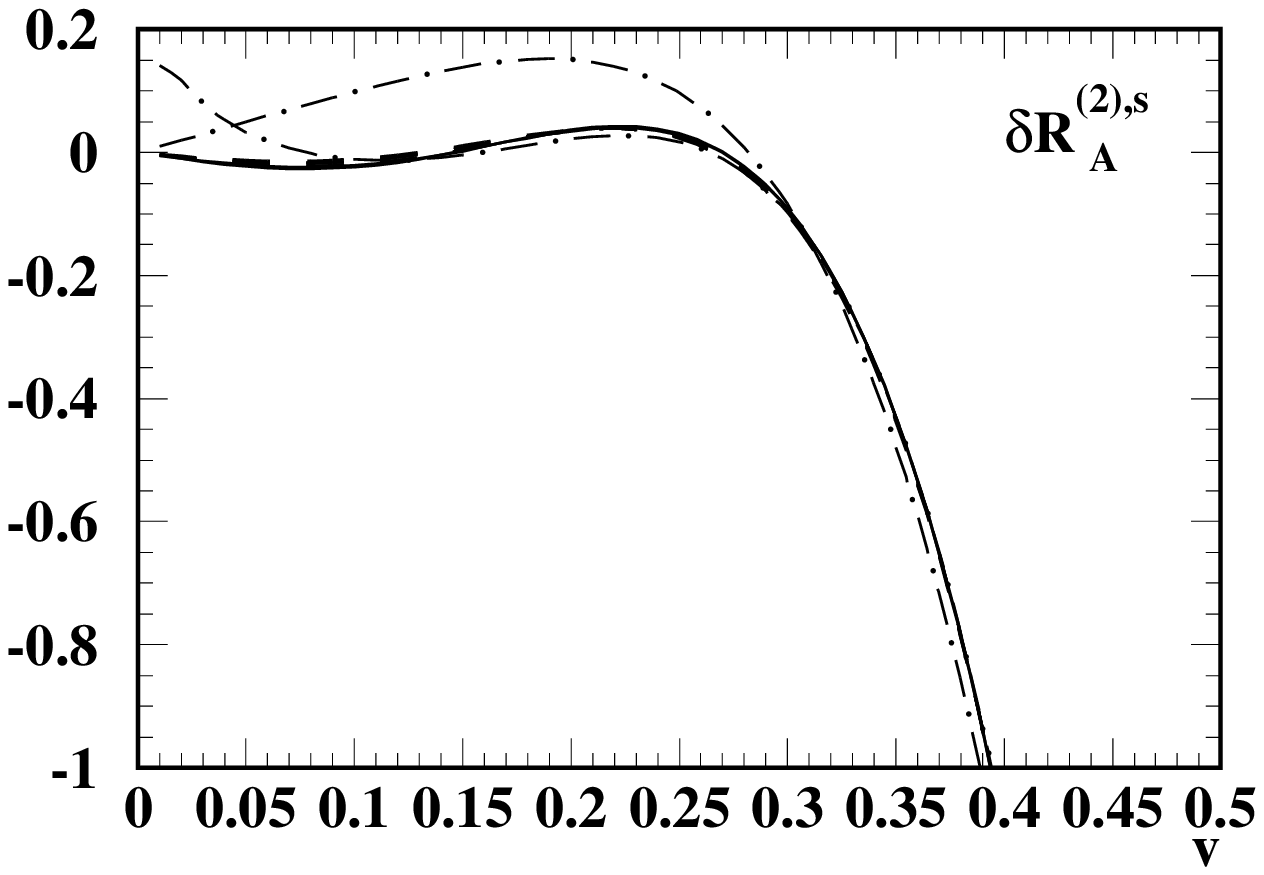}
   \\
   \epsfxsize=7.0cm
   \epsffile[110 290 460 540]{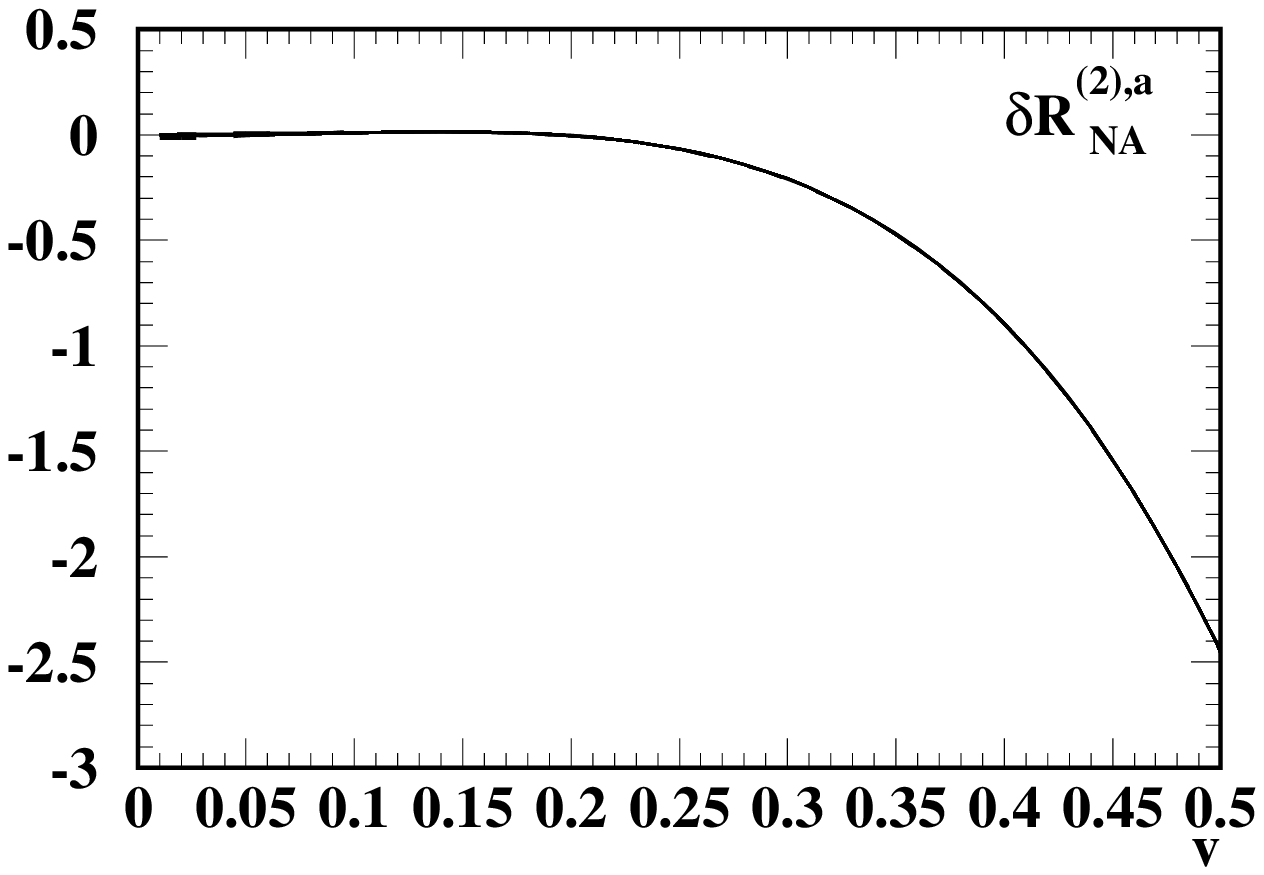}
   &
   \epsfxsize=7.0cm
   \epsffile[110 290 460 540]{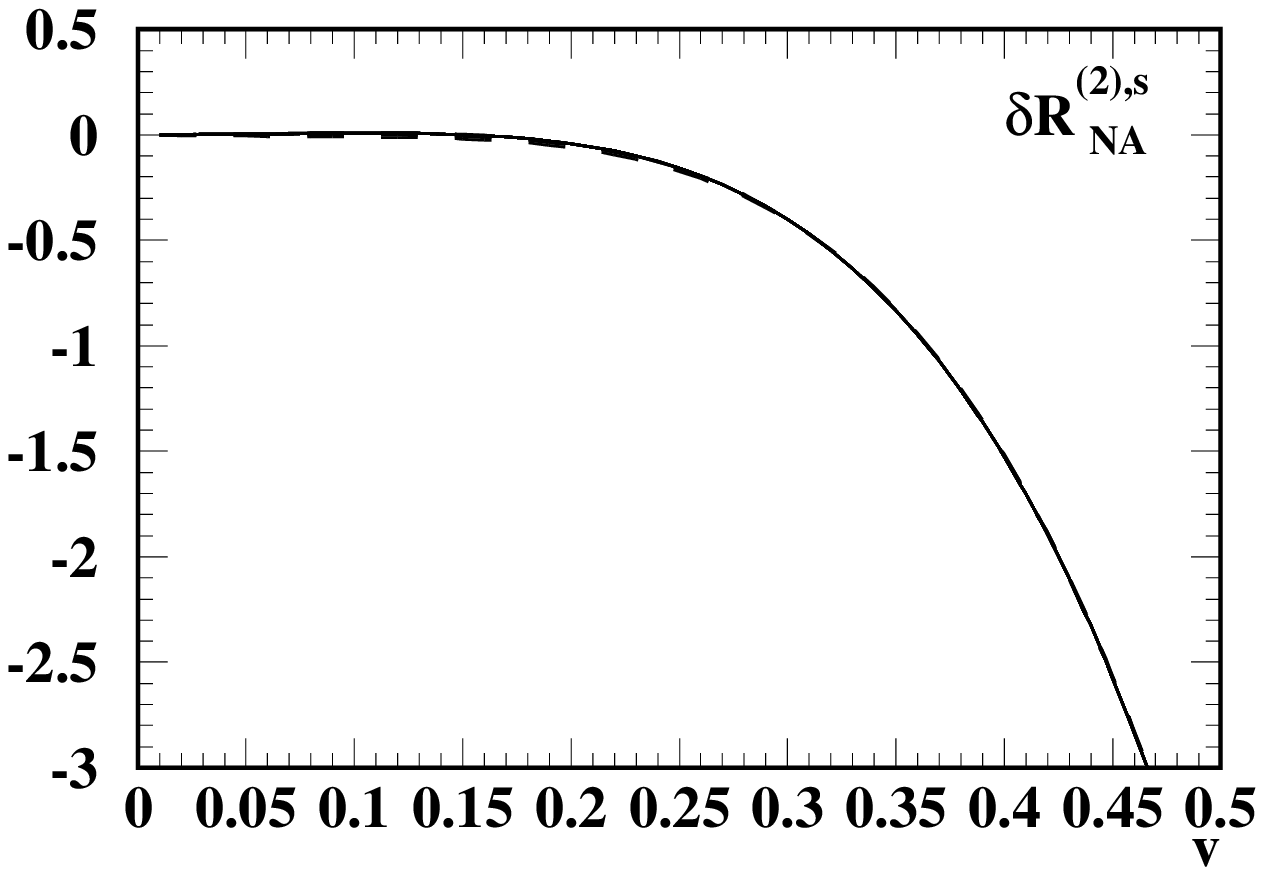}
   \\
   \epsfxsize=7.0cm
   \epsffile[110 290 460 540]{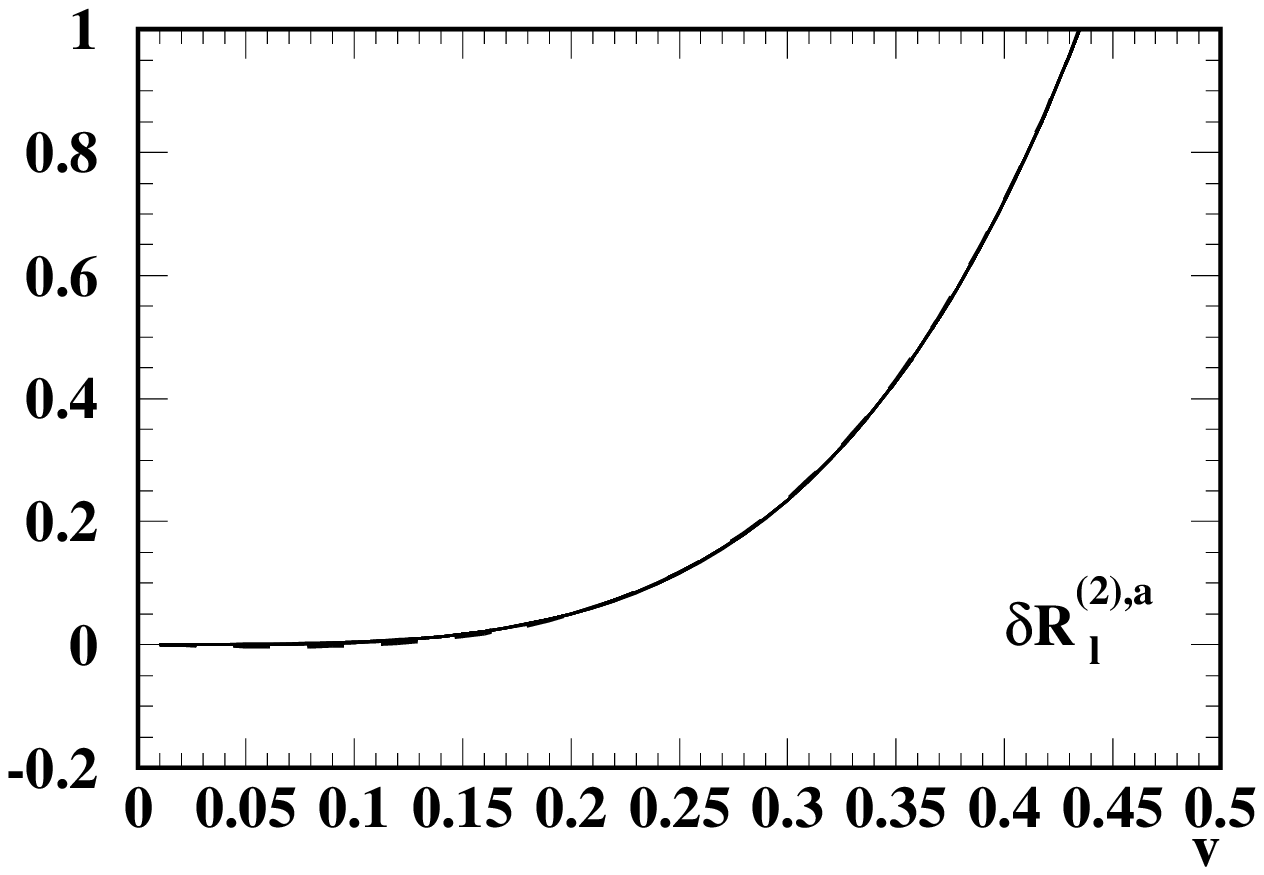}
   &
   \epsfxsize=7.0cm
   \epsffile[110 290 460 540]{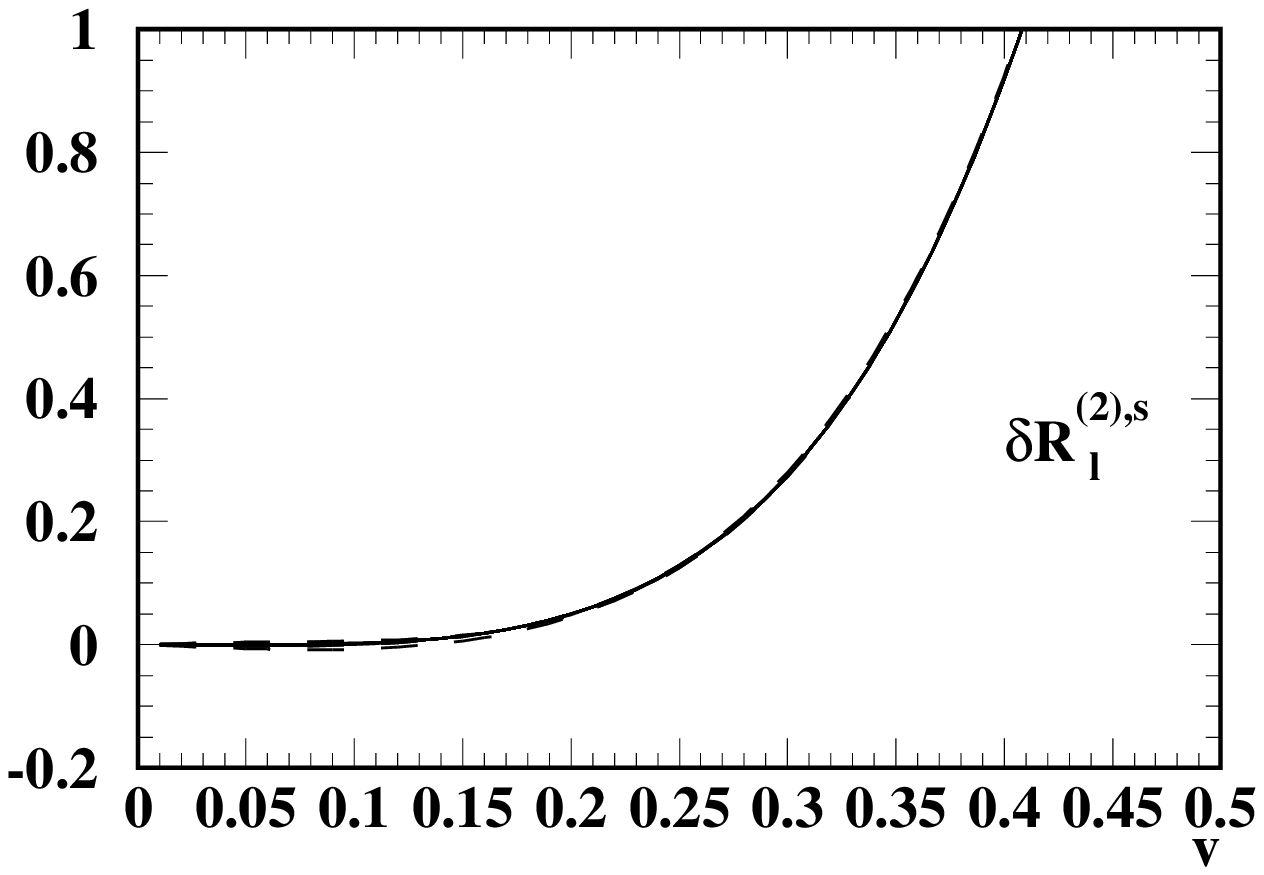}
 \end{tabular}
 \caption{\label{figasvsub} $R^{(2),a}$ and $R^{(2),s}$ plotted against $v$.
          The leading threshold terms are subtracted.
          The same notation as in Fig.~\ref{figvpvsub} is adopted.}
 \end{center}
\end{figure}

In this presentation it is not possible to notice any difference between
the different Pad\'e approximants. Minor differences can be seen 
after the leading terms at threshold are subtracted.
This can be seen in Figs.~\ref{figvpvsub} and \ref{figasvsub}. It
should be stressed that
the vertical scale is expanded by up to
a factor 100 in comparison with
Figs.~\ref{figvpv} and \ref{figasv}. 
The following notation is adopted: All Pad\'e approximations containing
information up to $C_6$ are plotted as a dashed line and the higher ones 
as full lines. The obvious exceptions are represented by a dash-dotted 
line and the exact results are drawn as dotted curves.
 
The vector case is already discussed in \cite{CheKueSte96}. 
The inclusion of $C_8$ into the analysis shows a further
stabilization of the results. The plot for the
abelian part in Fig.~\ref{figvpvsub}
contains altogether 14 Pad\'e approximations. Eight of them
contain information up to $C_6$ (dashed lines), and six contain also
information from $C_7$ and $C_8$.
These latter six lines coincide even on the expanded scale. 
The dash-dotted curve belongs to the $[2/5]$ result and contains a pole
for $\omega\approx1.06$.
In the case of the non-abelian contribution 15 Pad\'e
approximations are plotted. Again the dashed lines belong to
the lower order results. The dash-dotted curves are the results of two
Pad\'e approximants which have poles close to $\omega=1$
($[4/3]: \omega\approx1.07$ and $[2/5]: \omega\approx1.06$).

For the pseudo-scalar correlator we find similar results
concerning the behaviour of the Pad\'e approximations when
more information is included. 
The abelian contribution for the pseudo-scalar case contains
17 different results. The dash-dotted lines differ from the
remaining ones significantly. The corresponding Pad\'e approximants
are $[3/2]$ and $[2/5]$.
A spread between the different
Pad\'e approximants can also be observed for the non-abelian contribution
to the vector and pseudo-scalar cases. In both cases, however,
convergence is visible if more information is included
into the construction procedure. In the plot for 
$\delta R^{(2),p}_{NA}(s)$, e.g., the dash-dotted line correspond to the 
Pad\'e approximation $[2/3]$ containing only the first three
moments for $q^2\to0$. 
If this curve is ignored the spread is much less dramatic and the
difference between the remaining Pad\'e approximations shown is
very tiny and completely negligible.
The excellent agreement for the fermionic contribution 
in the pseudo-scalar case is 
comparable to the one for the vector correlator. In both cases the exact 
results
\cite{HoaTeu96},
plotted as a dotted line,
is indistinguishable from the approximations.

\begin{table}[t]
{\footnotesize
\renewcommand{\arraystretch}{1.2}
\begin{center}
\begin{tabular}{|l|r||l|r|}
\hline
&&&\\[-4mm]
  P.A.  &  $R^{(2),a}_A$  & P.A. & $R^{(2),s}_A$   \\
\hline
\hline
$[3/2]$ & 29.56 & $[3/3]$ & 31.31\\
$[2/3]$ & 29.59 & $[4/2]$ & 31.31\\
$[3/3]$ & 29.64 & $[2/4]$ & 31.44\\
$[4/2]$ & 29.69 & $[3/4]$ & 31.38\\
$[2/4]$ & 29.63 & $[5/2]$ & 31.39\\              
$[2/5]$ & 29.77 & $[2/5]$ & 31.39\\
$[3/5]$ & 29.96 & $[6/3]$ & 31.22\\
$[6/2]$ & 30.14 & $[3/6]$ & 31.28\\
$[2/6]$ & 29.82 &&\\ 
$[5/4]$ & 31.23 &&\\ 
$[4/5]$ & 31.71 &&\\
$[6/3]$ & 32.55 &&\\
\hline
\hline
exact & 31.75 & exact & 31.75 \\
\hline
\end{tabular}
\end{center}
}
\caption{\label{tabthras}
Comparison of the leading term at threshold for $R^{(2),\delta}_A$
$(\delta=a,s)$ with the exact expression. All Pad\'e approximants (P.A.)
contain two terms from the high energy region as input. Only the
number of moments from the expansion $q^2\to 0$ is different.}
\end{table}

Coming to the axial-vector and scalar case we would like to remind
the reader that for these correlators  
no singularities are present in the limit $v\to0$.
However, also here it is instructive to subtract the leading terms
and look closer to the remainders $\delta R^{(2),a}$ and $\delta R^{(2),s}$.
As can be seen by comparing Fig.~\ref{figasv}
and Fig.~\ref{figasvsub} the reduction in the scale lies between
a factor two and ten.
Also the very smooth behaviour of the subtracted results 
near threshold is clearly visible. One recognizes that, 
e.g. the remainders of the non-abelian and light-fermion
contributions shown in Fig.~\ref{figasvsub} are zero 
almost up to $v\approx 0.2$.

Let us now consider the threshold behaviour of the abelian contribution.
As mentioned in Section~\ref{seckinreg}
the corrections start with a term linear in $v$.
We are now in the position to compare the Pad\'e results with
the exact expressions.
In Tab.~\ref{tabthras} the numerical values of the coefficients
of both the expansion and the exact result is shown.
Although the analytically known terms are not incorporated into
the approximation method they are very well reproduced
by our method.

In the abelian part of the scalar correlator there are two
Pad\'e approximants which differ significantly (dash-dotted lines)
from the other eight results. 
One of them is a low-order Pad\'e approximation containing only
the information up to $C_2$ and the other one ($[4/3]$)
has a pole close to $\omega=1$ $(1.02)$ which is reflected in the 
enhancement in the vicinity of the threshold.

Both the non-abelian and fermionic contributions show an excellent 
agreement between the different Pad\'e approximations. We
should mention that at least 14 approximations are plotted and for 
$R_l^{(2),a}(s)$ and $R_l^{(2),s}(s)$ in addition the exact results are
included. Again no differences are visible.

\begin{figure}[ht]
 \begin{center}
 \begin{tabular}{cc}
   \leavevmode
   \epsfxsize=6.5cm
   \epsffile[110 290 460 540]{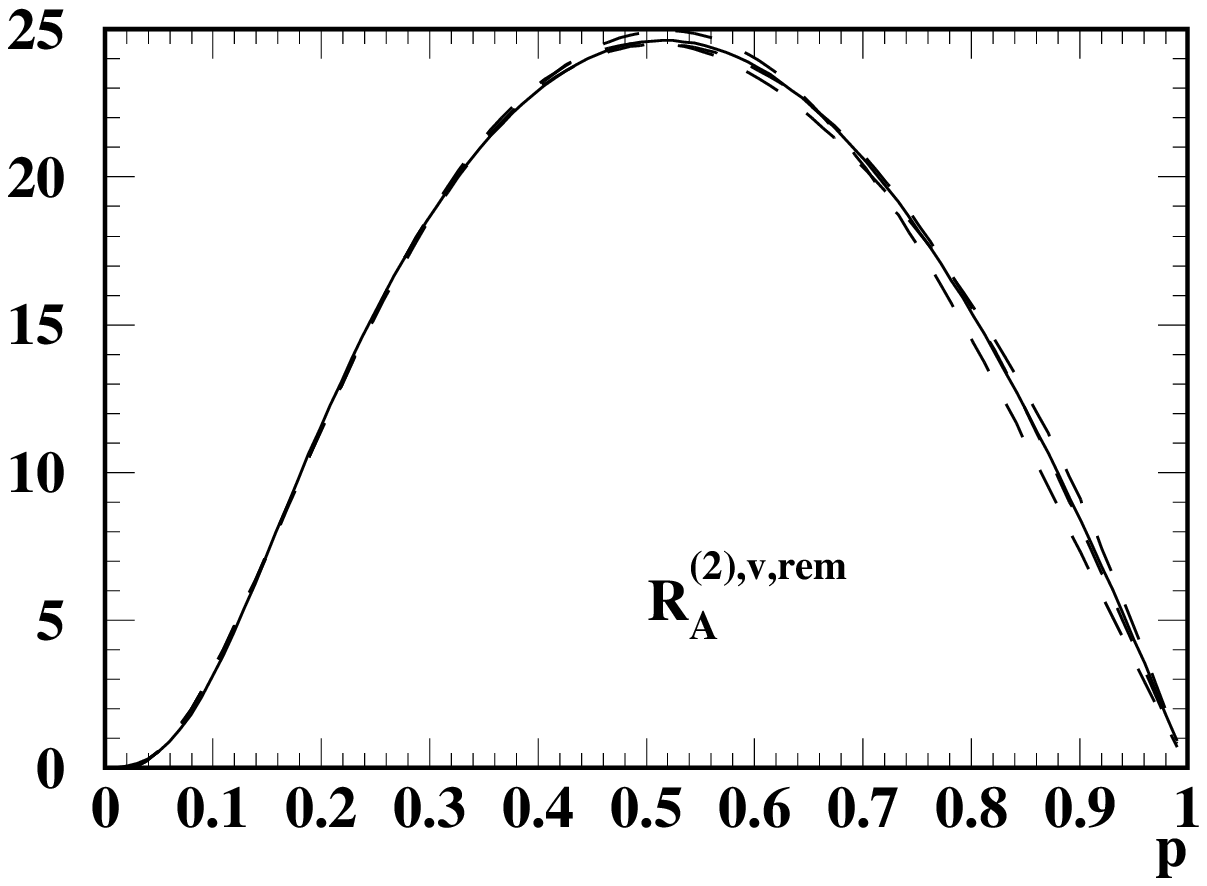}
   &
   \epsfxsize=6.5cm
   \epsffile[110 290 460 540]{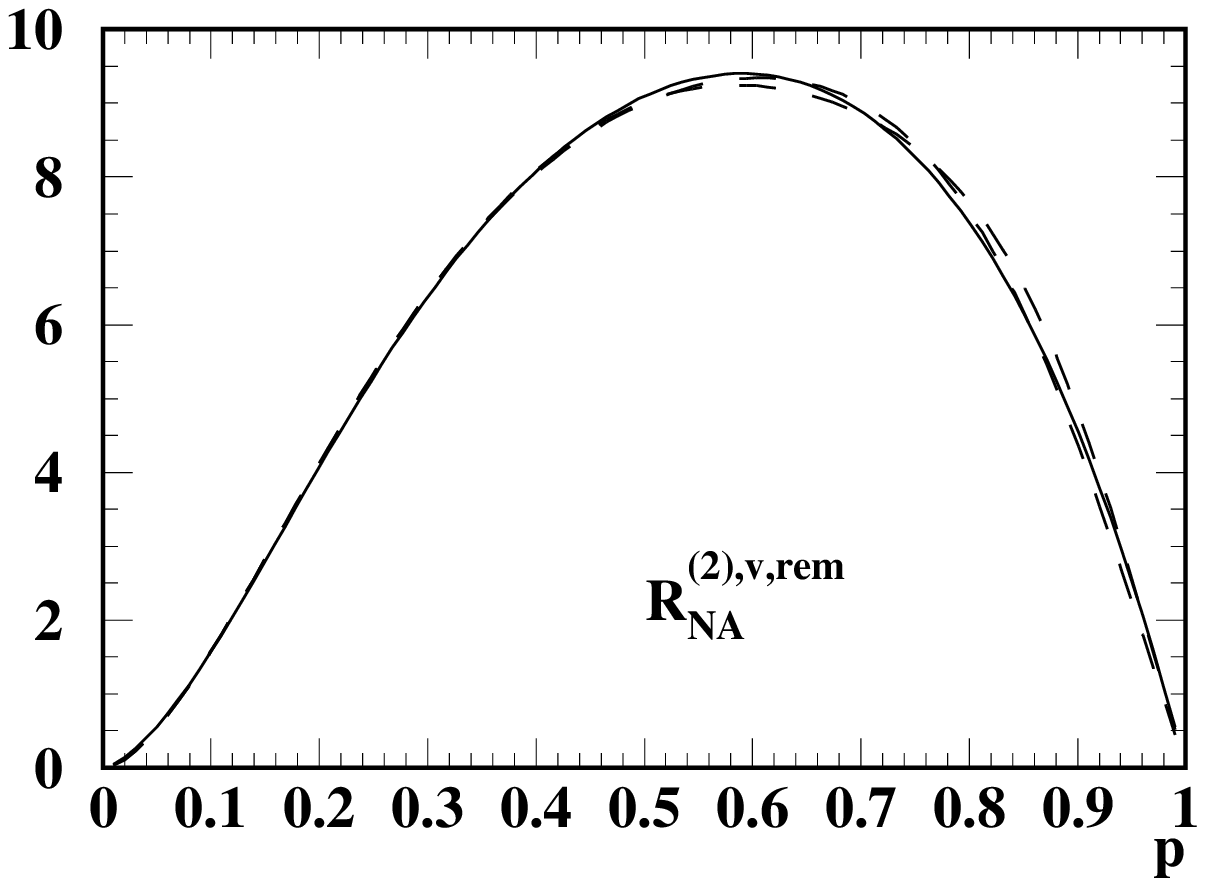}
   \\
   \epsfxsize=6.5cm
   \epsffile[110 290 460 540]{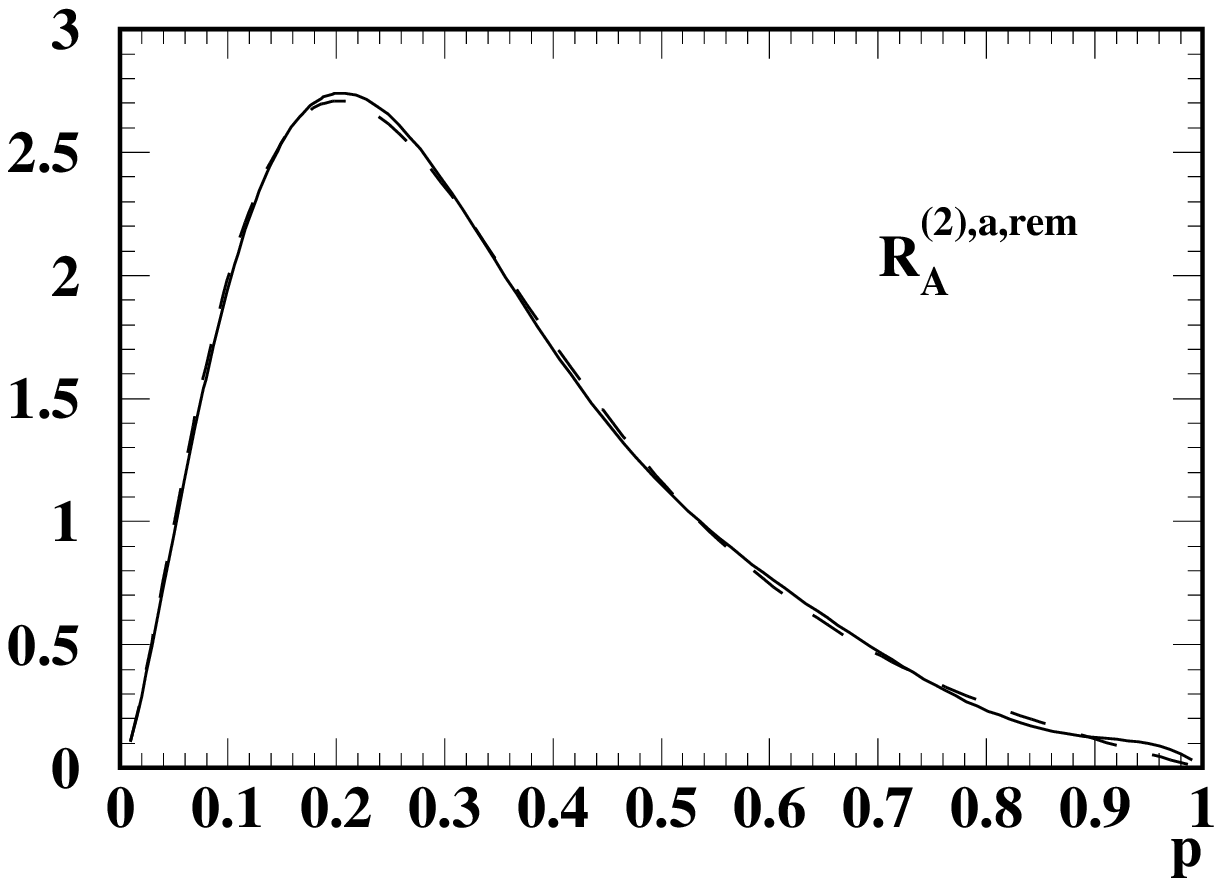}
   &
   \epsfxsize=6.5cm
   \epsffile[110 290 460 540]{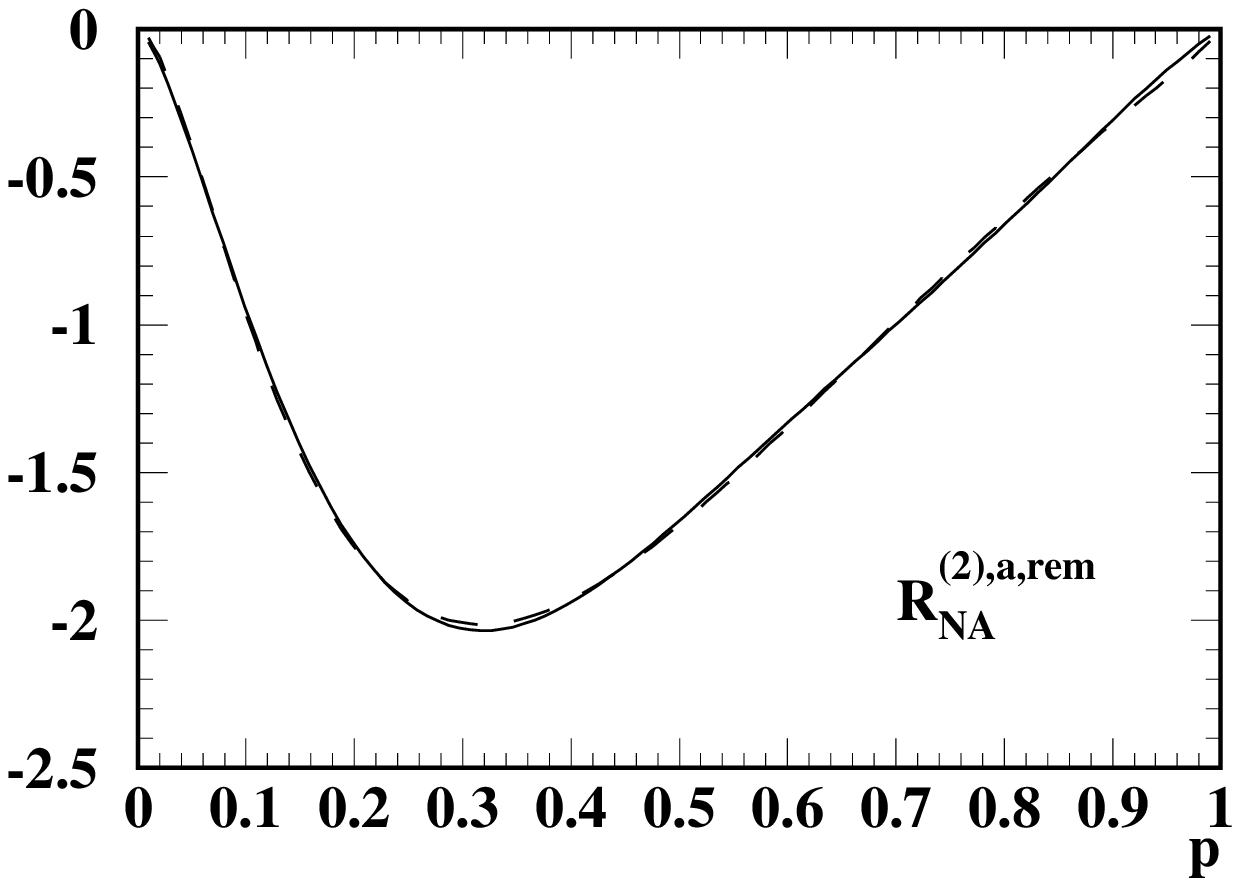}
   \\
   \epsfxsize=6.5cm
   \epsffile[110 290 460 540]{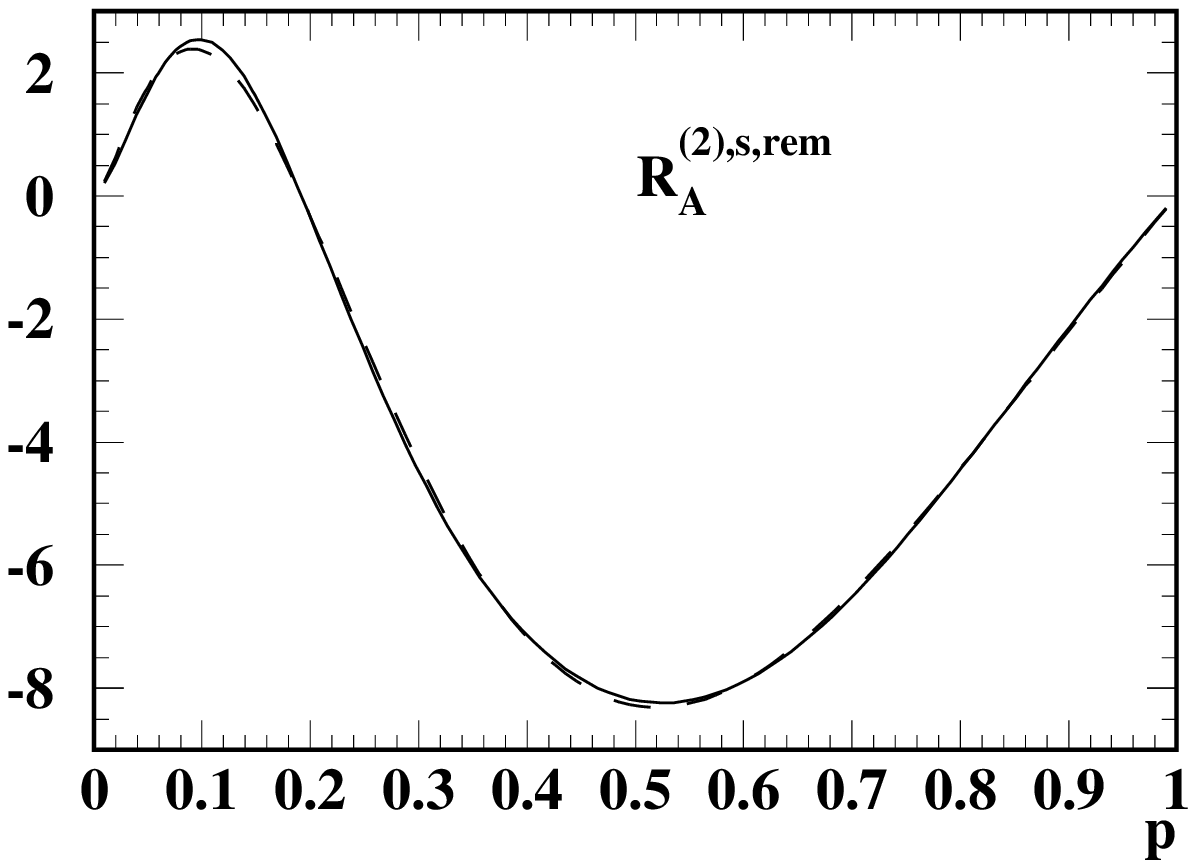}
   &
   \epsfxsize=6.5cm
   \epsffile[110 290 460 540]{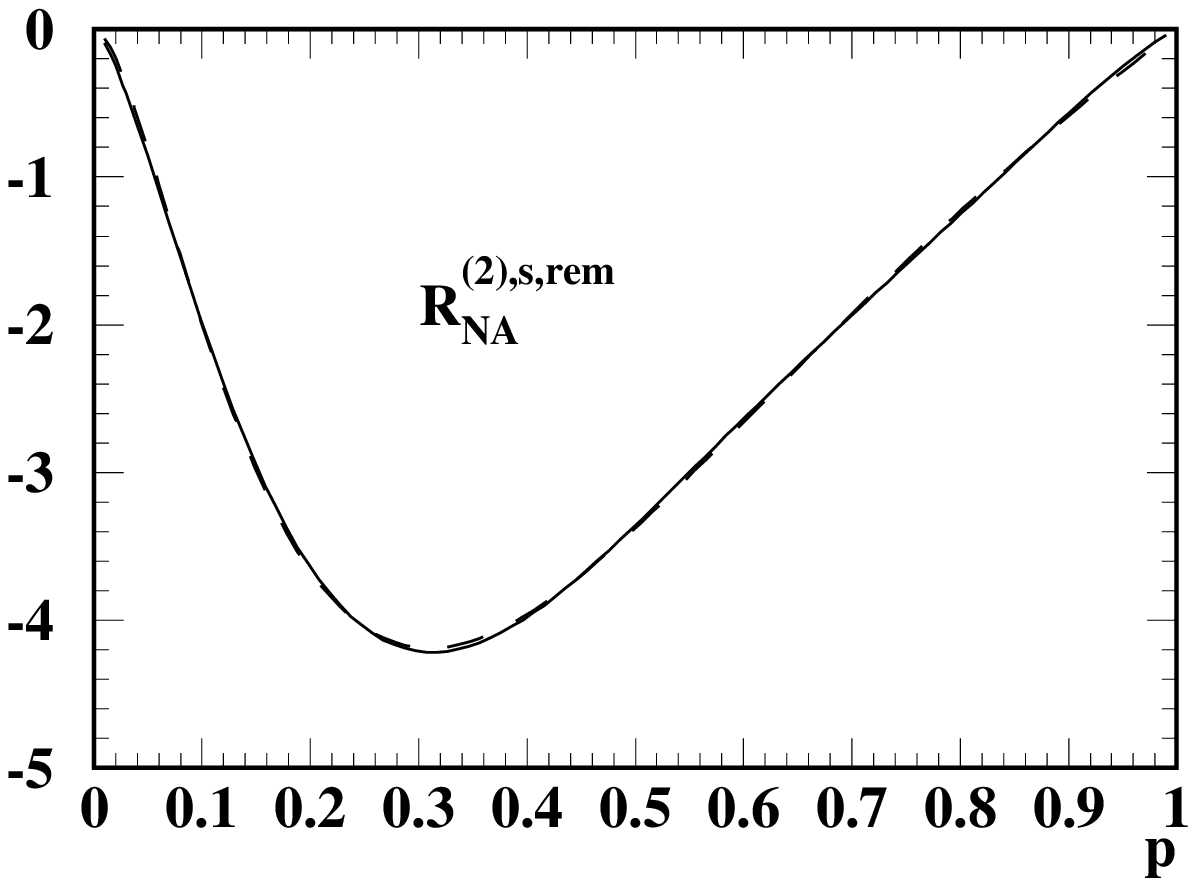}
   \\
   \epsfxsize=6.5cm
   \epsffile[110 290 460 540]{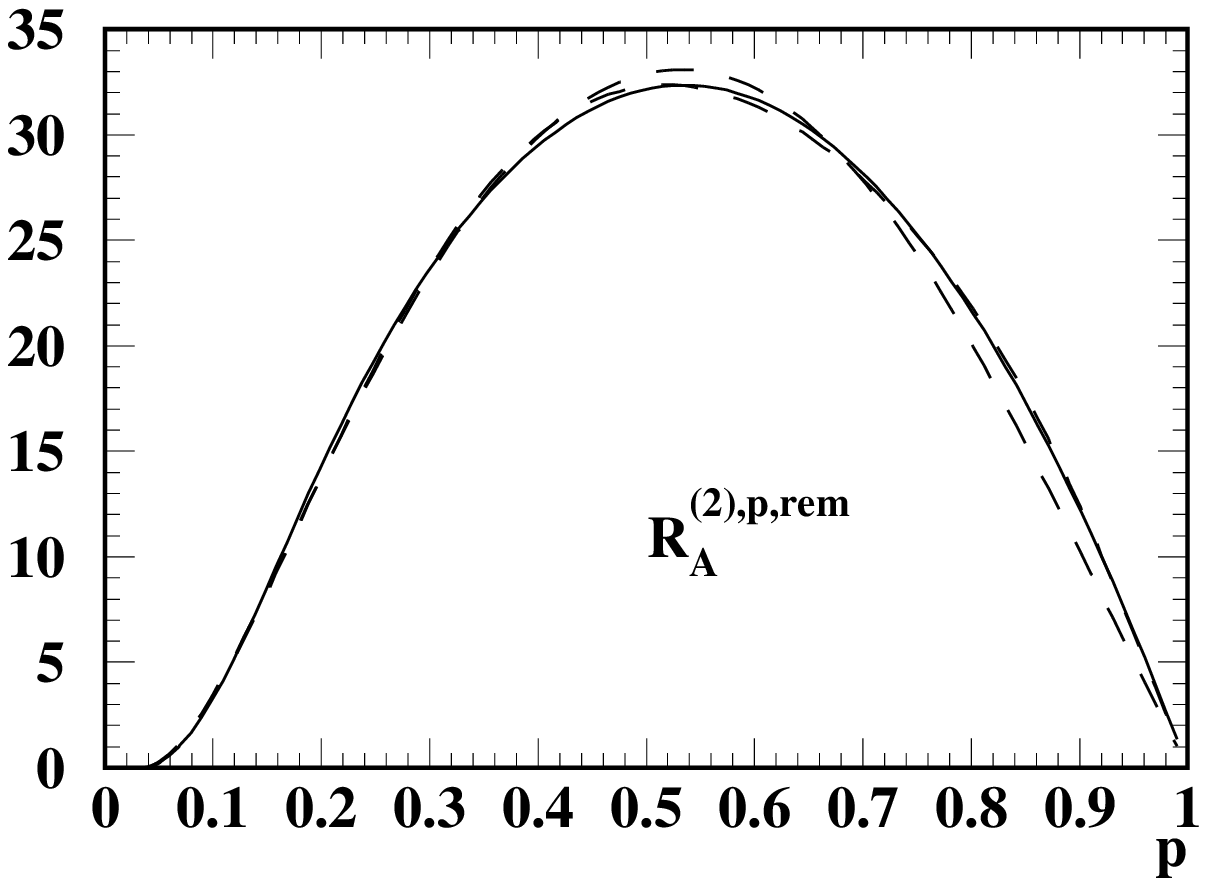}
   &
   \epsfxsize=6.5cm
   \epsffile[110 290 460 540]{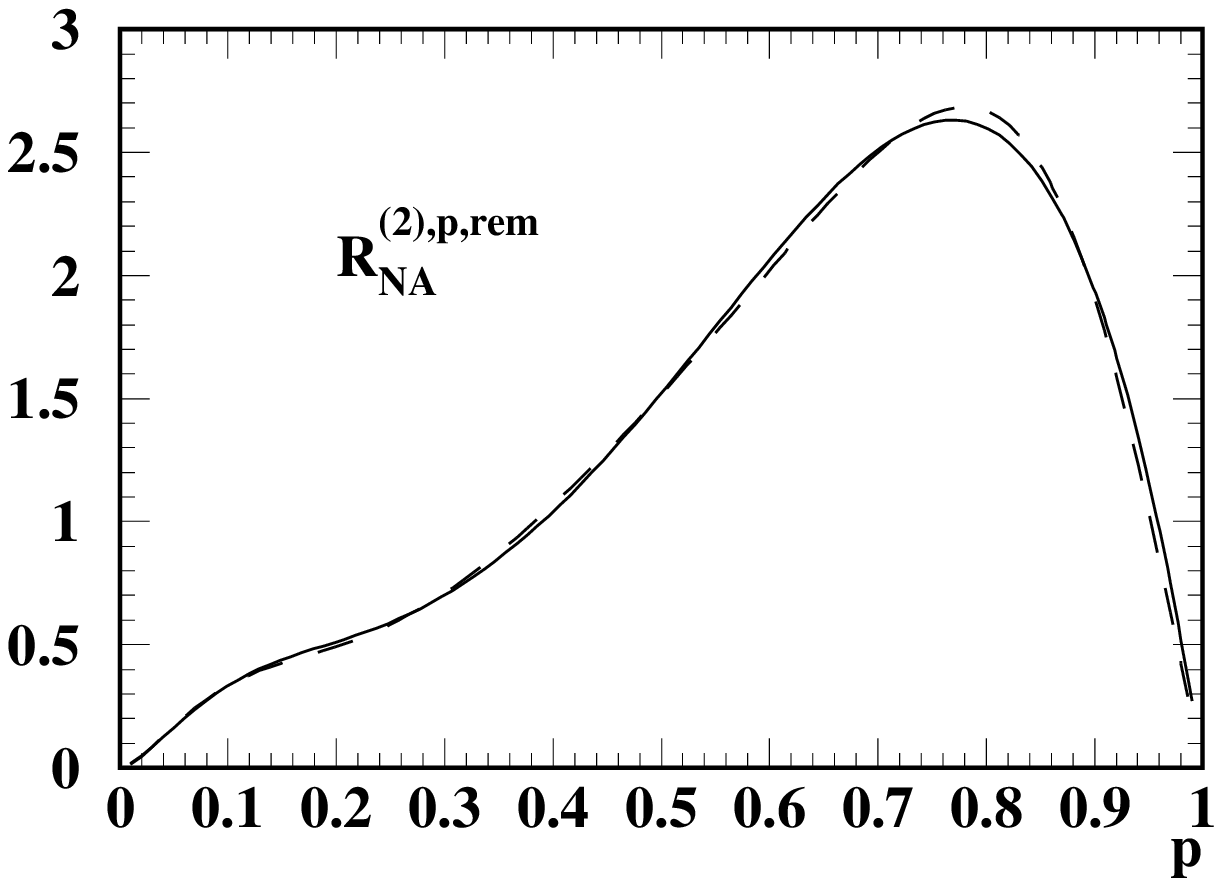}
 \end{tabular}
 \caption{\label{figappr} The remainder $R^{(2),\delta,rem}_x$, 
          $x\in\{A,NA\}$ is plotted for different Pad\'e approximants
          (dashed lines) together with the fit (solid line).}
 \end{center}
\end{figure}

Finally we present handy approximation formulae for the 
abelian and non-abelian contributions. The procedure 
used to get them is described in \cite{CheKueSte96}.
There the approximation formulae for the vector case are already 
listed. For completeness we repeat them at this point:
\begin{eqnarray}
R_A^{(2),v} &=& \frac{(1-v^2)^4}{v}\frac{3\pi^4}{8}
              - 4 R^{(1),v}
                +v\frac{2619}{64}-v^3\frac{2061}{64}
                +\frac{81}{8}\left(1-v^2\right)\ln\frac{1-v}{1+v}
\nonumber\\
&&\mbox{} 
-198\left(\frac{m^2}{s}\right)^{3/2} \left(v^4-2v^2\right)^6 
\nonumber\\
&&\mbox{} 
+100 p^{3/2} (1-p) \left[
2.21\, P_0(p)
-1.57\, P_1(p)
+0.27\, P_2(p)
\right],
\label{appfora}
\\
R_{NA}^{(2),v} &=& R_{g}^{(2),v}\Big|_{\xi=4}
               + v\frac{351}{32} - v^3\frac{297}{32}
\nonumber\\
&&\mbox{}  
-18\left(\frac{m^2}{s}\right)^{3/2}  \left(v^4-2v^2\right)^4
\nonumber\\
&&\mbox{}  
+50 p^{3/2} (1-p) \left[
1.73\, P_0(p)
-1.24\, P_1(p)
+0.64\, P_2(p)
\right],
\label{appfornaxi}
\\[4mm]
R_A^{(2),a} &=&
  -\frac{585}{32}v 
  + 18 v^3 
  + \left(\left(27 v - 27 v^3\right)\left(1-\ln2\right)\right)\zeta(2) 
  + \left(\frac{135}{8}\left(v - v^3\right)\right)\zeta(3)
\nonumber\\
&&\mbox{}  
  + \left(-\frac{189}{32}\left(1-v^2\right)\right)\ln\frac{1-v}{1+v} 
  + \left(-\frac{81}{16}\left(v - v^3\right)\right)\ln^2\frac{1-v}{1+v}
\nonumber\\
&&\mbox{}  
+50 p^{3/2} (1-p) \left[
 11.97\, P_0(p)
-25.37\, P_1(p)
+20.67\, P_2(p)
\right.
\nonumber\\
&&
\left.
\qquad
\mbox{} 
-9.048\, P_3(p)
+1.85\, P_4(p)
\right],
\\
R_{NA}^{(2),a} &=& R_{g}^{(2),a}\Big|_{\xi=4}
-\frac{9}{16}v 
+\frac{9}{4}v^3
+\left(\left(-\frac{135}{8}
       +\frac{27}{2}\ln2\right)\left(v-v^3\right)\right)\zeta(2)
\nonumber\\
&&\mbox{}  
+    \left(\frac{27}{16}\left(v-v^3\right)\right)\zeta(3)
+\left(-\frac{135}{16}\left(1-v^2\right)\right)\ln\frac{1-v}{1+v}
\nonumber\\
&&\mbox{}  
+50 p^{3/2} (1-p) \left[
-1.88 P_0(p)
+3.31 P_1(p)
-1.96 P_2(p)
+0.483 P_3(p)
\right],
\\[4mm]
R_A^{(2),s} &=&
-\frac{1125}{64}v + \frac{1779}{64}v^3
+ \left(\frac{189}{4}v 
               -\frac{261}{4}v^3
               +\left(-27v
                      +45v^3
                \right)\ln2
          \right) \zeta(2)
\nonumber\\
&&\mbox{} 
+\left(\frac{189}{8}v - \frac{279}{8}v^3\right)\zeta(3)
+\left(-\frac{63}{8} + \frac{297}{16}v^2\right)\ln\frac{1 - v}{1 + v}
\nonumber\\
&&\mbox{} 
+\left(-\frac{243}{16}v + \frac{297}{16}v^3\right)\ln^2\frac{1 - v}{1 + v}
\nonumber\\
&&\mbox{} 
+ 240\left(\frac{m^2}{s}\right)^{3/2}\left(v^4-2v^2\right)^4
\nonumber\\
&&\mbox{} 
+ 50p^{3/2}(1 - p)
\left[
1.30\,P_0(p)
-4.37\,P_1(p)
+3.58\,P_2(p)
-0.91\,P_3(p)
\right],
\\
R_{NA}^{(2),s} &=& R_{g}^{(2),s}\Big|_{\xi=4}
+ \frac{99}{32}v + \frac{147}{32}v^3
+ \left(-\frac{135}{8}v + \frac{225}{8}v^3
        +\left(\frac{27}{2}v - \frac{45}{2}v^3\right)\ln2\right)\zeta(2)
\nonumber\\
&&\mbox{} 
+ \left(-\frac{27}{16}v + \frac{9}{16}v^3\right)\zeta(3)
+ \left(-\frac{45}{4} + \frac{135}{8}v^2\right)\ln\frac{1-v}{1+v}
\nonumber\\
&&\mbox{} 
+ 50p^{3/2}(1 - p)
\left[
-3.94\,P_0(p)
+6.97\,P_1(p)
-4.11\,P_2(p)
+1.00\,P_3(p)
\right],
\\[4mm]
R_A^{(2),p} &=&
+ \frac{(1-v^2)^4}{v}\frac{\pi^4}{4}
- 3 R^{(1),p}
+ \frac{2763}{64}v 
- \frac{813}{64}v^3
+ \left(-\frac{9}{4}v - \frac{63}{4}v^3 
\right.
\nonumber\\
&&
\left.
\mbox{} 
        +\left(9v + 9v^3\right)\ln2\right)\zeta(2)
+ \left(-\frac{27}{8}v - \frac{63}{8}v^3\right)\zeta(3)
\nonumber\\
&&\mbox{} 
+ \left(\frac{81}{4} + \frac{63}{16}v^2\right)\ln\frac{1-v}{1+v}
+ \left(-\frac{27}{16}v + \frac{81}{16}v^3\right)\ln^2\frac{1-v}{1+v}
\nonumber\\
&&\mbox{} 
 -300 \left(\frac{m^2}{s}\right)^{3/2}\left(v^4-2v^2\right)^6
\nonumber\\
&&\mbox{} 
+ 100p^{3/2}(1 - p)
\left[
2.79\,P_0(p)
-1.83\,P_1(p)
+0.419\,P_2(p)
\right],
\\
R_{NA}^{(2),p} &=& R_{g}^{(2),p}\Big|_{\xi=4}
+ \frac{459}{32}v 
- \frac{213}{32}v^3
+       \left(\frac{45}{8}-\frac{9}{2}\ln2\right)
        \left(v + v^3\right)\zeta(2) 
\nonumber\\
&&\mbox{} 
+ \left(-\frac{27}{16}v + \frac{9}{16}v^3\right)\zeta(3)
+ \frac{45}{8}v^2\ln\frac{1-v}{1+v}
\nonumber\\
&&\mbox{} 
+ 12\left(\frac{m^2}{s}\right)^{3/2}\left(v^4-2v^2\right)^4
\nonumber\\
&&\mbox{} 
+ 50p^{3/2}(1 - p)
\left[
0.354\,P_0(p)
-0.251\,P_1(p)
+0.456\,P_2(p)
\right],
\end{eqnarray}
where $p=(1-v)/(1+v)$, $\zeta(2)=\pi^2/6$, $\zeta(3)\approx1.20206$ 
and $P_i(p)$ are the Legendre polynoms:
\begin{eqnarray}
&&
P_0(p)=1,\,\,\, P_1(p)=p,\,\,\, P_2(p)=-\frac{1}{2}+\frac{3}{2}p^2,
\\
&&
P_3(p)=-\frac{3}{2}p+\frac{5}{2}p^3,
\,\,\,
P_4(p)=\frac{3}{8}-\frac{15}{4}p^2+\frac{35}{8}p^4.
\end{eqnarray}
$R_{g}^{(2),\delta}$ is the exact result for the
gluonic double-bubble to be reconstructed from
the fermionic contribution
\cite{CheHoaKueSteTeu96}:
\begin{eqnarray}
R_g^{(2),\delta}\Big|_{\xi=4} &=& 
-\frac{11}{4} R_l^{(2),\delta} - \frac{2}{3} R^{(1),\delta}.
\end{eqnarray}
For some cases the degree of the polynomial used for the fit has to be 
increased in order to end up with reasonable approximations.
The first lines of the result contain 
the exactly known high energy and threshold contributions.
The proceeding lines represent the numerically small remainder,
$R_x^{(2),\delta,rem}$ with $x\in\{A,NA\}$,
which is plotted in Fig.~\ref{figappr} together with
the result from the Pad\'e approximation.


\section{\label{seccon}Conclusions and summary}

The vacuum polarization function has been evaluated in order
$\alpha_s^2$ for vector, axial-vector, scalar and pseudo-scalar currents. 
The results take full account of the quark mass and are applicable between
the production threshold and the high energy region. The method is based
on the Pad\'e approximation and uses the leading terms at high energies
and at threshold plus the lowest eight coefficients of the Taylor series of
$\Pi^\delta(q^2)$ around zero. The stability of this approximation has been
verified and excellent agreement between the present result and the
predictions based on the high energy expansion is observed. These results
can be used to evaluate the cross section for top pair production in
electron positron annihilation through the vector and axial-vector 
current and the decay rate of a scalar or pseudo-scalar Higgs boson 
into top quarks in the full kinematical region and in next-to-leading order.


\vspace{1em}

\centerline{\bf Acknowledgments}
\smallskip\noindent
We would like to thank A.H. Hoang and T. Teubner for interesting
discussions and for providing us with the analytic results 
for $R_l^{(2),\delta}$ prior to publication which were
crucial for our tests of the approximation methods.
\noindent


\vspace{5ex}                  
\noindent
{\Large \bf Appendix}

\renewcommand {\theequation}{\Alph{section}.\arabic{equation}}
\begin{appendix}

\setcounter{equation}{0}
\section{\label{appmsbar}$\overline{\mbox{MS}}$ definition of the
         polarization functions}

In this appendix we present the missing pieces needed to express
the polarization functions, $\Pi^\delta(q^2)$, in the
$\overline{\mbox{MS}}$ scheme which means that in the expression
obtained after the renormalization of $\alpha_s$ and $m$
only the poles are subtracted.
Expressing the results still in terms of the on-shell mass, $m$, 
$\bar{\Pi}^\delta(q^2)$ reads:
\begin{eqnarray}
\bar{\Pi}^\delta(q^2) &=& \frac{3}{16\pi^2}
\left[
\bar{C}^\delta_{-1}\frac{1}{z} + \bar{C}^\delta_0
\right]
+\Pi^\delta(q^2),
\end{eqnarray}
where the bar only refers to the overall renormalization. For the
different cases we get
($\bar{C}^v_0$ is already listed in \cite{CheKueSte96}):
\begin{eqnarray}
\bar{C}^v_0 &=&
\frac{4}{3} \lnmum + 
    \frac{\alpha_s}{\pi} C_F 
\left(\frac{15}{4} + \lnmum
\right)
\nonumber\\&&\mbox{}
 + 
\left(\frac{\alpha_s}{\pi}\right)^2 
\Bigg[
      C_F^2 
\left(
\frac{77}{144} 
-\frac{1}{8} \lnmum 
+          (5 - 8 \ln2) \zeta(2) 
+ \frac{1}{48} \zeta(3)
\right)
\nonumber\\&&\mbox{}
 + 
          C_F C_A 
\left(
\frac{14977}{2592} 
+ \frac{157}{36} \lnmum 
+    \frac{11}{24} \lnmum^2 
+   (-\frac{4}{3} + 4 \ln2) \zeta(2) 
+ \frac{127}{96} \zeta(3)
\right)
\nonumber\\&&\mbox{}
 + 
      C_F T n_l 
\left(
      -\frac{917}{648} 
-\frac{14}{9} \lnmum         
      -\frac{1}{6} \lnmum^2 
-\frac{4}{3} \zeta(2)
\right)
\nonumber\\&&\mbox{}
 + 
          C_F T 
\left(
-\frac{695}{162} 
-\frac{14}{9} \lnmum 
-\frac{1}{6} \lnmum^2 
+ \frac{8}{3} \zeta(2) 
+ \frac{7}{16} \zeta(3)
\right)
\Bigg]
,\\
\bar{C}^v_{-1} &=& 0,
\\
\bar{C}^a_0 &=&
-\frac{4}{3}
 +  \frac{4}{3} \lnmum
 +     \frac{\alpha_s}{\pi} C_F    \left(\frac{67}{36}
 +  \lnmum\right)
\nonumber\\&&\mbox{}
 +     \left(\frac{\alpha_s}{\pi}\right)^2     
\Bigg[ C_F^2 \left(\frac{131}{54}
 -\frac{1}{8} \lnmum
 +           (5 - 8 \ln2) \zeta(2)
 +  \frac{115}{288} \zeta(3)\right)
\nonumber\\&&\mbox{}
 +           C_F C_A 
\left(
\frac{4081}{432}
 +  \frac{71}{27} \lnmum
 +              \frac{11}{24} \lnmum^2
 +              (-\frac{4}{3} +  4 \ln2) \zeta(2)
 -\frac{883}{576} \zeta(3)
\right)
\nonumber\\&&\mbox{}
 +        C_F T n_l \left(
-\frac{149}{72}
 -\frac{25}{27} \lnmum
 +              -\frac{1}{6} \lnmum^2
 -\frac{4}{3} \zeta(2)\right)
\nonumber\\&&\mbox{}
 +           C_F T \left(
-\frac{55}{12}
 -\frac{25}{27} \lnmum
               -\frac{1}{6} \lnmum^2
 +  \frac{8}{3} \zeta(2)
               -\frac{7}{48} \zeta(3)\right)
\Bigg]
,\\
\bar{C}^a_{-1} &=&
-2 \lnmum
 +      \frac{\alpha_s}{\pi} C_F      
\left(
-\frac{33}{8}
 +  \frac{3}{2} \lnmum
 +          \frac{3}{2} \lnmum^2
\right)
\nonumber\\&&\mbox{}
 +       \left(\frac{\alpha_s}{\pi}\right)^2       
\Bigg[ C_F^2 
\left(
    \frac{529}{64}
 -\frac{3}{2} B_4
 +             \frac{13}{2} \lnmum
              -\frac{15}{16} \lnmum^2
              -\frac{3}{4} \lnmum^3
\right.\nonumber\\&&\left.\mbox{}\quad
 +  \left(-\frac{15}{2} + 12 \ln2\right)
    \left(1- \lnmum\right) \zeta(2)
 +  \left(-12 -\frac{3}{2} \lnmum\right) \zeta(3)
 +             \frac{27}{4} \zeta(4)
\right)
\nonumber\\&&\mbox{}
 +  C_FC_A 
\left(
-\frac{1039}{96}
 +   \frac{3}{4} B_4
 +                \frac{143}{48} \lnmum
 +                \frac{109}{24} \lnmum^2
 +                \frac{11}{12} \lnmum^3
\right.\nonumber\\&&\left.\mbox{}\quad
 +  \left(2 - 6 \ln2\right)\left(1 - \lnmum\right) \zeta(2)
 +  \left(\frac{23}{6} +  \frac{3}{4} \lnmum\right) \zeta(3)
                 -\frac{27}{8} \zeta(4)
\right)
\nonumber\\&&\mbox{}
 +          C_F T n_l 
\left(
\frac{35}{12}
 -\frac{7}{12} \lnmum
                 -\frac{4}{3} \lnmum^2
                 -\frac{1}{3} \lnmum^3
 +     \left(2 - 2 \lnmum\right) \zeta(2)
 +  \frac{4}{3} \zeta(3)
\right)
\nonumber\\&&\mbox{}
             +  C_F T 
\left(
\frac{59}{12}
 -\frac{43}{12} \lnmum
                 -\frac{4}{3} \lnmum^2
                 -\frac{1}{3} \lnmum^3
 + \left(-4 +  4 \lnmum \right) \zeta(2)
 +  \frac{7}{12} \zeta(3)
\right)
\Bigg]
,\\
\bar{C}^s_0 &=&
-\frac{4}{3}
 +  2 \lnmum
 +     \frac{\alpha_s}{\pi} C_F    
\left(
\frac{41}{8}
 -\frac{3}{2} \lnmum
 +        -\frac{3}{2} \lnmum^2
\right)
\nonumber\\&&\mbox{}
 +     \left(\frac{\alpha_s}{\pi}\right)^2     
\Bigg[ C_F^2 
\left(
 -\frac{505}{64}
+\frac{3}{2} B_4 
            -\frac{13}{2} \lnmum
 +           \frac{15}{16} \lnmum^2
 +           \frac{3}{4} \lnmum^3
\right.\nonumber\\&&\left.\mbox{}\quad
 +  \left(\frac{25}{2} - 20 \ln2 \right)
    \left(1 - \frac{3}{5} \lnmum\right) \zeta(2)
 +  \left(\frac{93}{8}+  \frac{3}{2} \lnmum\right) \zeta(3)
            -\frac{27}{4} \zeta(4)
\right)
\nonumber\\&&\mbox{}
 +           C_F C_A 
\left(
   \frac{5429}{288}
-\frac{3}{4}B_4 
               -\frac{33}{16} \lnmum
               -\frac{109}{24} \lnmum^2
               -\frac{11}{12} \lnmum^3
\right.\nonumber\\&&\left.\mbox{}\quad
 +  \left(-\frac{10}{3} +  10 \ln2 \right)
    \left(1 -\frac{3}{5} \lnmum\right)      \zeta(2)
 + \left(-\frac{175}{48}      -\frac{3}{4} \lnmum\right)\zeta(3)
 +              \frac{27}{8} \zeta(4)
\right)
\nonumber\\&&\mbox{}
 +        C_F T n_l 
\left(
-\frac{191}{36}
 +  \frac{1}{4} \lnmum
 +              \frac{4}{3} \lnmum^2
 +              \frac{1}{3} \lnmum^3
 +        \left(-\frac{10}{3}+  2 \lnmum\right) \zeta(2)
 +              -\frac{4}{3} \zeta(3)
\right)
\nonumber\\&&\mbox{}
 +           C_F T 
\left(
-\frac{733}{72}
 +  \frac{13}{4} \lnmum
 +              \frac{4}{3} \lnmum^2
 +              \frac{1}{3} \lnmum^3
 +     \left(\frac{20}{3} - 4 \lnmum \right) \zeta(2)
 +              -\frac{49}{48} \zeta(3)
\right)
\Bigg]
,\nonumber\\ 
\\
\bar{C}^s_{-1} &=&
-1 - 3 \lnmum
 +       \frac{\alpha_s}{\pi} C_F      
\left(
-\frac{9}{2}
 +  9 \lnmum
 +          \frac{9}{2} \lnmum^2
\right)
\nonumber\\&&\mbox{}
 +       \left(\frac{\alpha_s}{\pi}\right)^2       
\Bigg[ C_F^2 
\left(
   \frac{1673}{64}
-3 B_4
 +             \frac{45}{8} \lnmum
              -\frac{207}{16} \lnmum^2
              -\frac{9}{2} \lnmum^3
\right.\nonumber\\&&\left.\mbox{}\quad
 + \left(-\frac{15}{4} + 6\ln2\right)
   \left(1 - 6 \lnmum \right) \zeta(2) 
- 27 \zeta(3)
 +  \frac{27}{2} \zeta(4)
\right)
\nonumber\\&&\mbox{}
 +             C_A C_F 
\left(
-\frac{2641}{192}
 +   \frac{3}{2} B_4
 +                \frac{163}{8} \lnmum
 +                \frac{251}{16} \lnmum^2
 +                \frac{11}{4} \lnmum^3
\right.\nonumber\\&&\left.\mbox{}\quad
 +    \left(1 - 3 \ln2\right)
      \left(1 - 6  \lnmum\right) \zeta(2)
 +                10 \zeta(3)
 -\frac{27}{4} \zeta(4)
\right)
\nonumber\\&&\mbox{}
 +          C_F T n_l 
\left(
\frac{161}{48}
 -\frac{11}{2} \lnmum
                 -\frac{19}{4} \lnmum^2 
-               \lnmum^3
 +       \left(1 - 6 \lnmum\right) \zeta(2)
 +  4 \zeta(3)
\right)
\nonumber\\&&\mbox{}
 +             C_F T 
\left(
\frac{99}{16}
 -\frac{23}{2} \lnmum
                 -\frac{19}{4} \lnmum^2 
-               \lnmum^3
 +            \left(-2 +  12 \lnmum\right) \zeta(2)
                 -\frac{7}{2} \zeta(3)
\right)
\Bigg]
,\\
\bar{C}^p_0 &=&
 2 \lnmum
 +     \frac{\alpha_s}{\pi} C_F    
\left(
\frac{33}{8}
 -\frac{3}{2} \lnmum
 +        -\frac{3}{2} \lnmum^2
\right)
\nonumber\\&&\mbox{}
 +     \left(\frac{\alpha_s}{\pi}\right)^2     
\Bigg[ C_F^2 
\left(
 -\frac{529}{64}
 +\frac{3}{2} B_4
 +           -\frac{13}{2} \lnmum
 +           \frac{15}{16} \lnmum^2
 +           \frac{3}{4} \lnmum^3
\right.\nonumber\\&&\left.\mbox{}\quad
 +  \left(\frac{15}{2} - 12 \ln2\right)
    \left(1 - \lnmum\right) \zeta(2)
 +  \left(12 +  \frac{3}{2} \lnmum\right) \zeta(3)
 +           -\frac{27}{4} \zeta(4)
\right)
\nonumber\\&&\mbox{}
           +  C_A C_F 
\left(
\frac{1039}{96}
    -\frac{3}{4} B_4
               -\frac{143}{48} \lnmum
               -\frac{109}{24} \lnmum^2
               -\frac{11}{12} \lnmum^3
\right.\nonumber\\&&\left.\mbox{}\quad
 + \left(-2 + +  6 \ln2 \right)\left(1 - \lnmum\right) \zeta(2)
 + \left(-\frac{23}{6} - \frac{3}{4} \lnmum\right) \zeta(3)
 +              \frac{27}{8} \zeta(4)
\right)
\nonumber\\&&\mbox{}
 +        C_F T n_l 
\left(
-\frac{35}{12}
 +  \frac{7}{12} \lnmum
 +              \frac{4}{3} \lnmum^2
 +              \frac{1}{3} \lnmum^3
 +              \left(-2 +  2 \lnmum\right) \zeta(2)
 -\frac{4}{3} \zeta(3)
\right)
\nonumber\\&&\mbox{}
           +  C_F T 
\left(
-\frac{59}{12}
 +  \frac{43}{12} \lnmum
 +              \frac{4}{3} \lnmum^2
 +              \frac{1}{3} \lnmum^3
 +              \left(4 - 4 \lnmum\right) \zeta(2)
 -\frac{7}{12} \zeta(3)
\right)
\Bigg]
,\\
\bar{C}^p_{-1} &=&
-1 - \lnmum
 +       \frac{\alpha_s}{\pi} C_F      
\left(
-\frac{1}{2}
 +  3 \lnmum
 +          \frac{3}{2} \lnmum^2
\right)
\nonumber\\&&\mbox{}
 +       \left(\frac{\alpha_s}{\pi}\right)^2       
\Bigg[ C_F^2 
\left(
   \frac{1409}{192}
-B_4
 +  \frac{15}{8} \lnmum
              -\frac{69}{16} \lnmum^2
              -\frac{3}{2} \lnmum^3
\right.\nonumber\\&&\left.\mbox{}\quad
 + \left(\frac{15}{4} -  6 \ln2\right)
   \left(1 + 2 \lnmum\right) \zeta(2) 
 - 9 \zeta(3)
 +  \frac{9}{2} \zeta(4)
\right)
\nonumber\\&&\mbox{}
             +  C_F C_A 
\left(
 -\frac{129}{64}
 +\frac{1}{2}B_4
 +                \frac{185}{24} \lnmum
 +                \frac{251}{48} \lnmum^2
 +                \frac{11}{12} \lnmum^3
\right.\nonumber\\&&\left.\mbox{}\quad
 +  \left(-1 +  3 \ln2\right)
    \left(1 + 2 \lnmum\right) \zeta(2)
 +                \frac{10}{3} \zeta(3)
 -\frac{9}{4} \zeta(4)
\right)
\nonumber\\&&\mbox{}
 +  C_F T n_l 
\left(
\frac{19}{48}
                 -\frac{13}{6} \lnmum
                 -\frac{19}{12} \lnmum^2
                 -\frac{1}{3} \lnmum^3
 +  \left(-1 - 2 \lnmum\right) \zeta(2)
 +  \frac{4}{3} \zeta(3)
\right)
\nonumber\\&&\mbox{}
 +          C_F T 
\left(
\frac{107}{48}
 -\frac{13}{6} \lnmum
                 -\frac{19}{12} \lnmum^2
                 -\frac{1}{3} \lnmum^3
 +                \left(2 +  4 \lnmum\right) \zeta(2)
 -\frac{14}{3} \zeta(3)
\right)
\Bigg],
\end{eqnarray}
with $\lnmum=\ln\mu^2/m^2$.
$\zeta$ is Riemanns zeta-function with the values
$\zeta(2)=\pi^2/6$,
$\zeta(3)\approx1.20206$,
$\zeta(4)=\pi^4/90$,
and $B_4\approx-1.76280$ is a numerical constant typical for
three-loop tadpole integrals
\cite{Bro92}. 

\setcounter{equation}{0}
\section{\label{appdn}Analytic results for $D_n^\delta$}

For completeness we present in this appendix the analytical results for
$D_0^a$ and $D_1^a$.

\begin{eqnarray}
D_0^{(0),a} &=&
\frac{32}{9}  - \frac{4}{3}\ln\frac{-q^2}{m^2}
,\\
D_1^{(0),a} &=&
- 2 + 2\ln\frac{-q^2}{m^2}
,\\
D_0^{(1),a} &=&
    \frac{49}{18} 
  - 4\zeta(3)
  - \ln\frac{-q^2}{m^2}
,\\
D_1^{(1),a} &=&
  - \frac{21}{4} 
  + 6\zeta(3)
  + \frac{3}{2}\ln\frac{-q^2}{m^2}
  - \frac{3}{2}\ln^2\frac{-q^2}{m^2}
,\\
   D_0^{(2),a} &=&
         C_F^2 \,\Bigg(
          - {953\over 216}
          + 8\,\zeta(2)\,\ln 2
          - 5\,\zeta(2)
          - {1891\over 288}\,\zeta(3)
          + 10\,\zeta(5)
          + {1\over 8}\,\ln{-q^2\over m^2}
         \Bigg) 
\nonumber\\&&\mbox{}
       + C_F\,C_A \,\Bigg(
            {19729\over 2592}
          - 4\,\zeta(2)\,\ln 2
          + {4\over 3}\,\zeta(2)
          + {11\over 3}\,\zeta(3)\,\ln{-q^2\over \mu^2}
          - {709\over 64}\,\zeta(3)
\nonumber\\&&\mbox{}\quad
          - {5\over 3}\,\zeta(5)
          + {11\over 12}\,\ln{-q^2\over m^2}\,\ln{-q^2\over \mu^2}
          - {71\over 27}\,\ln{-q^2\over m^2}
          - {11\over 24}\,\ln^{2}{-q^2\over m^2}
          - {539\over 216}\,\ln{-q^2\over \mu^2}
         \Bigg) 
\nonumber\\&&\mbox{}
       + C_F\,T\,n_l \,\Bigg(
          - {295\over 81}
          + {4\over 3}\,\zeta(2)
          - {4\over 3}\,\zeta(3)\,\ln{-q^2\over \mu^2}
          + {38\over 9}\,\zeta(3)
\nonumber\\&&\mbox{}\quad
          - {1\over 3}\,\ln{-q^2\over m^2}\,\ln{-q^2\over \mu^2}
          + {25\over 27}\,\ln{-q^2\over m^2}
          + {1\over 6}\,\ln^{2}{-q^2\over m^2}
          + {49\over 54}\,\ln{-q^2\over \mu^2}
         \Bigg) 
\nonumber\\&&\mbox{}
       + C_F\,T \,\Bigg(
          - {731\over 648}
          - {8\over 3}\,\zeta(2)
          - {4\over 3}\,\zeta(3)\,\ln{-q^2\over \mu^2}
          + {629\over 144}\,\zeta(3)
\nonumber\\&&\mbox{}\quad
          - {1\over 3}\,\ln{-q^2\over m^2}\,\ln{-q^2\over \mu^2}
          + {25\over 27}\,\ln{-q^2\over m^2}
          + {1\over 6}\,\ln^{2}{-q^2\over m^2}
          + {49\over 54}\,\ln{-q^2\over \mu^2}
         \Bigg) 
,\\
   D_1^{(2),a} &=&
        C_F^2 \,\Bigg(
           {111\over 32}
          + 12\,\zeta(2)\,\ln 2\,\ln{-q^2\over m^2}
          - 24\,\zeta(2)\,\ln 2
          - {15\over 2}\,\zeta(2)\,\ln{-q^2\over m^2}
          + 15\,\zeta(2)
\nonumber\\
&&\mbox{}\quad
          - {15\over 2}\,\zeta(3)\,\ln{-q^2\over m^2}
          + {87\over 4}\,\zeta(3)
          - 9\,\zeta(4)
          - {45\over 2}\,\zeta(5)
          + {3\over 2}\,B_4
          + {127\over 16}\,\ln{-q^2\over m^2}
\nonumber\\
&&\mbox{}\quad
          - {21\over 16}\,\ln^{2}{-q^2\over m^2}
          + {3\over 4}\,\ln^{3}{-q^2\over m^2}
         \Bigg) 
\nonumber\\
&&\mbox{}
       + C_F\,C_A \,\Bigg(
          - {1219\over 72}
          - 6\,\zeta(2)\,\ln 2\,\ln{-q^2\over m^2}
          + 12\,\zeta(2)\,\ln 2
          + 2\,\zeta(2)\,\ln{-q^2\over m^2}
          - 4\,\zeta(2)
\nonumber\\
&&\mbox{}\quad
          - {3\over 4}\,\zeta(3)\,\ln{-q^2\over m^2}
          - {11\over 2}\,\zeta(3)\,\ln{-q^2\over \mu^2}
          + {223\over 12}\,\zeta(3)
          + {9\over 2}\,\zeta(4)
          - {15\over 4}\,\zeta(5)
          - {3\over 4}\,B_4
\nonumber\\
&&\mbox{}\quad
          - {11\over 8}\,\ln{-q^2\over m^2}\,\ln{-q^2\over \mu^2}
          + {227\over 48}\,\ln{-q^2\over m^2}
          + {11\over 8}\,\ln^{2}{-q^2\over m^2}\,\ln{-q^2\over \mu^2}
          - {19\over 6}\,\ln^{2}{-q^2\over m^2}
\nonumber\\
&&\mbox{}\quad
          - {11\over 12}\,\ln^{3}{-q^2\over m^2}
          + {77\over 16}\,\ln{-q^2\over \mu^2}
         \Bigg) 
\nonumber\\
&&\mbox{}
       + C_F\,T\,n_l \,\Bigg(
           {217\over 36}
          + 2\,\zeta(2)\,\ln{-q^2\over m^2}
          - 4\,\zeta(2)
          + 2\,\zeta(3)\,\ln{-q^2\over \mu^2}
          - {20\over 3}\,\zeta(3)
\nonumber\\
&&\mbox{}\quad
          + {1\over 2}\,\ln{-q^2\over m^2}\,\ln{-q^2\over \mu^2}
          - {19\over 12}\,\ln{-q^2\over m^2}
          - {1\over 2}\,\ln^{2}{-q^2\over m^2}\,\ln{-q^2\over \mu^2}
          + {5\over 6}\,\ln^{2}{-q^2\over m^2}
\nonumber\\
&&\mbox{}\quad
          + {1\over 3}\,\ln^{3}{-q^2\over m^2}
          - {7\over 4}\,\ln{-q^2\over \mu^2}
         \Bigg) 
\nonumber\\
&&\mbox{}
       + C_F\,T \,\Bigg(
          - {155\over 36}
          - 4\,\zeta(2)\,\ln{-q^2\over m^2}
          + 8\,\zeta(2)
          + 2\,\zeta(3)\,\ln{-q^2\over \mu^2}
          - {23\over 12}\,\zeta(3)
\nonumber\\
&&\mbox{}\quad
          + {1\over 2}\,\ln{-q^2\over m^2}\,\ln{-q^2\over \mu^2}
          + {17\over 12}\,\ln{-q^2\over m^2}
          - {1\over 2}\,\ln^{2}{-q^2\over m^2}\,\ln{-q^2\over \mu^2}
          + {5\over 6}\,\ln^{2}{-q^2\over m^2}
\nonumber\\
&&\mbox{}\quad
          + {1\over 3}\,\ln^{3}{-q^2\over m^2}
          - {7\over 4}\,\ln{-q^2\over \mu^2}
         \Bigg), 
\end{eqnarray}
with $\zeta(5)\approx1.03693$.
$B_4$ appears in our results because
of the normalization condition $\Pi^a(0)=0$.


\setcounter{equation}{0}
\section{\label{appcn}Analytic results for $C_n^\delta$}

In this appendix we list the first eight 
moments for $q^2\to0$ expressed in terms of the on-shell mass, $m$,
in analytic form for the four correlators.

\begin{eqnarray}
\Pi^{(0),v} &=&
\frac{3}{16\pi^2}\bigg\{
  \frac{16}{15}z
+ \frac{16}{35}z^2
+ \frac{256}{945}z^3
+ \frac{128}{693}z^4
+ \frac{2048}{15015}z^5 
+ \frac{2048}{19305}z^6
+ \frac{65536}{765765}z^7 
\nonumber\\&&\mbox{}
+ \frac{16384}{230945}z^8
\bigg\}
+\ldots\,\,,
\nonumber\\
\Pi^{(1),v} &=&
\frac{3}{16\pi^2}\bigg\{  \frac{328}{81}z
+ \frac{1796}{675}z^2
+ \frac{999664}{496125}z^3
+ \frac{207944}{127575}z^4
+ \frac{1729540864}{1260653625}z^5
\nonumber\\
&&\mbox{}
+ \frac{21660988864}{18261468225}z^6
+ \frac{401009026048}{383490832725}z^7
+ \frac{633021048064}{676610809875}z^8
\bigg\}
+\ldots\,\,,
\nonumber\\
   C^{(2),v}_1 &=&
       C_F^2 \,\Bigg(
          - {8687\over 864}
          - {32\over 5}\,\zeta(2)\,\ln 2
          + 4\,\zeta(2)
          + {22781\over 1728}\,\zeta(3)
         \Bigg) 
\nonumber\\
&&\mbox{}
       + C_F\,C_A \,\Bigg(
           {127\over 192}
          + {902\over 243}\,\ln{\mu^2\over m^2}
          + {16\over 5}\,\zeta(2)\,\ln 2
          - {16\over 15}\,\zeta(2)
          + {1451\over 384}\,\zeta(3)
         \Bigg) 
\nonumber\\
&&\mbox{}
       + C_F\,T\,n_l \,\Bigg(
          - {142\over 243}
          - {328\over 243}\,\ln{\mu^2\over m^2}
          - {16\over 15}\,\zeta(2)
         \Bigg) 
\nonumber\\
&&\mbox{}
       + C_F\,T \,\Bigg(
          - {11407\over 2916}
          - {328\over 243}\,\ln{\mu^2\over m^2}
          + {32\over 15}\,\zeta(2)
          + {203\over 216}\,\zeta(3)
         \Bigg) 
,\nonumber\\
   C^{(2),v}_2 &=&
       C_F^2 \,\Bigg(
          - {223404289\over 1866240}
          - {192\over 35}\,\zeta(2)\,\ln 2
          + {24\over 7}\,\zeta(2)
          + {4857587\over 46080}\,\zeta(3)
         \Bigg) 
\nonumber\\
&&\mbox{}
       + C_F\,C_A \,\Bigg(
          - {1030213543\over 93312000}
          + {4939\over 2025}\,\ln{\mu^2\over m^2}
          + {96\over 35}\,\zeta(2)\,\ln 2
          - {32\over 35}\,\zeta(2)
\nonumber\\
&&\mbox{}
          + {723515\over 55296}\,\zeta(3)
         \Bigg) 
       + C_F\,T\,n_l \,\Bigg(
          - {40703\over 60750}
          - {1796\over 2025}\,\ln{\mu^2\over m^2}
          - {32\over 35}\,\zeta(2)
         \Bigg) 
\nonumber\\
&&\mbox{}
       + C_F\,T \,\Bigg(
          - {1520789\over 414720}
          - {1796\over 2025}\,\ln{\mu^2\over m^2}
          + {64\over 35}\,\zeta(2)
          + {14203\over 18432}\,\zeta(3)
         \Bigg) 
,\nonumber\\
   C^{(2),v}_3 &=&
       C_F^2 \,\Bigg(
          - {885937890461\over 1161216000}
          - {512\over 105}\,\zeta(2)\,\ln 2
          + {64\over 21}\,\zeta(2)
          + {33067024499\over 51609600}\,\zeta(3)
         \Bigg) 
\nonumber\\
&&\mbox{}
       + C_F\,C_A \,\Bigg(
          - {95905830011197\over 1706987520000}
          + {2749076\over 1488375}\,\ln{\mu^2\over m^2}
          + {256\over 105}\,\zeta(2)\,\ln 2
\nonumber\\
&&\mbox{}
          - {256\over 315}\,\zeta(2)
          + {5164056461\over 103219200}\,\zeta(3)
         \Bigg) 
\nonumber\\
&&\mbox{}
       + C_F\,T\,n_l \,\Bigg(
          - {9703588\over 17364375}
          - {999664\over 1488375}\,\ln{\mu^2\over m^2}
          - {256\over 315}\,\zeta(2)
         \Bigg) 
\nonumber\\
&&\mbox{}
       + C_F\,T \,\Bigg(
          - {83936527\over 23328000}
          - {999664\over 1488375}\,\ln{\mu^2\over m^2}
          + {512\over 315}\,\zeta(2)
          + {12355\over 13824}\,\zeta(3)
         \Bigg) 
,\nonumber\\
   C^{(2),v}_4 &=&
       C_F^2 \,\Bigg(
          - {269240669884818833\over 61451550720000}
          - {1024\over 231}\,\zeta(2)\,\ln 2
          + {640\over 231}\,\zeta(2)
\nonumber\\
&&\mbox{}
          + {1507351507033\over 412876800}\,\zeta(3)
         \Bigg) 
       + C_F\,C_A \,\Bigg(
          - {36675392331131681\over 158018273280000}
          + {571846\over 382725}\,\ln{\mu^2\over m^2}
\nonumber\\
&&\mbox{}
          + {512\over 231}\,\zeta(2)\,\ln 2
          - {512\over 693}\,\zeta(2)
          + {1455887207647\over 7431782400}\,\zeta(3)
         \Bigg) 
\nonumber\\
&&\mbox{}
       + C_F\,T\,n_l \,\Bigg(
          - {54924808\over 120558375}
          - {207944\over 382725}\,\ln{\mu^2\over m^2}
          - {512\over 693}\,\zeta(2)
         \Bigg) 
\nonumber\\
&&\mbox{}
       + C_F\,T \,\Bigg(
          - {129586264289\over 35831808000}
          - {207944\over 382725}\,\ln{\mu^2\over m^2}
          + {1024\over 693}\,\zeta(2)
          + {2522821\over 2359296}\,\zeta(3)
         \Bigg) 
,\nonumber\\
   C^{(2),v}_5 &=&
       C_F^2 \,\Bigg(
          - {360248170450504167133\over 15209258803200000}
          - {4096\over 1001}\,\zeta(2)\,\ln 2
          + {2560\over 1001}\,\zeta(2)
\nonumber\\
&&\mbox{}
          + {939939943788973\over 47687270400}\,\zeta(3)
         \Bigg) 
       + C_F\,C_A \,\Bigg(
          - {21883348499544169357\over 23658847027200000}
\nonumber\\
&&\mbox{}
          + {432385216\over 343814625}\,\ln{\mu^2\over m^2}
          + {2048\over 1001}\,\zeta(2)\,\ln 2
          - {2048\over 3003}\,\zeta(2)
          + {14724562345079\over 19074908160}\,\zeta(3)
         \Bigg) 
\nonumber\\
&&\mbox{}
       + C_F\,T\,n_l \,\Bigg(
          - {4881989801536\over 13104494431875}
          - {1729540864\over 3781960875}\,\ln{\mu^2\over m^2}
          - {2048\over 3003}\,\zeta(2)
         \Bigg) 
\nonumber\\
&&\mbox{}
       + C_F\,T \,\Bigg(
          - {512847330943\over 139087872000}
          - {1729540864\over 3781960875}\,\ln{\mu^2\over m^2}
\nonumber\\
&&\mbox{}
          + {4096\over 3003}\,\zeta(2)
          + {1239683\over 983040}\,\zeta(3)
         \Bigg) 
,\nonumber\\
   C^{(2),v}_6 &=&
       C_F^2 \,\Bigg(
          - {64959156551995419148501103\over 529285210649395200000}
          - {8192\over 2145}\,\zeta(2)\,\ln 2
          + {1024\over 429}\,\zeta(2)
\nonumber\\
&&\mbox{}
          + {330704075360938001\over 3238841548800}\,\zeta(3)
         \Bigg) 
\nonumber\\
&&\mbox{}
       + C_F\,C_A \,\Bigg(
          - {4826864658245605658856745531\over 1317342772469012889600000}
          + {5415247216\over 4980400425}\,\ln{\mu^2\over m^2}
\nonumber\\
&&\mbox{}
          + {4096\over 2145}\,\zeta(2)\,\ln 2
          - {4096\over 6435}\,\zeta(2)
          + {580922571682067161\over 190443883069440}\,\zeta(3)
         \Bigg) 
\nonumber\\
&&\mbox{}
       + C_F\,T\,n_l \,\Bigg(
          - {151249070952032\over 493552701717075}
          - {21660988864\over 54784404675}\,\ln{\mu^2\over m^2}
          - {4096\over 6435}\,\zeta(2)
         \Bigg) 
\nonumber\\
&&\mbox{}
       + C_F\,T \,\Bigg(
          - {3411069430668887863\over 899847347503104000}
          - {21660988864\over 54784404675}\,\ln{\mu^2\over m^2}
\nonumber\\
&&\mbox{}
          + {8192\over 6435}\,\zeta(2)
          + {1760922667\over 1207959552}\,\zeta(3)
         \Bigg) 
,\nonumber\\
   C^{(2),v}_7 &=&
       C_F^2 \,\Bigg(
          - {571365897351090627148045413923471\over 
             927409311818185074278400000}
          - {131072\over 36465}\,\zeta(2)\,\ln 2
\nonumber\\
&&\mbox{}
          + {16384\over 7293}\,\zeta(2)
          + {13386367971827490465799\over 26118018249523200}\,\zeta(3)
         \Bigg) 
\nonumber\\
&&\mbox{}
       + C_F\,C_A \,\Bigg(
          - {7342721436809271685822267340249\over 505859624628100949606400000}
          + {100252256512\over 104588408925}\,\ln{\mu^2\over m^2}
\nonumber\\
&&\mbox{}
          + {65536\over 36465}\,\zeta(2)\,\ln 2
          - {65536\over 109395}\,\zeta(2)
          + {14019414333929589373\over 1160800811089920}\,\zeta(3)
         \Bigg) 
\nonumber\\
&&\mbox{}
       + C_F\,T\,n_l \,\Bigg(
          - {13125091764358528\over 51823033680292875}
          - {401009026048\over 1150472498175}\,\ln{\mu^2\over m^2}
          - {65536\over 109395}\,\zeta(2)
         \Bigg) 
\nonumber\\
&&\mbox{}
       + C_F\,T \,\Bigg(
          - {7927736038867601807\over 2024656531881984000}
          - {401009026048\over 1150472498175}\,\ln{\mu^2\over m^2}
\nonumber\\
&&\mbox{}
          + {131072\over 109395}\,\zeta(2)
          + {4497899939\over 2717908992}\,\zeta(3)
         \Bigg) 
,\nonumber\\
   C^{(2),v}_8 &=&
       C_F^2 \,\Bigg(
          - {190302182417255312898886115648452691\over 
             63063833203636585050931200000}
\nonumber\\
&&\mbox{}
          - {786432\over 230945}\,\zeta(2)\,\ln 2
          + {98304\over 46189}\,\zeta(2)
          + {31209476560803609727258477\over 12432176686773043200}\,\zeta(3)
         \Bigg) 
\nonumber\\
&&\mbox{}
       + C_F\,C_A \,\Bigg(
          - {11413196924379471880248867066065741\over 
             198256619379886265047449600000}
          + {393216\over 230945}\,\zeta(2)\,\ln 2
\nonumber\\
&&\mbox{}
          - {131072\over 230945}\,\zeta(2)
          + {24302541873458280280067\over 507435783133593600}\,\zeta(3)
          + {158255262016\over 184530220875}\,\ln{\mu^2\over m^2}
         \Bigg) 
\nonumber\\
&&\mbox{}
       + C_F\,T\,n_l \,\Bigg(
          - {65233327834094144\over 310874926094357625}
          - {633021048064\over 2029832429625}\,\ln{\mu^2\over m^2}
          - {131072\over 230945}\,\zeta(2)
         \Bigg) 
\nonumber\\
&&\mbox{}
       + C_F\,T \,\Bigg(
          - {23818697864446985668391\over 5874203484500262912000}
          - {633021048064\over 2029832429625}\,\ln{\mu^2\over m^2}
\nonumber\\
&&\mbox{}
          + {262144\over 230945}\,\zeta(2)
          + {286122897977\over 154618822656}\,\zeta(3)
         \Bigg), 
\end{eqnarray}

\begin{eqnarray}
\Pi^{(0),a} &=&
\frac{3}{16\pi^2}\bigg\{  \frac{8}{15}z 
+ \frac{16}{105}z^2 
+ \frac{64}{945}z^3
+ \frac{128}{3465}z^4 
+ \frac{1024}{45045}z^5
+ \frac{2048}{135135}z^6
+ \frac{8192}{765765}z^7
\nonumber\\
&&\mbox{}
+ \frac{16384}{2078505}z^8
\bigg\}
+\ldots\,\,,
\nonumber\\
\Pi^{(1),a} &=&
\frac{3}{16\pi^2}\bigg\{  \frac{689}{405}z 
+ \frac{3382}{4725}z^2 
+ \frac{196852}{496125}z^3 
+ \frac{12398216}{49116375}z^4 
+ \frac{318252608}{1820944125}z^5
\nonumber\\
&&\mbox{}
+ \frac{655479040}{5113211103}z^6
+ \frac{639246915968}{6519344156325}z^7 
+ \frac{38821601949952}{501368610117375}z^8
\bigg\}
+\ldots\,\,,
\nonumber\\
   C^{(2),a}_1 &=&
       C_F^2 \,\Bigg(
           {2237369\over 51840}
          - {16\over 5}\,\zeta(2)\,\ln 2
          + 2\,\zeta(2)
          - {1164013\over 34560}\,\zeta(3)
         \Bigg) 
\nonumber\\
&&\mbox{}
       + C_F\,C_A \,\Bigg(
           {3226373\over 311040}
          + {8\over 5}\,\zeta(2)\,\ln 2
          - {8\over 15}\,\zeta(2)
          - {494867\over 69120}\,\zeta(3)
          + {7579\over 4860}\,\ln{\mu^2\over m^2}
         \Bigg) 
\nonumber\\
&&\mbox{}
       + C_F\,T\,n_l \,\Bigg(
          - {137\over 810}
          - {8\over 15}\,\zeta(2)
          - {689\over 1215}\,\ln{\mu^2\over m^2}
         \Bigg) 
\nonumber\\
&&\mbox{}
       + C_F\,T \,\Bigg(
          - {433669\over 186624}
          + {16\over 15}\,\zeta(2)
          + {10493\over 13824}\,\zeta(3)
          - {689\over 1215}\,\ln{\mu^2\over m^2}
         \Bigg) 
,\nonumber\\
   C^{(2),a}_2 &=&
       C_F^2 \,\Bigg(
           {7672813249\over 26127360}
          - {64\over 35}\,\zeta(2)\,\ln 2
          + {8\over 7}\,\zeta(2)
          - {2349181181\over 9676800}\,\zeta(3)
         \Bigg) 
\nonumber\\
&&\mbox{}
       + C_F\,C_A \,\Bigg(
           {47328042151\over 1306368000}
          + {32\over 35}\,\zeta(2)\,\ln 2
          - {32\over 105}\,\zeta(2)
          - {188251393\over 6451200}\,\zeta(3)
\nonumber\\
&&\mbox{}
          + {18601\over 28350}\,\ln{\mu^2\over m^2}
         \Bigg) 
       + C_F\,T\,n_l \,\Bigg(
          - {1097\over 6750}
          - {32\over 105}\,\zeta(2)
          - {3382\over 14175}\,\ln{\mu^2\over m^2}
         \Bigg) 
\nonumber\\
&&\mbox{}
       + C_F\,T \,\Bigg(
          - {27450553\over 17418240}
          + {64\over 105}\,\zeta(2)
          + {19579\over 36864}\,\zeta(3)
          - {3382\over 14175}\,\ln{\mu^2\over m^2}
         \Bigg) 
,\nonumber\\
   C^{(2),a}_3 &=&
       C_F^2 \,\Bigg(
           {111399585201971\over 58525286400}
          - {128\over 105}\,\zeta(2)\,\ln 2
          + {16\over 21}\,\zeta(2)
          - {979995241517\over 619315200}\,\zeta(3)
         \Bigg) 
\nonumber\\
&&\mbox{}
       + C_F\,C_A \,\Bigg(
           {2930267790199843\over 20483850240000}
          + {64\over 105}\,\zeta(2)\,\ln 2
          - {64\over 315}\,\zeta(2)
\nonumber\\
&&\mbox{}
          - {146653533139\over 1238630400}\,\zeta(3)
          + {541343\over 1488375}\,\ln{\mu^2\over m^2}
         \Bigg) 
\nonumber\\
&&\mbox{}
       + C_F\,T\,n_l \,\Bigg(
          - {5058122\over 52093125}
          - {64\over 315}\,\zeta(2)
          - {196852\over 1488375}\,\ln{\mu^2\over m^2}
         \Bigg) 
\nonumber\\
&&\mbox{}
       + C_F\,T \,\Bigg(
          - {26031430073\over 20901888000}
          + {128\over 315}\,\zeta(2)
          + {4411519\over 8847360}\,\zeta(3)
          - {196852\over 1488375}\,\ln{\mu^2\over m^2}
         \Bigg) 
,\nonumber\\
   C^{(2),a}_4 &=&
       C_F^2 \,\Bigg(
           {308356223383353917\over 27590492160000}
          - {1024\over 1155}\,\zeta(2)\,\ln 2
          + {128\over 231}\,\zeta(2)
\nonumber\\
&&\mbox{}
          - {197037714570097\over 21194342400}\,\zeta(3)
         \Bigg) 
       + C_F\,C_A \,\Bigg(
           {1275464959378469537\over 2212255825920000}
\nonumber\\
&&\mbox{}
          + {512\over 1155}\,\zeta(2)\,\ln 2
          - {512\over 3465}\,\zeta(2)
          - {109692872248273\over 228898897920}\,\zeta(3)
          + {3099554\over 13395375}\,\ln{\mu^2\over m^2}
         \Bigg) 
\nonumber\\
&&\mbox{}
       + C_F\,T\,n_l \,\Bigg(
          - {2710286584\over 46414974375}
          - {512\over 3465}\,\zeta(2)
          - {12398216\over 147349125}\,\ln{\mu^2\over m^2}
         \Bigg) 
\nonumber\\
&&\mbox{}
       + C_F\,T \,\Bigg(
          - {731128794367\over 689762304000}
          + {1024\over 3465}\,\zeta(2)
          + {1432739\over 2949120}\,\zeta(3)
          - {12398216\over 147349125}\,\ln{\mu^2\over m^2}
         \Bigg) 
,\nonumber\\
   C^{(2),a}_5 &=&
       C_F^2 \,\Bigg(
           {4277005531832013845390021\over 69597568283443200000}
          - {2048\over 3003}\,\zeta(2)\,\ln 2
\nonumber\\
&&\mbox{}
          + {1280\over 3003}\,\zeta(2)
          - {1014170497519835231\over 19837904486400}\,\zeta(3)
         \Bigg) 
\nonumber\\
&&\mbox{}
       + C_F\,C_A \,\Bigg(
           {58119452968289341424539\over 24983742460723200000}
          + {1024\over 3003}\,\zeta(2)\,\ln 2
\nonumber\\
&&\mbox{}
          - {1024\over 9009}\,\zeta(2)
          - {2193462351270763\over 1133594542080}\,\zeta(3)
          + {79563152\over 496621125}\,\ln{\mu^2\over m^2}
         \Bigg) 
\nonumber\\
&&\mbox{}
       + C_F\,T\,n_l \,\Bigg(
          - {226047457424\over 6309571393125}
          - {1024\over 9009}\,\zeta(2)
          - {318252608\over 5462832375}\,\ln{\mu^2\over m^2}
         \Bigg) 
\nonumber\\
&&\mbox{}
       + C_F\,T \,\Bigg(
          - {5461272114191\over 5796790272000}
          + {2048\over 9009}\,\zeta(2)
          + {30020447\over 62914560}\,\zeta(3)
\nonumber\\
&&\mbox{}
          - {318252608\over 5462832375}\,\ln{\mu^2\over m^2}
         \Bigg) 
,\nonumber\\
   C^{(2),a}_6 &=&
       C_F^2 \,\Bigg(
           {359745448810293562716400230493\over 1114674653627626291200000}
          - {8192\over 15015}\,\zeta(2)\,\ln 2
\nonumber\\
&&\mbox{}
          + {1024\over 3003}\,\zeta(2)
          - {23241579953084394919\over 86565401395200}\,\zeta(3)
         \Bigg) 
\nonumber\\
&&\mbox{}
       + C_F\,C_A \,\Bigg(
           {24694796807630112104602086197\over 2634685544938025779200000}
          + {4096\over 15015}\,\zeta(2)\,\ln 2
\nonumber\\
&&\mbox{}
          - {4096\over 45045}\,\zeta(2)
          - {76150305462878641\over 9766352977920}\,\zeta(3)
          + {163869760\over 1394512119}\,\ln{\mu^2\over m^2}
         \Bigg) 
\nonumber\\
&&\mbox{}
       + C_F\,T\,n_l \,\Bigg(
          - {381648296450416\over 17274344560097625}
          - {4096\over 45045}\,\zeta(2)
          - {655479040\over 15339633309}\,\ln{\mu^2\over m^2}
         \Bigg) 
\nonumber\\
&&\mbox{}
       + C_F\,T \,\Bigg(
          - {1548825962112515819\over 1799694695006208000}
          + {8192\over 45045}\,\zeta(2)
\nonumber\\
&&\mbox{}
          + {1134854351\over 2415919104}\,\zeta(3)
          - {655479040\over 15339633309}\,\ln{\mu^2\over m^2}
         \Bigg) 
,\nonumber\\
   C^{(2),a}_7 &=&
       C_F^2 \,\Bigg(
           {2295850065917186141074716812133631\over 
             1401418515636368556687360000}
          - {16384\over 36465}\,\zeta(2)\,\ln 2
\nonumber\\
&&\mbox{}
          + {2048\over 7293}\,\zeta(2)
          - {2420469632151392380640363\over 1776025240967577600}\,\zeta(3)
         \Bigg) 
\nonumber\\
&&\mbox{}
       + C_F\,C_A \,\Bigg(
         {117926764107372779240781607407103\over 3127132224973714961203200000}
          + {8192\over 36465}\,\zeta(2)\,\ln 2
\nonumber\\
&&\mbox{}
          - {8192\over 109395}\,\zeta(2)
          - {111434031814012253905813\over 3552050481935155200}\,\zeta(3)
          + {159811728992\over 1778002951725}\,\ln{\mu^2\over m^2}
         \Bigg) 
\nonumber\\
&&\mbox{}
       + C_F\,T\,n_l \,\Bigg(
          - {2357049928630816\over 176198314512995775}
          - {8192\over 109395}\,\zeta(2)
\nonumber\\
&&\mbox{}
          - {639246915968\over 19558032468975}\,\ln{\mu^2\over m^2}
         \Bigg) 
       + C_F\,T \,\Bigg(
          - {705492229082574766543\over 881130522675039436800}
\nonumber\\
&&\mbox{}
          + {16384\over 109395}\,\zeta(2)
          + {161018056831\over 347892350976}\,\zeta(3)
          - {639246915968\over 19558032468975}\,\ln{\mu^2\over m^2}
         \Bigg) 
,\nonumber\\
   C^{(2),a}_8 &=&
       C_F^2 \,\Bigg(
           {36666382863813217294681656413671975999\over 
         4526581805505470438100172800000}
          - {262144\over 692835}\,\zeta(2)\,\ln 2
\nonumber\\
&&\mbox{}
          + {32768\over 138567}\,\zeta(2)
          - {3890931550494737377107721691\over 577405539452348006400}\,\zeta(3)
         \Bigg) 
\nonumber\\
&&\mbox{}
       + C_F\,C_A \,\Bigg(
           {489815334595084347787765229172106365989\over 
         3231979409130905892803523379200000}
          + {131072\over 692835}\,\zeta(2)\,\ln 2
\nonumber\\
&&\mbox{}
          - {131072\over 2078505}\,\zeta(2)
          - {524355086420656861203887\over 4158983477447884800}\,\zeta(3)
\nonumber\\
&&\mbox{}
          + {9705400487488\over 136736893668375}\,\ln{\mu^2\over m^2}
         \Bigg) 
\nonumber\\
&&\mbox{}
       + C_F\,T\,n_l \,\Bigg(
          - {1762232386535569216\over 230358320235919000125}
          - {131072\over 2078505}\,\zeta(2)
\nonumber\\
&&\mbox{}
          - {38821601949952\over 1504105830352125}\,\ln{\mu^2\over m^2}
         \Bigg) 
       + C_F\,T \,\Bigg(
          - {84212007346306764915559\over 111609866205504995328000}
\nonumber\\
&&\mbox{}
          + {262144\over 2078505}\,\zeta(2)
          + {1058200490221\over 2319282339840}\,\zeta(3)
          - {38821601949952\over 1504105830352125}\,\ln{\mu^2\over m^2}
         \Bigg), 
\end{eqnarray}

\begin{eqnarray}
\Pi^{(0),s} &=&
\frac{3}{16\pi^2}\bigg\{  \frac{4}{5}z 
+ \frac{8}{35}z^2 
+ \frac{32}{315}z^3
+ \frac{64}{1155}z^4
+ \frac{512}{15015}z^5
+ \frac{1024}{45045}z^6
+ \frac{4096}{255255}z^7
\nonumber\\&&\mbox{}
+ \frac{8192}{692835}z^8
\bigg\}
+\ldots\,\,,
\nonumber\\
\Pi^{(1),s} &=&
\frac{3}{16\pi^2}\bigg\{  \frac{61}{135}z
+ \frac{1223}{1575}z^2
+ \frac{86246}{165375}z^3 
+ \frac{5845948}{16372125}z^4 
+ \frac{155689024}{606981375}z^5
\nonumber\\
&&\mbox{}
+ \frac{1637544448}{8522018505}z^6
+ \frac{323629508032}{2173114718775}z^7 
+ \frac{19824721740416}{167122870039125}z^8
\bigg\}
+\ldots\,\,,
\nonumber\\
   C^{(2),s}_1 &=&
        C_F^2 \,\Bigg(
           {413\over 30}
          - {1645\over 144}\,\zeta(3)
         \Bigg) 
       + C_F\,C_A \,\Bigg(
          - {191\over 648}
          - {59\over 32}\,\zeta(3)
          + {671\over 1620}\,\ln{\mu^2\over m^2}
         \Bigg) 
\nonumber\\
&&\mbox{}
       + C_F\,T\,n_l \,\Bigg(
           {119\over 135}
          - {61\over 405}\,\ln{\mu^2\over m^2}
         \Bigg) 
       + C_F\,T \,\Bigg(\!
          - {60559\over 77760}
          + {1435\over 1152}\,\zeta(3)
          - {61\over 405}\,\ln{\mu^2\over m^2}
         \Bigg) 
,\nonumber\\
   C^{(2),s}_2 &=&
       C_F^2 \,\Bigg(
           {627541597\over 10886400}
          - {48\over 35}\,\zeta(2)\,\ln 2
          + {6\over 7}\,\zeta(2)
          - {1074607\over 23040}\,\zeta(3)
         \Bigg) 
\nonumber\\
&&\mbox{}
       + C_F\,C_A \,\Bigg(
          + {991366223\over 108864000}
          + {24\over 35}\,\zeta(2)\,\ln 2
          - {8\over 35}\,\zeta(2)
          - {110107\over 15360}\,\zeta(3)
\nonumber\\
&&\mbox{}
          + {13453\over 18900}\,\ln{\mu^2\over m^2}
         \Bigg) 
       + C_F\,T\,n_l \,\Bigg(
          + {797\over 47250}
          - {8\over 35}\,\zeta(2)
          - {1223\over 4725}\,\ln{\mu^2\over m^2}
         \Bigg) 
\nonumber\\
&&\mbox{}
       + C_F\,T \,\Bigg(
          - {1685773\over 1161216}
          + {16\over 35}\,\zeta(2)
          + {9107\over 12288}\,\zeta(3)
          - {1223\over 4725}\,\ln{\mu^2\over m^2}
         \Bigg) 
,\nonumber\\
   C^{(2),s}_3 &=&
       C_F^2 \,\Bigg(
            {1619371436071\over 4064256000}
          - {128\over 105}\,\zeta(2)\,\ln 2
          + {16\over 21}\,\zeta(2)
          - {405607027\over 1228800}\,\zeta(3)
         \Bigg) 
\nonumber\\
&&\mbox{}
       + C_F\,C_A \,\Bigg(
            {11448730350251\over 284497920000}
          + {64\over 105}\,\zeta(2)\,\ln 2
          - {64\over 315}\,\zeta(2)
          - {566787803\over 17203200}\,\zeta(3)
\nonumber\\
&&\mbox{}
          + {474353\over 992250}\,\ln{\mu^2\over m^2}
         \Bigg) 
       + C_F\,T\,n_l \,\Bigg(
          - {1146421\over 17364375}
          - {64\over 315}\,\zeta(2)
          - {86246\over 496125}\,\ln{\mu^2\over m^2}
         \Bigg) 
\nonumber\\
&&\mbox{}
       + C_F\,T \,\Bigg(
          - {694040519\over 497664000}
          + {128\over 315}\,\zeta(2)
          + {978439\over 1474560}\,\zeta(3)
          - {86246\over 496125}\,\ln{\mu^2\over m^2}
         \Bigg) 
,\nonumber\\
   C^{(2),s}_4 &=&
       C_F^2 \,\Bigg(
            {147161013073070141\over 56330588160000}
          - {384\over 385}\,\zeta(2)\,\ln 2
          + {48\over 77}\,\zeta(2)
\nonumber\\
&&\mbox{}
          - {224204681453\over 103219200}\,\zeta(3)
         \Bigg) 
       + C_F\,C_A \,\Bigg(
            {35969257153127519\over 202790117376000}
          + {192\over 385}\,\zeta(2)\,\ln 2
\nonumber\\
&&\mbox{}
          - {64\over 385}\,\zeta(2)
          - {18221998757\over 123863040}\,\zeta(3)
          + {1461487\over 4465125}\,\ln{\mu^2\over m^2}
         \Bigg) 
\nonumber\\
&&\mbox{}
       + C_F\,T\,n_l \,\Bigg(
          - {930573962\over 15471658125}
          - {64\over 385}\,\zeta(2)
          - {5845948\over 49116375}\,\ln{\mu^2\over m^2}
         \Bigg) 
\nonumber\\
&&\mbox{}
       + C_F\,T \,\Bigg(
          - {590888856583\over 459841536000}
          + {128\over 385}\,\zeta(2)
          + {1262219\over 1966080}\,\zeta(3)
          - {5845948\over 49116375}\,\ln{\mu^2\over m^2}
         \Bigg) 
,\nonumber\\
   C^{(2),s}_5 &=&
       C_F^2 \,\Bigg(
           {45884811924398978440541\over 2899898678476800000}
          - {4096\over 5005}\,\zeta(2)\,\ln 2
          + {512\over 1001}\,\zeta(2)
\nonumber\\
&&\mbox{}
          - {13283992935869\over 1009254400}\,\zeta(3)
         \Bigg) 
       + C_F\,C_A \,\Bigg(
           {945084080119306598357\over 1230260045414400000}
\nonumber\\
&&\mbox{}
          + {2048\over 5005}\,\zeta(2)\,\ln 2
          - {2048\over 15015}\,\zeta(2)
          - {5414889135283\over 8477736960}\,\zeta(3)
          + {38922256\over 165540375}\,\ln{\mu^2\over m^2}
         \Bigg) 
\nonumber\\
&&\mbox{}
       + C_F\,T\,n_l \,\Bigg(
          - {21726270352\over 485351645625}
          - {2048\over 15015}\,\zeta(2)
          - {155689024\over 1820944125}\,\ln{\mu^2\over m^2}
         \Bigg) 
\nonumber\\
&&\mbox{}
       + C_F\,T \,\Bigg(
          - {413776570931\over 347163328512}
          + {4096\over 15015}\,\zeta(2)
          + {1660607\over 2621440}\,\zeta(3)
\nonumber\\
&&\mbox{}
          - {155689024\over 1820944125}\,\ln{\mu^2\over m^2}
         \Bigg) 
,\nonumber\\
   C^{(2),s}_6 &=&
       C_F^2 \,\Bigg(
            {397501731663152341632983791\over 4423312117569945600000}
          - {2048\over 3003}\,\zeta(2)\,\ln 2
          + {1280\over 3003}\,\zeta(2)
\nonumber\\
&&\mbox{}
          - {1977406903785590041\over 26450539315200}\,\zeta(3)
         \Bigg)   
       + C_F\,C_A \,\Bigg(
            {717522378440002995500293379\over 219557128744835481600000}
\nonumber\\
&&\mbox{}
          + {1024\over 3003}\,\zeta(2)\,\ln 2
          - {1024\over 9009}\,\zeta(2)
          - {12326391884997959\over 4534378168320}\,\zeta(3)
\nonumber\\
&&\mbox{}
          + {409386112\over 2324186865}\,\ln{\mu^2\over m^2}
         \Bigg) 
       + C_F\,T\,n_l \,\Bigg(
          - {7250780973536\over 230324594134635}
          - {1024\over 9009}\,\zeta(2)
\nonumber\\
&&\mbox{}
          - {1637544448\over 25566055515}\,\ln{\mu^2\over m^2}
         \Bigg) 
       + C_F\,T \,\Bigg(
          - {51604525307586967\over 46146017820672000}
          + {2048\over 9009}\,\zeta(2)
\nonumber\\
&&\mbox{}
          + {506059663\over 805306368}\,\zeta(3)
          - {1637544448\over 25566055515}\,\ln{\mu^2\over m^2}
         \Bigg) 
,\nonumber\\
   C^{(2),s}_7 &=&
       C_F^2 \,\Bigg(
            {2781508068462396120688370396051\over 5724748838383858483200000}
          - {49152\over 85085}\,\zeta(2)\,\ln 2
\nonumber\\
&&\mbox{}
          + {6144\over 17017}\,\zeta(2)
          - {159953731628328432443\over 395727549235200}\,\zeta(3)
         \Bigg) 
\nonumber\\
&&\mbox{}
       + C_F\,C_A \,\Bigg(
            {19658778113043866943074074991111\over 
             1433268936446286023884800000}
          + {24576\over 85085}\,\zeta(2)\,\ln 2
\nonumber\\
&&\mbox{}
          - {8192\over 85085}\,\zeta(2)
          - {2207504939742233011\over 193466801848320}\,\zeta(3)
          + {80907377008\over 592667650575}\,\ln{\mu^2\over m^2}
         \Bigg) 
\nonumber\\
&&\mbox{}
       + C_F\,T\,n_l \,\Bigg(
          - {6298337396620816\over 293663857521659625}
          - {8192\over 85085}\,\zeta(2)
          - {323629508032\over 6519344156325}\,\ln{\mu^2\over m^2}
         \Bigg) 
\nonumber\\
&&\mbox{}
       + C_F\,T \,\Bigg(
          - {111176247094824256811\over 104896490794647552000}
          + {16384\over 85085}\,\zeta(2)
\nonumber\\
&&\mbox{}
          + {36189456601\over 57982058496}\,\zeta(3)
          - {323629508032\over 6519344156325}\,\ln{\mu^2\over m^2}
         \Bigg) 
,\nonumber\\
   C^{(2),s}_8 &=&
       C_F^2 \,\Bigg(
            {90860323801590559420949997562702411\over 
             35925252424646590778572800000}
          - {114688\over 230945}\,\zeta(2)\,\ln 2
\nonumber\\
&&\mbox{}
          + {14336\over 46189}\,\zeta(2)
          - {8719171444685991398931083\over 4144058895591014400}\,\zeta(3)
         \Bigg) 
\nonumber\\
&&\mbox{}
       + C_F\,C_A \,\Bigg(
            {2791491385643572306216795357083768311\over 
         48969384986831907466720051200000}
          + {57344\over 230945}\,\zeta(2)\,\ln 2
\nonumber\\
&&\mbox{}
          - {57344\over 692835}\,\zeta(2)
          - {891255560853790732189\over 18793917893836800}\,\zeta(3)
          + {4956180435104\over 45578964556125}\,\ln{\mu^2\over m^2}
         \Bigg) 
\nonumber\\
&&\mbox{}
       + C_F\,T\,n_l \,\Bigg(
          - {216441517065785056\over 15357221349061266675}
          - {57344\over 692835}\,\zeta(2)
\nonumber\\
&&\mbox{}
          - {19824721740416\over 501368610117375}\,\ln{\mu^2\over m^2}
         \Bigg) 
       + C_F\,T \,\Bigg(
          - {7534267707657422828683\over 7440657747033666355200}
\nonumber\\
&&\mbox{}
          + {114688\over 692835}\,\zeta(2)
          + {479255846237\over 773094113280}\,\zeta(3)
          - {19824721740416\over 501368610117375}\,\ln{\mu^2\over m^2}
         \Bigg),
\end{eqnarray}

\begin{eqnarray}
\Pi^{(0),p} &=&
\frac{3}{16\pi^2}\bigg\{  \frac{4}{3}z 
+ \frac{8}{15}z^2 
+ \frac{32}{105}z^3
+ \frac{64}{315}z^4
+ \frac{512}{3465}z^5
+ \frac{1024}{9009}z^6
+ \frac{4096}{45045}z^7
\nonumber\\&&\mbox{}
+ \frac{8192}{109395}z^8
\bigg\}
+\ldots\,\,,
\nonumber\\
\Pi^{(1),p} &=&
\frac{3}{16\pi^2}\bigg\{  \frac{7}{3}z
+ \frac{353}{135}z^2
+ \frac{10054}{4725}z^3
+ \frac{96668}{55125}z^4
+ \frac{24281408}{16372125}z^5
\nonumber\\
&&\mbox{}
+ \frac{4203369152}{3277699425}z^6
+ \frac{1781242688}{1578151575}z^7
+ \frac{312784060544}{310444959825}z^8
\bigg\}
+\ldots\,\,,
\nonumber\\
   C^{(2),p}_1 &=&
        C_F^2 \,\Bigg(
          - {401\over 144}
          + {439\over 96}\,\zeta(3)
         \Bigg) 
       + C_F\,C_A \,\Bigg(
          - {3385\over 864}
          + {329\over 192}\,\zeta(3)
          + {77\over 36}\,\ln{\mu^2\over m^2}
         \Bigg) 
\nonumber\\
&&\mbox{}
       + C_F\,T\,n_l \,\Bigg(
           {25\over 27}
          - {7\over 9}\,\ln{\mu^2\over m^2}
         \Bigg) 
       + C_F\,T \,\Bigg(
           {7\over 27}
          + {7\over 8}\,\zeta(3)
          - {7\over 9}\,\ln{\mu^2\over m^2}
         \Bigg) 
,\nonumber\\
   C^{(2),p}_2 &=&
        C_F^2 \,\Bigg(
          - {1100707\over 17280}
          - {16\over 5}\,\zeta(2)\,\ln 2
          + 2\,\zeta(2)
          + {681359\over 11520}\,\zeta(3)
         \Bigg) 
\nonumber\\
&&\mbox{}
       + C_F\,C_A \,\Bigg(
          - {1137479\over 103680}
          + {8\over 5}\,\zeta(2)\,\ln 2
          - {8\over 15}\,\zeta(2)
          + {28969\over 2560}\,\zeta(3)
          + {3883\over 1620}\,\ln{\mu^2\over m^2}
         \Bigg) 
\nonumber\\
&&\mbox{}
       + C_F\,T\,n_l \,\Bigg(
          - {287\over 810}
          - {8\over 15}\,\zeta(2)
          - {353\over 405}\,\ln{\mu^2\over m^2}
         \Bigg) 
\nonumber\\
&&\mbox{}
       + C_F\,T \,\Bigg(
          - {98605\over 62208}
          + {16\over 15}\,\zeta(2)
          + {1253\over 4608}\,\zeta(3)
          - {353\over 405}\,\ln{\mu^2\over m^2}
         \Bigg) 
,\nonumber\\
   C^{(2),p}_3 &=&
       C_F^2 \,\Bigg(
          - {22191983083\over 43545600}
          - {128\over 35}\,\zeta(2)\,\ln 2
          + {16\over 7}\,\zeta(2)
          + {1390832179\over 3225600}\,\zeta(3)
         \Bigg) 
\nonumber\\
&&\mbox{}
       + C_F\,C_A \,\Bigg(
          - {4990621717\over 87091200}
          + {64\over 35}\,\zeta(2)\,\ln 2
          - {64\over 105}\,\zeta(2)
          + {107917807\over 2150400}\,\zeta(3)
\nonumber\\
&&\mbox{}
          + {55297\over 28350}\,\ln{\mu^2\over m^2}
         \Bigg) 
       + C_F\,T\,n_l \,\Bigg(
          - {1687\over 3375}
          - {64\over 105}\,\zeta(2)
          - {10054\over 14175}\,\ln{\mu^2\over m^2}
         \Bigg) 
\nonumber\\
&&\mbox{}
       + C_F\,T \,\Bigg(
          - {36123823\over 17418240}
          + {128\over 105}\,\zeta(2)
          + {10045\over 36864}\,\zeta(3)
          - {10054\over 14175}\,\ln{\mu^2\over m^2}
         \Bigg) 
,\nonumber\\
   C^{(2),p}_4 &=&
       C_F^2 \,\Bigg(
          - {329691878962513\over 97542144000}
          - {128\over 35}\,\zeta(2)\,\ln 2
          + {16\over 7}\,\zeta(2)
          + {581996570819\over 206438400}\,\zeta(3)
         \Bigg) 
\nonumber\\
&&\mbox{}
       + C_F\,C_A \,\Bigg(
          - {571511627867983\over 2275983360000}
          + {64\over 35}\,\zeta(2)\,\ln 2
          - {64\over 105}\,\zeta(2)
\nonumber\\
&&\mbox{}
          + {17445959641\over 82575360}\,\zeta(3)
          + {265837\over 165375}\,\ln{\mu^2\over m^2}
         \Bigg) 
\nonumber\\
&&\mbox{}
       + C_F\,T\,n_l \,\Bigg(
          - {8198894\over 17364375}
          - {64\over 105}\,\zeta(2)
          - {96668\over 165375}\,\ln{\mu^2\over m^2}
         \Bigg) 
\nonumber\\
&&\mbox{}
       + C_F\,T \,\Bigg(
          - {1077978107\over 464486400}
          + {128\over 105}\,\zeta(2)
          + {130123\over 327680}\,\zeta(3)
          - {96668\over 165375}\,\ln{\mu^2\over m^2}
         \Bigg) 
,\nonumber\\
   C^{(2),p}_5 &=&
       C_F^2 \,\Bigg(
          - {9089416219983580783\over 450644705280000}
          - {4096\over 1155}\,\zeta(2)\,\ln 2
          + {512\over 231}\,\zeta(2)
\nonumber\\
&&\mbox{}
          + {213469483642711\over 12716605440}\,\zeta(3)
         \Bigg) 
       + C_F\,C_A \,\Bigg(
          - {771845002398293227\over 737418608640000}
\nonumber\\
&&\mbox{}
          + {2048\over 1155}\,\zeta(2)\,\ln 2
          - {2048\over 3465}\,\zeta(2)
          + {66603161317883\over 76299632640}\,\zeta(3)
          + {6070352\over 4465125}\,\ln{\mu^2\over m^2}
         \Bigg) 
\nonumber\\
&&\mbox{}
       + C_F\,T\,n_l \,\Bigg(
          - {6409976752\over 15471658125}
          - {2048\over 3465}\,\zeta(2)
          - {24281408\over 49116375}\,\ln{\mu^2\over m^2}
         \Bigg) 
\nonumber\\
&&\mbox{}
       + C_F\,T \,\Bigg(
          - {329953898617\over 131383296000}
          + {4096\over 3465}\,\zeta(2)
          + {2222003\over 3932160}\,\zeta(3)
\nonumber\\
&&\mbox{}
          - {24281408\over 49116375}\,\ln{\mu^2\over m^2}
         \Bigg) 
,\nonumber\\
   C^{(2),p}_6 &=&
       C_F^2 \,\Bigg(
          - {372359772998064628281949\over 3314169918259200000}
          - {10240\over 3003}\,\zeta(2)\,\ln 2
          + {6400\over 3003}\,\zeta(2)
\nonumber\\
&&\mbox{}
          + {618116373887820433\over 6612634828800}\,\zeta(3)
         \Bigg) 
       + C_F\,C_A \,\Bigg(
          - {107211161626223001664831\over 24983742460723200000}
\nonumber\\
&&\mbox{}
          + {5120\over 3003}\,\zeta(2)\,\ln 2
          - {5120\over 9009}\,\zeta(2)
          + {115687685688677\over 32388415488}\,\zeta(3)
          + {1050842288\over 893918025}\,\ln{\mu^2\over m^2}
         \Bigg) 
\nonumber\\
&&\mbox{}
       + C_F\,T\,n_l \,\Bigg(
          - {12132112100624\over 34071685522875}
          - {5120\over 9009}\,\zeta(2)
\nonumber\\
&&\mbox{}
          - {4203369152\over 9833098275}\,\ln{\mu^2\over m^2}
         \Bigg) 
       + C_F\,T \,\Bigg(
          - {2290986762786311\over 852128169984000}
          + {10240\over 9009}\,\zeta(2)
\nonumber\\
&&\mbox{}
          + {9445897\over 12582912}\,\zeta(3)
          - {4203369152\over 9833098275}\,\ln{\mu^2\over m^2}
         \Bigg) 
,\nonumber\\
   C^{(2),p}_7 &=&
       C_F^2 \,\Bigg(
          - {221521574638803295862282113747\over 371558217875875430400000}
          - {16384\over 5005}\,\zeta(2)\,\ln 2
\nonumber\\
&&\mbox{}
          + {2048\over 1001}\,\zeta(2)
          + {1729993541168029561\over 3487983206400}\,\zeta(3)
         \Bigg) 
\nonumber\\
&&\mbox{}
       + C_F\,C_A \,\Bigg(
          - {15338320757467109893990945403\over 878228514979341926400000}
          + {8192\over 5005}\,\zeta(2)\,\ln 2
\nonumber\\
&&\mbox{}
          - {8192\over 15015}\,\zeta(2)
          + {263558316028764511\over 18137512673280}\,\zeta(3)
          + {445310672\over 430404975}\,\ln{\mu^2\over m^2}
         \Bigg) 
\nonumber\\
&&\mbox{}
       + C_F\,T\,n_l \,\Bigg(
          - {64850722258864\over 213263513087625}
          - {8192\over 15015}\,\zeta(2)
          - {1781242688\over 4734454725}\,\ln{\mu^2\over m^2}
         \Bigg) 
\nonumber\\
&&\mbox{}
       + C_F\,T \,\Bigg(
          - {190892441981633663\over 66655359074304000}
          + {16384\over 15015}\,\zeta(2)
\nonumber\\
&&\mbox{}
          + {253247865\over 268435456}\,\zeta(3)
          - {1781242688\over 4734454725}\,\ln{\mu^2\over m^2}
         \Bigg) 
,\nonumber\\
   C^{(2),p}_8 &=&
       C_F^2 \,\Bigg(
          - {1018252563630160440365157797976011\over 
            333671075151516323020800000}
          - {114688\over 36465}\,\zeta(2)\,\ln 2
\nonumber\\
&&\mbox{}
          + {14336\over 7293}\,\zeta(2)
          + {214705361130392874134587\over 84572630522265600}\,\zeta(3)
         \Bigg) 
\nonumber\\
&&\mbox{}
       + C_F\,C_A \,\Bigg(
          - {2366402466662694875682083064373\over 33429013094957108428800000}
          + {57344\over 36465}\,\zeta(2)\,\ln 2
\nonumber\\
&&\mbox{}
          - {57344\over 109395}\,\zeta(2)
          + {1106800878920761371869\over 18793917893836800}\,\zeta(3)
          + {78196015136\over 84666807225}\,\ln{\mu^2\over m^2}
         \Bigg) 
\nonumber\\
&&\mbox{}
       + C_F\,T\,n_l \,\Bigg(
          - {10872485544378464\over 41951979645951375}
          - {57344\over 109395}\,\zeta(2)
          - {312784060544\over 931334879475}\,\ln{\mu^2\over m^2}
         \Bigg) 
\nonumber\\
&&\mbox{}
       + C_F\,T \,\Bigg(
          - {4465014818406761701847\over 1468550871125065728000}
          + {114688\over 109395}\,\zeta(2)
\nonumber\\
&&\mbox{}
          + {132010901659\over 115964116992}\,\zeta(3)
          - {312784060544\over 931334879475}\,\ln{\mu^2\over m^2}
         \Bigg) 
\end{eqnarray}

For the vector case the first seven moments were already presented in
\cite{CheKueSte96}. All other results are new.

\end{appendix}



\begin{thebibliography}{99}

\bibitem{CKKRep}
K.G. Chetyrkin, J.H. K\"uhn and A. Kwiatkowski,
{\it Phys. Rept.} {\bf 277} (1996) 189.

\bibitem{GorKatLar91SurSam91}
S.G. Gorishny, A.L. Kataev and S.A. Larin, 
{\it Phys. Lett.} {\bf B 259} (1991) 144;\\
L.R. Surguladze and M.A. Samuel,  
{\it Phys. Rev. Lett.} {\bf 66} (1991) 560;
(E) ibid., 2416;\\
K.G. Chetyrkin, {\it Phys. Lett.} {\bf B 391} (1997) 402.

\bibitem{Che96} 
K.G. Chetyrkin, {\it Phys. Lett.} {\bf B 390} (1997) 309.

\bibitem{CheHarKueSte97}
K.G. Chetyrkin, R. Harlander, J.H. K\"uhn and M. Steinhauser,
MPI/PhT/97-012, TTP97-11, hep-ph/9704222.

\bibitem{HarSte97}
R. Harlander and M. Steinhauser,
MPI/PhT/97-013, TTP97-12, hep-ph/9704436.

\bibitem{CheKueSte96}
K.G. Chetyrkin, J.H. K\"uhn and M. Steinhauser,  
{\it Phys. Lett.} {\bf B 371} (1996) 93;
{\it Nucl. Phys.} {\bf B 482} (1996) 213.

\bibitem{CheTka81}
F.V. Tkachov, {\it Phys. Lett.} {\bf B 100} (1981) 65;\\
K.G. Chetyrkin and F.V. Tkachov, {\it Nucl. Phys.} {\bf B 192} (1981) 159.

\bibitem{axpssc}
S.G. Gorishny, A.L. Kataev and S.A. Larin
{\it Nuovo Cim.} {\bf 92A} (1986) 119;\\
S.G. Gorishny, A.L. Kataev and S.A. Larin and L.R. Surguladze,
{\it Mod. Phys. Lett.} {\bf A 5} (1990) 2703;\\
K.G. Chetyrkin and A. Kwiatkowski, {\it Z. Phys.} {\bf C 59} (1993) 525;\\
L.R. Surguladze, {\it Phys.\ Lett.} {\bf B 338} (1994) 229;\\
L.R. Surguladze, {\it Phys.\ Lett.} {\bf B 341} (1994) 60;\\
K.G. Chetyrkin, C.A. Dominguez, D. Pirjol and K. Schilcher;
{\it Phys. Rev.} {\bf D 51} (1995) 5090;\\
K.G. Chetyrkin and A. Kwiatkowski, {\it Nucl. Phys.} {\bf B 461} (1996) 3;\\
L.R. Surguladze, {\it Phys. Rev.} {\bf D 54} (1996) 2118;\\
K.G. Chetyrkin and J.H. K\"uhn, MPI/PhT/96-084, hep-ph/9609202.

\bibitem{VerFORM}
J.A.M. Vermaseren, {\it Symbolic Manipulation with FORM},
(Computer Algebra Netherlands, Amsterdam, 1991).

\bibitem{Bro92}
D.J. Broadhurst, {\it Z. Phys.} {\bf C 54} (1992) 54.


\bibitem{JerLaeZer82}
J. Jers\'ak, E. Laermann and P. Zerwas,
{\it Phys. Rev.} {\bf D 25} (1982) 1218;\\
L.J. Reinders, H.  Rubinstein and S. Yazaki,
{\it Nucl. Phys.} {\bf B 186} (1981) 109.

\bibitem{DreHik90}
M. Drees and K. Hikasa, {\it Phys. Lett.} {\bf B 240} (1990) 455;
(E) ibid. {\bf B 262} (1991) 497.

\bibitem{HoaKueTeu95}
A.H. Hoang, J.H. K\"uhn and T. Teubner, 
{\it Nucl. Phys.} {\bf B 452} (1995) 173.

\bibitem{FadKho91}
L.D. Landau and E.M. Lifschitz, 
{\it Lehrbuch der Theoretischen Physik III, Quantenmechanik},
\S 36, Eq. (24);\\
A. Messiah, {\it Quantenmechanik 1}, Eq. (B32/2);\\
V.S. Fadin and V.A. Khoze, {\it Yad. Fiz.} {\bf 53} (1991) 1118.

\bibitem{HoaTeu96}
A.H. Hoang and T. Teubner, {\it private communication.}

\bibitem{Mel96}
K. Melnikov, {\it Phys. Rev.} {\bf D 53} (1996) 5020.

\bibitem{CheHoaKueSteTeu96}
K.G. Chetyrkin, A.H. Hoang, J.H. K\"uhn, M. Steinhauser and T. Teubner,
{\it Phys. Lett.} {\bf B 384} (1996) 233.

\bibitem{BaiBro95}
P.A. Baikov and D.J. Broadhurst, 
presented at the 
{\it 4th International Workshop on Software Engineering and Artificial
Intelligence for High Energy and Nuclear Physics (AIHENP95), 
Pisa, Italy, 3-8 April 1995}. 
Published in Pisa AIHENP (1995) 167.

\bibitem{pi1v}
G. K\"all\'en and A. Sabry, {\it K. Dan. Vidensk. Selsk. Mat.-Fys. Medd.} 
{\bf 29} (1955) No. 17;\\ 
R. Barbieri and E. Remiddi, 
{\it Nuovo Cim.} {\bf 13A} (1973) 99;\\
B.A. Kniehl, {\it Nucl. Phys.} {\bf B 347} (1990) 65;\\
D.J. Broadhurst, J. Fleischer and O.V. Tarasov, 
{\it Z. Phys.} {\bf C 60} (1993) 287.

\bibitem{pi1p}
A. Djouadi and P. Gambino,
{\it Phys. Rev.} {\bf D 51} (1995) 218; 
(E) ibid. {\bf D 53} (1996) 4111;\\
see also: D.J. Broadhurst, {\it Phys. Lett.} {\bf B 101} (1981) 423.

\bibitem{CheKue94}
K.G. Chetyrkin and J.H. K\"uhn, {\it Nucl. Phys.} {\bf B 432} (1994) 337.

\end{thebibliography}
\end{document}